\documentclass[aps,prc,tightenlines,showpacs,superscriptaddress,nofootinbib]{revtex4-1}
\usepackage{amsmath,amssymb,amsfonts}
\usepackage[dvips]{graphicx,color}
\usepackage{dcolumn}
\usepackage{bm}
\usepackage{bbm}

\usepackage{ulem}

\newcommand{\blue}[1]{#1}

\newcommand{\ket}[1]{|{#1}\rangle}
\newcommand{\bra}[1]{\langle{#1}|}

\newcommand{\slas}[1]{\not\!{#1}}

\newcommand{\unitfour}{\mathbbm{1}_{4}}

\newcommand{\Slash}[1]{\ooalign{\hfil/\hfil\crcr$#1$}}
\def\gtap{\ \raise.3ex\hbox{$>$\kern-.75em\lower1ex\hbox{$\sim$}}\ }
\def\ltap{\ \raise.3ex\hbox{$<$\kern-.75em\lower1ex\hbox{$\sim$}}\ }

\begin{document}


\title{Dynamical coupled-channels model for neutrino-induced meson
productions in resonance region}
\author{S. X. Nakamura}
\affiliation{Department of Physics, Osaka University, Toyonaka, Osaka 560-0043, Japan}
\author{H. Kamano}
\affiliation{Research Center for Nuclear Physics, Osaka University, Ibaraki, Osaka 567-0047, Japan}
\author{T. Sato}
\affiliation{Department of Physics, Osaka University, Toyonaka, Osaka 560-0043, Japan}

\begin{abstract}
A dynamical coupled-channels (DCC) model for neutrino-nucleon
reactions in the resonance region is developed.
Starting from the DCC model that we have previously developed through an
analysis of $\pi N, \gamma N\to \pi N, \eta N, K\Lambda, K\Sigma$
reaction data for $W\le 2.1$~GeV, we extend
the model of the vector current
to $Q^2\le$ 3.0~(GeV/$c$)$^2$ by analyzing electron-induced reaction
data for both proton and neutron targets.
We derive axial-current matrix elements that are related to
the $\pi N$ interactions of the DCC model
through the Partially Conserved Axial Current (PCAC)
relation.
Consequently, the interference pattern between resonant
and non-resonant amplitudes is uniquely determined.
We calculate cross sections for neutrino-induced meson productions, and
compare them with available data.
Our result for the single-pion production reasonably agrees with the data.
We also make a comparison with the double-pion production data.
Our model is the first DCC model that
can give the double-pion production cross sections in the resonance region.
We also make comparison of our result with other existing models to
reveal an 
importance of testing the models
in the light of PCAC and electron reaction data.
The DCC model developed here will be a useful input for constructing a
 neutrino-nucleus reaction model and a neutrino event generator
for analyses of neutrino experiments.
\end{abstract}


\maketitle

\section{Introduction}

An experimental observation of
the leptonic CP violation and 
a determination of
the neutrino mass hierarchy will be central
issues in forthcoming next-generation neutrino oscillation experiments.
Recent findings of relatively large
$\theta_{13}$~\cite{t2k,dayabay,reno,dchooz} have boosted the
momentum of the neutrino physics community towards addressing these issues.
For a success of the next-generation experiments,
a more precise interpretation of data will be necessary.
This means that
more precise knowledge of neutrino-nucleus reactions is critically important
because neutrinos are detected in the experiments through observing
remnants of the neutrino-nucleus reactions.

Neutrino experiments
utilize neutrinos in a wide energy range, 
and therefore the relevant neutrino-nucleus reactions have various 
microscopic reaction mechanisms depending on the kinematics.
For a relatively low-energy neutrino ($E_\nu\ltap$1~GeV) relevant to,
e.g., the T2K~\cite{t2k-2}, MiniBooNE~\cite{miniboone-2}, and
nuPRISM~\cite{nuprism} experiments, 
the dominant reaction mechanisms are the quasi-elastic (QE) knockout of a
nucleon, and quasi-free excitation of the $\Delta(1232)$
resonance followed by a decay into a $\pi N$ final state. 
For a higher-energy neutrino ($2\ltap E_\nu\ltap$4~GeV) relevant to,
e.g., the Miner$\nu$a~\cite{minerva-2} and future DUNE~\cite{dune} experiments, 
a large portion of data are from 
higher resonance excitations and deep inelastic scattering (DIS).
In order to understand the neutrino-nucleus reactions of these
different characteristics, obviously, it is important to combine different
expertise.
For example, nuclear theorists and neutrino experimentalists recently organized a
collaboration at the J-PARC branch of the KEK theory
center~\cite{collab,unified} to tackle this challenging problem.

In this work,
we focus on studying the neutrino reactions in the resonance region where
the total hadronic energy $W$ extends, $m_N+m_\pi < W \ltap 2$~GeV; 
$m_N$ ($m_\pi$) is the nucleon (pion) mass.
Furthermore, 
as a step toward developing a neutrino-nucleus reaction model,
we will be concerned with the neutrino reaction on a single nucleon. 
In the resonance region, 
particularly between the $\Delta(1232)$ and DIS regions,
we are still in the stage of developing a
single nucleon model that is a basic ingredient to construct a
neutrino-nucleus reaction model. 

First we discuss experimental data that are crucial to
determine form factors associated with axial $N$-$N^*$ ($N^*$: nucleon resonance) transitions.
Because of small cross sections, neutrino data are rather scarce. 
Available data are from old bubble chamber experiments at the Argonne
National Laboratory (ANL)~\cite{anl} and
the Brookhaven National Laboratory (BNL)~\cite{bnl};
hydrogen and deuterium targets were used in the experiments.
The single pion production data for $E_\nu\ltap$2~GeV are particularly
useful,
and theoretical models are confronted with the data to fix (or test) the
strength of the predominant axial $N$-$\Delta(1232)$ transition.
However, there has been the well-known discrepancy between the two
datasets from ANL and BNL by $\sim$10\%.
This discrepancy is reflected in the uncertainty of the axial $N\Delta$
coupling, leading to theoretical uncertainty for neutrino-nucleus
reaction cross sections.
Regarding this, an interesting progress has been reported recently in
Ref.~\cite{reanalysis}.
In the reference, the authors tried to avoid the neutrino
flux uncertainty of the old bubble chamber experiments.
They took advantage of the fact that the ratio
$\sigma(1\pi)/\sigma(0 \pi)$ is fairly unaffected by the neutrino flux
uncertainty, and that $\sigma(0 \pi)$ on the deuterium is relatively
well understood.
Here, we denoted 
the cross section for the single-$\pi$ production
[no $\pi$-emission process, mainly QE] by 
$\sigma(1\pi)$ [$\sigma(0 \pi)$].
Multiplying 
$\sigma(1\pi)/\sigma(0 \pi)$ from the two experiments by 
$\sigma(0 \pi)$ from the GENIE 2.8~\cite{genie},
they obtained $\sigma(1\pi)$ for the two experiments.
They found that the newly obtained $\sigma(1\pi)$
are both fairly close to the original ANL data.
Once this result is established, theoretical uncertainty associated with
the strength of the axial $N\Delta$ coupling will be significantly
reduced.

Another interesting analysis relevant to ANL and BNL data has been
conducted by one of the present authors
and his coworkers in Ref.~\cite{wsl}.
They examined effects of the final state interactions (FSI) 
on cross sections for the single pion production off the deuteron.
They found that the orthogonality between the deuteron and final $pn$
scattering wave functions significantly reduces the cross sections.
Thus the ANL and BNL data from deuterium target would need more careful analysis
with the FSI taken into account.
While this kind of reanalysis is important, still the available
data are rather scarce. 
It is highly desirable to have new data that are more precise and abundant.
There is an idea~\cite{wilking} to upgrade the near detector of the T2K experiment
to use D$_2$O as a target and study this important elementary process.

Regarding theoretical models, several models have been developed
for neutrino-nucleon reactions in the resonance region;
particularly
the $\Delta(1232)$ region has been extensively studied because of
its importance.
Those models can be categorized into three classes according to their
dynamical contents. 
The first class gives amplitudes as a sum of
Breit-Wigner functions that represent resonant contributions.
An example of this type developed recently is found in
Ref.~\cite{lalakulich}, and they considered
$\Delta(1232)~3/2^+$,
$N(1535)~1/2^-$, $N(1440)~1/2^+$ and $N(1520)~3/2^-$ resonances.
The second class of models considers tree-level non-resonant mechanisms
in addition to resonant ones of the Breit-Wigner type.
For example,
the authors of
Refs.~\cite{valencia1,valencia2,indiana1,indiana2} derived tree-level
non-resonant mechanisms from a chiral Lagrangian, and combined them with 
$\Delta(1232)$ of the Breit-Wigner type. 
The model of Ref.~\cite{valencia2} has been 
further extended by including the $N(1520)~3/2^-$
resonance~\cite{valencia3}.
Meanwhile, the authors of Ref.~\cite{giessen} developed a model that
contains all 4-star resonances with masses below 1.8~GeV, and included
rather phenomenological non-resonant contributions.
A model of the third class additionally takes account of the hadronic rescattering,
so that the unitarity of the amplitude is satisfied.
One of the present authors has developed such a model that works for the
$\Delta(1232)$ region~\cite{sul,msl}.
A more complete list of existing models for the resonance region can be found,
e.g., in Ref.~\cite{AHN}.

Although substantial efforts have been made recently as seen above, 
there still remain conceptual and/or practical problems in the existing
models as follows:
\begin{enumerate}
\item
We point out that 
reactions in the resonance region are
multi-channel processes in nature.
The relevant channels such as 
$\pi N, \pi\pi N, \eta N, K\Lambda, K\Sigma$ 
are strongly coupled with each other through the strong interaction.
Therefore, the multi-channel couplings required by the unitarity
are essential physics to be considered.
However, no existing model takes account of this.

\item
The neutrino-induced double pion production over the entire resonance
     region has not been seriously studied previously.
As known from the photo- and electron-reactions, cross sections for the
double-pion production are
comparable or even larger than those for the single-pion production
around and beyond the second resonance region, and a similar tendency is
expected for the neutrino reactions.
Therefore, it is important to have a good estimate for the double-pion
production rate.
Although some neutrino-induced double-pion production models have been
developed previously~\cite{biswas,adjei,spain-pipin}, they are supposed to work
for the threshold region only, a very limited kinematics in the whole resonance region.
Thus for a practical calculation of neutrino reactions, 
for example, the authors of Ref.~\cite{LGM} extrapolated a DIS model to
the resonance region, and gradually switched on its contribution for
$W>1.6$~GeV, thereby simulating the double-pion contributions.
This is not a well-justified procedure, and obviously the situation for
the double-pion production models is unsatisfactory.

\item
Interference between resonant and non-resonant amplitudes are not 
well under control for the axial-current in most of the
previous models.
This is due to the fact that the axial-current was not constructed in a
manner consistent with the $\pi N$ interaction model. 
More detailed discussion on this will be given later in Sec.~\ref{sec:axial-nstar}.
\end{enumerate}

Our goal here is to develop a neutrino-nucleon reaction model in the
resonance region by overcoming the problems discussed above.
In order to do so, the best available option would be to work with a
coupled-channels model.
In the last few years, we have developed a dynamical coupled-channels (DCC)
model to analyze $\pi N, \gamma N\to \pi N, \eta N, K\Lambda, K\Sigma$
reaction data for a study of the baryon spectroscopy~\cite{knls13}.
In there, we have shown that 
the model is successful in giving
a reasonable fit to a large amount ($\sim$ 23,000 data points) of the data.
The model also has been shown to give a reasonable prediction for
the pion-induced double pion productions~\cite{kamano-pipin}.
Thus the DCC model seems a promising starting point for developing a
neutrino-reaction model in the resonance region.
At $Q^2=0$, we already have made an extension of the DCC model to the
neutrino sector by invoking
the PCAC (Partially Conserved Axial Current) hypothesis~\cite{knls12}.
At this particular kinematics, the cross section is given by the
divergence of the axial-current amplitude that is related to the $\pi N$ amplitude
through the PCAC relation.
However, for describing the neutrino reactions in the whole kinematical
region ($Q^2\ne 0$), a dynamical model for the vector- and
axial-currents is needed.

Here are what we need to do in practice for extending the DCC model to
cover the neutrino reactions.
Regarding the vector current, we already have fixed the amplitude for the
proton target at $Q^2$=0 in our previous analysis~\cite{knls13}.
The remaining task is to determine the $Q^2$-dependence of the vector
couplings, i.e., form factors.
This can be done by analyzing data for the single pion
electroproduction and inclusive electron scattering.
A similar analysis also needs to be done with the neutron target model.
By combining the vector current amplitudes for the proton and neutron targets, we
can do the isospin separation of the vector current.
This is a necessary step before calculating neutrino processes.
As for the axial-current matrix elements at $Q^2$=0,
we derive them so that the consistency, required by the PCAC relation,
with the DCC $\pi N$ interaction model is maintained.
As a result of this derivation, 
the interference pattern between the resonant and non-resonant
amplitudes are uniquely fixed within our model;
this is an advantage of our approach.
For the $Q^2$-dependence of the axial-current matrix elements,
we still inevitably need to employ a simple ansatz due to
the lack of experimental information.
This is a limitation shared by all the existing neutrino-reaction models
in the resonance region.

With the vector- and axial-current models as described above, 
we calculate the neutrino-induced meson productions in the resonance
region.
We compare our numerical results with available data for
single-pion, double-pion, and Kaon productions.
Particularly, comparison with the double-pion production data is
made for the first time with the relevant resonance contributions 
and coupled-channels effects taken into account;
the previous double-pion production models did not include
resonant contributions~\cite{biswas,adjei},
or include a $N(1440)~1/2^+$ resonance contribution only~\cite{spain-pipin}.
Also, a comparison with other models would be interesting. 
Thus we will compare structure functions from the DCC model with those
from the models of Refs.~\cite{lalakulich,RS,RS2}.

The organization of this paper is as follows:
In Sec.~\ref{sec:xsc}, we present cross section formulae for the
neutrino-induced meson productions.
Then in Sec.~\ref{sec:dcc}, we 
start with an overview of the DCC model, and 
discuss the hadronic vector- and
axial-current amplitudes of the DCC model.
We present our analysis of electron-induced reaction data in
Sec.~\ref{sec:electron}.
Then we present 
numerical results for the neutrino reactions in
Sec.~\ref{sec:results}, followed by a conclusion in
Sec.~\ref{sec:conclusion}.
Mathematical expressions for 
matrix elements of
the non-resonant axial-current, 
resonant axial-current, 
and resonant vector-current 
are collected in Appendices~\ref{app:axial}, \ref{app:axial-nstar},
 and \ref{app:vector-nstar}, respectively.
We also tabulate parameters associated with the vector $N$-$N^*$ form
factors in Appendix~\ref{app:cn}.

\section{Cross section formulae}
\label{sec:xsc}


The weak interaction Lagrangian for charged-current (CC) processes is given by
\begin{equation}
   {\cal L}^{\rm CC} =
{G_F V_{ud}\over \sqrt{2}}\int d^3x [ J^{\rm CC}_\mu(x) l^{\rm CC\, \mu}(x) + {\rm h.c.} ]
\ ,
\label{eq:weak_lag}
\end{equation}
where $G_F=1.16637 \times 10^{-5}$ (GeV)$^{-2}$ and $V_{ud}=0.974$~\cite{pdg}.
The leptonic current is denoted by $l^{\rm CC}_\mu$, and is explicitly written as
\begin{eqnarray}
l^{\rm CC}_{\mu}(x)&=&\bar{\psi}_l(x)\gamma_\mu(1-\gamma_5)\psi_{\nu}(x) 
\ .
\end{eqnarray}
The hadronic current is
\begin{eqnarray}
J^{\rm CC}_\mu(x)=V^{+}_\mu(x) - A^{+}_\mu(x) \ ,
\label{eq:J}
\end{eqnarray}
where $V^{+}_\mu$ and $A^{+}_\mu$ are the vector and axial currents.
The superscript $+$ denotes the isospin raising operator.
For neutral-current (NC) processes, the Lagrangian is given by 
\begin{equation}
   {\cal L}^{{\rm NC}} =
{G_F\over \sqrt{2}} \int d^3 x J^{{\rm NC}}_\mu(x) l^{{\rm NC}\, \mu}(x) \ ,
\label{eq:weak_lag_nc}
\end{equation}
with the hadronic and the leptonic currents given by,
\begin{eqnarray}
\label{eq:Jnc}
J^{{\rm NC}}_\mu(x)&=&(1-\sin^2\theta_W)V^3_\mu(x) - \sin^2\theta_W V^s_\mu(x) - A^3_\mu(x) \ ,\\
l^{{\rm NC}}_{\mu}(x)&=&\bar{\psi}_\nu(x)\gamma_\mu(1-\gamma_5)\psi_{\nu}(x)
 \ ,
\end{eqnarray}
where $V^s_\mu$ is the isoscalar current, and $\sin^2\theta_W=0.23$~\cite{pdg}.


We are concerned with a meson production reaction in neutrino-nucleon scattering
$\nu(p_\nu) + N(p_N) \rightarrow l(p_l) + f(p_f)$,
where the variables in the parentheses are four-momenta of the
corresponding particles.
A charged lepton (neutrino) is denoted by $l$ for CC (NC) reaction.
A hadronic final state  ($f$) in our reaction model is one of two-body meson-baryon states
($\pi N,  \eta N, K\Lambda, K\Sigma$) or three-body $\pi\pi N$ states.
The double differential cross section with respect to the final lepton
distribution 
in the laboratory frame 
is given with a lepton tensor, $L^{\mu\nu}$, and a hadron tensor,
$W^{{\rm Y},N\to f}_{\mu\nu}$, as
\begin{eqnarray}
\frac{d^3\sigma_{\nu N \rightarrow l f}}{dE_l d\Omega_l}
 =  \frac{G_F^2 C_{\rm Y}}{4\pi^2} \frac{|\bm{p}_l|}{|\bm{p}_\nu|}
 L^{\mu\nu}
W^{{\rm Y},N\to f}_{\mu\nu} 
\ ,
\label{eq:dcross}
\end{eqnarray}
where Y=CC$\nu$, CC$\bar\nu$, NC for 
the neutrino CC, antineutrino CC, and NC reactions;
$C_{\rm Y}=V_{ud}^2$ for Y=CC$\nu$, CC$\bar\nu$ and 
$C_{\rm Y}=1$ for Y=NC.
We have also used $E_a$ as a notation for
the on-shell energy of a particle $a$.
The energy is related to the particle mass
($m_a$) by $E_a(p_a) =\sqrt{m_a^2 + \bm{p}_a^2}$. 
The leptonic tensor is
\begin{eqnarray}
L^{\mu\nu} 
&=& p^\mu_l p^\nu_\nu +p^{\nu}_l p^{\mu}_\nu
  -g^{\mu\nu}(p_\nu\cdot p_l)
 \pm i \epsilon^{\alpha\beta\mu\nu}p_{\nu,\alpha} p_{l,\beta} \ ,
 \label{lmunu}
\end{eqnarray}
where $\epsilon^{0123}= +1$ and $+(-)$ in the last term is for neutrino
(anti-neutrino) reactions.
The hadron tensor is given by
\begin{eqnarray}
W^{{\rm Y},N\to f}_{\mu\nu} & = &
 \sum_{s_f^z,p_f} \frac{1}{2}\sum_{s_N^z}(2\pi)^3 \frac{E_N}{m_N} \delta^{(4)}(p_N + q - p_f)
\langle f^{(-)}|J^{\rm Y}_\mu(0)|N\rangle \langle f^{(-)}|J^{\rm Y}_\nu(0)|N\rangle^*
\ .
\label{eq:hadron-tensor}
\end{eqnarray}
Here $\sum_{s_f^z,p_f}$ denotes the summation (integral) over the spin
(momenta) states of the final hadrons.
We used the momentum transfer $q^\mu$ defined by $q= (\omega, \bm{q})=p_\nu - p_l$.
The state vector $|N\rangle =|N(p_N,s_N^z,t_N^z)\rangle$ denotes
the initial nucleon state with momentum $p_N$ 
and the $z$-components of 
the nucleon spin ($s_N^z$) and isospin ($t_N^z$).
The state vector $\langle f^{(-)}|$ stands for the scattering final state
with the incoming wave boundary condition. 
Explicit form of the matrix element of the hadron current, 
$\langle f^{(-)}|J^{\rm Y}_\mu(0)|N\rangle$, will be specified in the next section.

\begin{figure}[t]
\includegraphics[width=0.4\textheight]{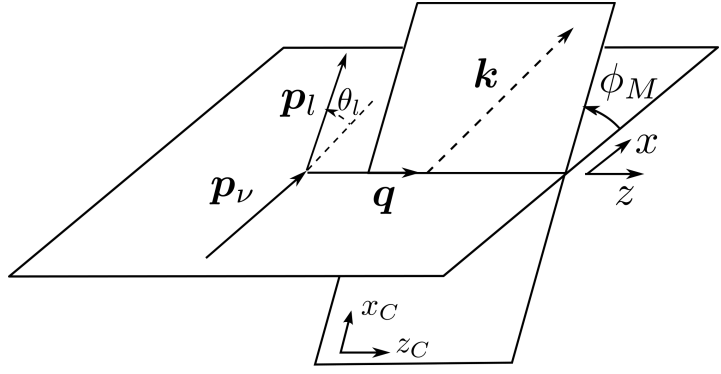}
\caption{Momentum variables and scattering angles for a neutrino-induced
 single meson production reaction.}
\label{fig:kinem}
\end{figure}

We express the hadron tensor for a two-body
meson-baryon ($MB$) final state,
$\nu(p_\nu) + N(p_N) \rightarrow l(p_l) + M(k)+ B(p)$,
in terms of the matrix element of the hadronic current in the center-of-mass
frame of the hadronic system (hCM). 
This expression is useful for working with reaction models
in which 
the matrix elements are most easily calculated in the hCM.
Also, the out-of-plane angle of the meson ($\phi_M$; see
Fig.~\ref{fig:kinem}) explicitly appears
in the cross section formula.
We choose a coordinate system such that 
the $xz$-plane coincides with
the lepton scattering plane, 
and the $z$-axis is along the momentum transfer $\bm{q}$.
Regarding the coordinate system in the hCM,  the meson is scattered in the $x_Cz_C$-plane 
and the direction of the  $z_C$-axis is chosen along  $\bm{q}$.
The angle between the lepton and hadron reaction planes is denoted by
$\phi_M$, and is shown along with our coordinate system in Fig.~\ref{fig:kinem}.
The hadron tensor for a two-body meson-baryon final state is given as
\begin{eqnarray}
W^{{\rm Y},N\to MB}_{\mu\nu}
  & = & \frac{(2\pi)^3}{2}\sum_{s_N^z,s_B^z} \int d\Omega_M^*
  \Lambda_\mu^{\ \lambda} \Lambda_\nu^{\ \sigma} 
  \frac{|{\bm{k}}^*| E_M(k^*)E_B(p^*) E_N(p_N^*) }{W m_N }
\nonumber \\
& \times & 
     \langle MB^{(-)}|J^{\rm Y}_\lambda(0)|N\rangle_{\rm hCM} \langle
     MB^{(-)}|J^{\rm Y}_\sigma(0)|N\rangle^*_{\rm hCM}
\ .
\label{eq:hadron-tensor-hcm}
\end{eqnarray}
Here the matrix element of the hadron current is evaluated
in hCM. The initial nucleon state in hCM is 
$|N\rangle =|N(p_N^*,s_N^z,t_N^z)\rangle$, while the final
meson-baryon state in hCM, 
$|MB\rangle= |M(k^*,t^z_M)B(p^*,s_B^z,t_B^z)\rangle$,
has the momentum $k^*$ ($p^*$), the isospin $z$-components $t_M^z$
($t_B^z$), and the spin $z$-component
0 ($s_B^z$) for the meson (baryon).
Here we assume the meson is a pseudo-scalar particle.
The momenta with the asterisk ($*$) are those in the hCM. 
Therefore $\bm{p}^* = - \bm{k}^*$ and $\bm{p}_N^* = - \bm{q}^*$. 
The quantity $W=\sqrt{(q+p_N)^2}$ is the invariant mass of the hadronic system.
We have also introduced the Lorentz transformation matrix,
$\Lambda^{\ \mu}_{\nu}$,
that transforms a vector in hCM to that in the laboratory frame.
Nonzero elements of $\Lambda^{\ \mu}_{\nu}$ are explicitly given as
\begin{eqnarray}
\Lambda^{\ 1}_{1}=\Lambda^{\ 2}_{2} = \cos\phi_M,\  
\Lambda^{\ 2}_{1}=-\Lambda^{\ 1}_{2} = -\sin\phi_M, \ 
\Lambda^{\ 0}_{0}=\Lambda^{\ 3}_{3} = \frac{m_N+\omega}{W}, \  
\Lambda^{\ 0}_{3}=\Lambda^{\ 3}_{0} = -\frac{|\bm{q}|}{W}.\nonumber \\
\end{eqnarray}

For a two-pion production,
$\nu(p_\nu) + N(p_N) \rightarrow l(p_l) + \pi(p_1) + \pi(p_2) + N(p_{N'})$,
a formal expression of the hadron tensor is given as
\begin{eqnarray}
W^{{\rm Y},N\to\pi\pi N}_{\mu\nu}
  & = & {\cal B}_{\pi\pi} \frac{(2\pi)^3}{2}\sum_{s_N^z,s_{N'}^z}
\int d\bm{p}_1d\bm{p}_2 d\bm{p}_{N'} \delta^{(4)}(q + p_N - p_1 - p_2 - p_{N'})\frac{E_N(p_N)}{m_N}
\nonumber \\
& &\times
\langle \pi\pi N^{(-)}| J^{\rm Y}_\mu(0)|N\rangle\langle \pi\pi N^{(-)}| J^{\rm Y}_\nu(0)|N\rangle^*
\ ,
\label{eq:W-two-pion}
\end{eqnarray}
where 
$|\pi\pi N\rangle =|\pi(p_1,t_1^z)\pi(p_2,t_2^z)N(p_{N'},s_{N'}^z,t_{N'}^z)\rangle$
is the  $\pi\pi N$ final state.  The isospin states of the pions
and the spin-isospin states of the final nucleon are denoted as 
$t_1^z,t_2^z$ and $s_{N'}^z,t_{N'}^z$, respectively.
For identical pions, we need a symmetry factor, 
${\cal B}_{\pi\pi}^{-1}=1 + \delta_{t_1^z,t_2^z}$.
An explicit formula of the hadron tensor for the two-pion final
state in our reaction model will be given in the next section.

Here we also introduce the inclusive cross section in which all the final hadronic
states are integrated over.
In the vanishing lepton mass limit, the inclusive cross section 
is given  in terms of three structure functions, 
$W^{\rm Y}_i(W,Q^2)$ 
($i$=1,2,3; ${\rm Y}$=CC$\nu$, CC$\bar\nu$, NC):
\begin{eqnarray}
\frac{d^3\sigma_{\nu N\to lX}}{dE_l d\Omega_l}
=\frac{G^2_F C_{\rm Y}}{2\pi^2}E_l^2
[2W^{\rm Y}_1 \sin^2\frac{\theta_l}{2}
+W^{\rm Y}_2 \cos^2\frac{\theta_l}{2}
\pm W^{\rm Y}_3 \frac{E_\nu+E_l}{m_N}\sin^2\frac{\theta_l}{2}] \,,
\label{eq:incl-cross2}
\end{eqnarray}
where 
$C_{\rm Y}$ has been introduced in Eq.~(\ref{eq:dcross});
the $+$ ($-$) sign in front of $W^{\rm Y}_3$ is for neutrino
(anti-neutrino) reactions,
and $\theta_l$ is the lepton scattering angle.
The structure functions can be written with the inclusive hadron tensor
evaluated in hCM 
and with $Q^2=-q^2$
as follows:
\begin{eqnarray}
W^{\rm Y}_1 &=& {1\over 2} \left(W^{\rm Y}_{xx}+W^{\rm Y}_{yy}\right) \ , 
\label{eq:W1}
\\
W^{\rm Y}_2 &=& {Q^2\over |\bm{q}|^2}\left(W^{\rm Y}_1 
+ {Q^2\over |\bm{q}^*|^2} 
\left\{  W^{\rm Y}_{00} + \left( {\omega^*\over Q^2}\right)^2 W^{\rm Y}_{\mu\nu}q^{*\mu} q^{*\nu}
+ 2 {\omega^*\over Q^2} {\rm Re}\left[ W^{\rm Y}_{0\mu}\right]q^{*\mu}
\right\} 
\right)  \ , 
\label{eq:W2}
\\
W^{\rm Y}_3 &=&  - {2 m_N\over |\bm{q}|} {\rm Im}[W^{\rm Y}_{xy}]
\ ,
\label{eq:W3}
\end{eqnarray}
and the dimensionless structure functions are conventionally defined by
\begin{eqnarray}
F^{\rm Y}_1 &=& m_N W^{\rm Y}_1\ , \quad
F^{\rm Y}_2 = \omega W^{\rm Y}_2\ , \quad
F^{\rm Y}_3 = \omega W^{\rm Y}_3\ .
\label{eq:F}
\end{eqnarray}
The inclusive cross section within our DCC model is the sum of the two-body
and three-body final states:
\begin{eqnarray}
\frac{d^3\sigma_{\nu N\to lX}}{d E_l d\Omega_l}
=\sum_{MB}\frac{d^3\sigma_{\nu N \to lMB}}{d E_l d\Omega_l}
+\sum_{\pi\pi N} {d^3\sigma_{\nu N \to l \pi\pi N}\over d E_l d\Omega_l}
\ ,
\label{incl-cross}
\end{eqnarray}
where the summation of $MB$ ($\pi\pi N$) runs over all charge states of
$\pi N, \eta N, K\Lambda, K\Sigma$ ($\pi\pi N$).

Finally, we give the differential cross section with respect to $W$ and
$Q^2$ for our later purpose:
\begin{eqnarray}
\frac{d^2\sigma_{\nu N \to lf}}{dW dQ^2}
= {2\pi W\over 2 m_N |\bm{p}_\nu||\bm{p}_l|}
\frac{d^3\sigma_{\nu N \to lf}}{dE_l d\Omega_l} \ .
\end{eqnarray}

\section{Hadronic current in Dynamical Coupled-Channels Model}
\label{sec:dcc}

\subsection{Overview of Dynamical Coupled-Channels Model}
\label{sec:rescatt}

\begin{figure}
\includegraphics[width=0.6\textheight]{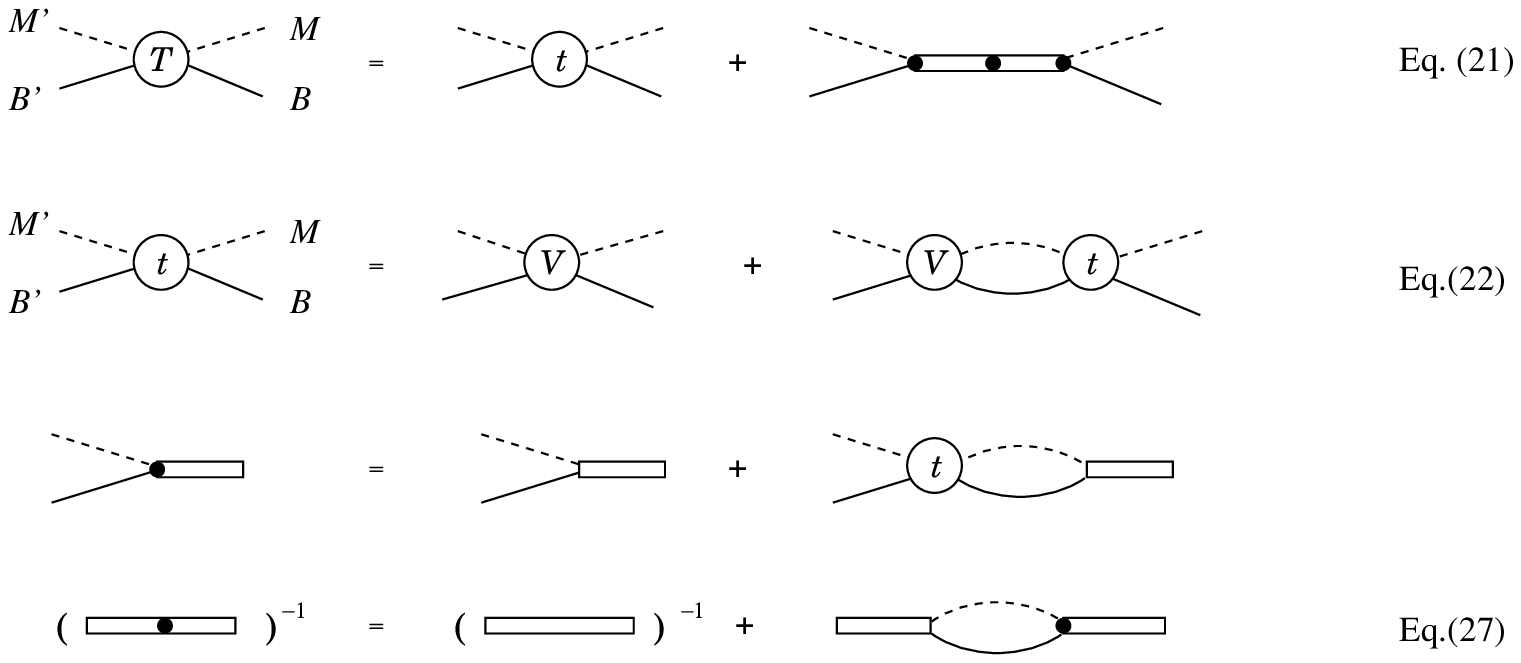}
\caption{(Color online)
Diagrammatic representation of the Lippmann-Schwinger equation. 
The corresponding equation numbers
in the text
are indicated on the right. 
The symbols $T$, $t$, and $V$ in the figures correspond to
$T^{(\pm)}(W)$ and 
$t^{(\pm)}_{\rm non-res}(W)$
in Eq.~(\ref{eq:t-matrix}), and 
$V^{(\pm)}(W)$ in Eq.~(\ref{eq:vpot}), respectively.
The third line represents the dressed $N^*$ decay vertex,
$\langle \chi^{(-)}|\Gamma|N^*\rangle$,
contained in Eq.~(\ref{eq:res-amp}).
The double lines represent $N^*$-propagators and 
the fourth line corresponds to Eq.~(\ref{eq:nstar-green}).
}
\label{fig:two-body}
\end{figure}

Here we present a brief and overall description of the DCC reaction
model~\cite{msl07},
which is followed by a more concrete and practical formulae
in the next subsections.
We start with an effective Hamiltonian for the meson-baryon system:
\begin{eqnarray}
H & = & H_0 + v + \Gamma  \ .
\label{eq:effectiveH}
\end{eqnarray}
The Fock space of the model consists of  
meson-baryon states ($\pi N, \eta N, K\Lambda,K\Sigma$ and $\pi\pi N$
states) and 'bare' excited states ($N^*,\Delta,\rho,\sigma$).
Here the bare $N^*$ state represents a quark core component of the
nucleon resonance; it is dressed by the meson cloud to form the
resonance.
We also include
$\pi\Delta, \rho N$ and $\sigma N$ states as
doorway states of the $\pi\pi N$ state.
The symbol $H_0$ is the free Hamiltonian of the particles, and $v$ is
non-resonant interactions among the  two-body meson-baryon states and
$\pi\pi$ states.
The non-resonant interactions are based on the
$s,t,u$-channel hadron-exchange mechanisms.
$\Gamma$ represents transitions between bare excited states
and two-body states such as $\Delta \leftrightarrow \pi N$.
The scattering amplitude ($T$-matrix element) is obtained
by solving the Lippmann-Schwinger equation 
as explained in detail in Ref.~\cite{msl07};
the Lippmann-Schwinger equation can be represented 
diagrammatically as shown in Fig.~\ref{fig:two-body}.
The two-body scattering $T$-matrix for 7 channels 
$\alpha \rightarrow \beta$
($\alpha, \beta=\pi N, \eta N, K\Lambda, K\Sigma, \pi\Delta, \rho N, \sigma N$)
is given by the sum of non-resonant and resonant $T$-matrix as
\begin{eqnarray}
\langle \beta |T^{(\pm)}(W)|\alpha\rangle & = & \langle \beta|t^{(\pm)}_{\rm non-res}(W)|\alpha\rangle 
+ \langle \beta|t^{(\pm)}_{\rm res}(W)|\alpha\rangle \ ,
\label{eq:t-matrix}
\end{eqnarray}
where the superscript $+(-)$ indicates the outgoing (incoming) boundary condition.
Each term of the above equation
corresponds to the diagram in Fig.~\ref{fig:two-body}
in the same ordering. 
The non-resonant $T$-matrix is obtained by solving the Lippmann-Schwinger equation,
\begin{eqnarray}
\langle \beta|t^{(\pm)}_{\rm non-res}(W)|\alpha\rangle & = & 
\langle \beta|
[ 
V^{(\pm)}(W) + V^{(\pm)}(W) G^{(\pm)}_{MB}(W)
t^{(\pm)}_{\rm non-res}(W) ]
|\alpha\rangle \ ,
\end{eqnarray}
where $G^{(\pm)}_{MB}(W)$ is Green function of a meson-baryon state;
this is the free Green function for the stable channels such as 
$MB=\pi N, \eta N, K\Lambda, K\Sigma$, 
while for the unstable 
$MB=\pi\Delta, \rho N,\sigma N$ channels 
the Green function contains the self-energy to account for the decay
into the $\pi\pi N$ channel.
The effective two-body interaction is the sum of
the interaction $v$ and the contribution of particle exchange 'Z-diagram'
that contains the $\pi\pi N$ intermediate state as
\begin{eqnarray}
V^{(\pm)}(W) = v + Z^{(\pm)}(W).
\label{eq:vpot}
\end{eqnarray}
Introducing a wave operator $\Omega^{(+/-)}$ 
and a scattering state $\chi^{(+/-)}$ 
for outgoing/incoming boundary condition as 
\begin{eqnarray}
\label{eq:wave-op}
\Omega^{(\pm)}(W) & = & 1 + G^{(\pm)}_{MB}(W) t^{(\pm)}_{\rm non-res}(W)\ , \\
\label{eq:chi}
|\chi_\alpha^{(\pm)}\rangle & = & \Omega^{(\pm)}|\alpha\rangle \ ,
\end{eqnarray}
the resonant $T$-matrix is given by
\begin{eqnarray}
\langle \beta|t^{(+)}_{\rm res}(W)|\alpha\rangle & = & \sum_{m,n}
\langle \chi_\beta^{(-)}|\Gamma|N^*_m\rangle
[D(W)]_{m,n}
 \langle N^*_n|\Gamma|\chi_\alpha^{(+)}\rangle \ ,
\label{eq:res-amp}
\end{eqnarray}
where the summation is taken over all the considered bare $N^*$ and
$\Delta$ states labelled by the indices, $m$ and $n$.
The $N^*$ propagator is denoted by 
$[D(W)]_{m,n}$,
and it can have nonzero off-diagonal elements through 
the rescattering as
\begin{eqnarray}
[D^{-1}(W)]_{m,n} & = & (W - m_m^0)\delta_{m,n} + 
 \langle N^*_m|\Gamma \Omega^{(+)}(W) G_{MB}^{(+)}(W) \Gamma|N^*_n\rangle
\ ,
\label{eq:nstar-green}
\end{eqnarray}
for $N^*_m$ and $N^*_n$ belonging to the same partial wave.
In the above equation, $m_m^0$ is the bare mass for $N^*_m$.
All the parameters appearing in the strong interaction Hamiltonian 
have already been determined in our previous analysis~\cite{knls13}.

Electroweak meson production reactions on a nucleon ($\gamma + N \rightarrow X$,
$e + N \rightarrow e' + X$ and $\nu + N \rightarrow l + X$) within the
DCC reaction model are described in terms of the matrix element of
the hadron current  $J^\mu$ which is, for example,  the current in Eq.~(\ref{eq:J}) 
for the CC neutrino reaction. In the DCC model, the hadron current consists of
non-resonant meson-production current and bare resonant current as
\begin{eqnarray}
J^\mu = j^\mu_{\rm non-res} + j^\mu_{\rm res} \ ,
\label{eq:jjj}
\end{eqnarray}
where we have suppressed the label ``Y''(=CC, NC) for simplicity, and we will do
the same to the symbols $J^\mu$ and $j^\mu$
 in the rest of this section.
The currents, $j^\mu_{\rm non-res}$ and $j^\mu_{\rm res}$, are the electroweak
counterparts of $v$ and $\Gamma$ in the strong interaction Hamiltonian.
Meson-baryon states produced by the current experience multiple
rescattering to form final states.
These scattering processes are described by $T$ matrix elements 
[Eq.~(\ref{eq:t-matrix})]
that satisfy the unitarity condition.
Thus the matrix element of the hadron current between nucleon and meson-baryon scattering
state 
$\langle\alpha^{(-)}|$ 
is given as
\begin{eqnarray}
\langle \alpha^{(-)} |J^\mu|N\rangle & = & \langle \chi_\alpha^{(-)}|j^\mu_{\rm non-res}|N\rangle \nonumber \\ 
& +& \sum_{m,n}\langle \chi_\alpha^{(-)}|\Gamma|N^*_m\rangle 
[D(W)]_{m,n}
\langle N^*_n|j^\mu_{\rm res} + \Gamma 
 \Omega^{(+)}(W) G^{(\pm)}_{MB} (W)
j^\mu_{\rm non-res}|N\rangle \ ,
\label{eq:current-me}
\end{eqnarray}
where the first term on the r.h.s. is purely from non-resonant dynamics
while the second term is due to excitations of nucleon resonances.
This matrix element is the input to calculate
the hadron tensor in Eq.~(\ref{eq:hadron-tensor}).
A more definite expression of the matrix element will be given in the
next subsection.
We note that not only the on-shell matrix elements but also their
off-energy-shell behavior are obtained in the DCC model.

Finally, a matrix element of the hadron current that contributes to 
the $\pi\pi N$ final state is given 
within the DCC model used in this work
as follows:
\begin{eqnarray}
\langle \pi\pi N^{(-)}|J^\mu|N\rangle 
& = & 
\langle \pi N|\Gamma|N\rangle G_{\pi N}^{(+)}(W)\langle \pi N^{(-)}|J^\mu|N\rangle \nonumber \\
& + &\langle \pi N|\Gamma|\Delta\rangle G_{\pi\Delta}^{(+)}(W)\langle \pi\Delta^{(-)}|J^\mu|N\rangle \nonumber \\
& + & \langle \pi \pi|\Gamma|\rho\rangle G_{\rho N}^{(+)}(W)\langle \rho N^{(-)}|J^\mu|N\rangle \nonumber \\
& + & \langle \pi \pi|\Gamma|\sigma\rangle G_{\sigma N}^{(+)}(W)\langle \sigma N^{(-)}|J^\mu|N\rangle 
\ ,
\label{eq:pipin-amp}
\end{eqnarray}
where \blue{$\langle MB^{(-)}|$} contains the hadronic
rescattering as has been defined in 
Eqs.~(\ref{eq:wave-op}), (\ref{eq:chi}) and (\ref{eq:current-me}), 
while
$\langle \pi N|$ and $\langle \pi \pi|$ are non-interacting states.

Through our previous analysis of the pion- and photon-induced meson production
reactions, we have already constructed a DCC model for the strong interaction and
the electromagnetic current 
of the proton at $Q^2$=0. 
Also, all the resonance parameters such as the masses and strong decay widths
have been extracted from the DCC model~\cite{knls13}.
In order to extend the model to
calculate neutrino-induced reaction cross sections,
our main task is to develop an axial-current model for $j^\mu_{\rm non-res}$ and
$j^\mu_{\rm res}$. 
We also need to determine the $Q^2$-dependence of the vector form
factors associated with $N$-$N^*$ transitions by analyzing 
both electron-proton and electron-neutron
reaction data.


\subsection{Formulae for numerical calculations}

Based on the discussion in the previous subsection, 
we here present formulae that are practically used in numerical calculations.
Particularly, we give expressions for matrix elements of the hadron
current that are directly plugged in the hadron tensor defined in Sec.~\ref{sec:xsc}.
In this subsection, all kinematical variables are those in hCM, and thus
we suppress the asterisk $({}^*)$ used in the previous section for simplicity.

Let us consider the matrix element,
defined in Eq.~(\ref{eq:current-me}),
for a meson
production off the nucleon induced by 
the hadron current $J^{\rm }_\mu$,
$\langle M(\bm k) B(-\bm k)^{(-)}|J^{\rm }_\mu (\bm{q},Q^2)| N(-\bm q) \rangle$,
where $MB$=$\pi N$, $\eta N$, $\pi\Delta$, $\rho N$, $\sigma N$,
$K\Lambda$, $K\Sigma$.
For calculating the matrix element, 
we find it convenience to
expand the matrix element in term of the
the helicity-$LSJ$ mixed representation;
$L$, $S$, and $J$ are the orbital angular momentum, total spin, and total
angular momentum, respectively, 
for the final $MB$ state.
For a detailed discussion on the $LSJ$- and mixed-representations we
employ, see Appendices~C and D in Ref.~\cite{knls13}.
Thus the matrix element of the hadron current is expressed
as follows:
\begin{eqnarray}
&&
\langle M(\bm k;s_{M}s_{M}^z t_{M}t_{M}^z) B(-\bm k;s_{B}s_{B}^z t_{B}t_{B}^z)^{(-)}
|J^{\rm }(\bm{q},Q^2)\cdot\epsilon(\lambda)
|N(-\bm q;s_{N}s_{N}^z t_{N}t_{N}^z) \rangle
\nonumber \\
&=& 
\sum_{LSJIL^z}
\sqrt{2J+1\over 4\pi}
(t_{M}t_{M}^z t_{B}t_{B}^z|II^z)
(t_{J}t_{J}^z t_{N}t_{N}^z|II^z)
(s_{M}s_{M}^z s_{B}s_{B}^z|SS^z)
\nonumber \\
&&\times
(LL^zSS^z|J s_N^z\!+\!m(\lambda))
Y_{LL^z}(\hat{k}) \, 
T_{MB,J^{\rm } N}(\lambda;k,q;W,Q^2) \ ,
\label{eq:j-expand}
\end{eqnarray}
where $\bm q$ is taken along the $z$-axis.
The spin (isospin) and its $z$-component for a particle $x$ are denoted by 
$s_x$ and $s_x^z$ ($t_x$ and $t_x^z$), respectively;
$t_J$ is the isospin of the hadron current.
The total isospin of the final $MB$ system is denoted by $I$.
The notation 
$(j_{1}j_{1}^z j_{2}j_{2}^z|j_3 j_3^z)$
stands for the Clebsch-Gordan coefficient.
Also, we denoted the polarization of the hadron current by
$\lambda=t,-1,0,1$ in the spherical basis
($t$: time component), and $\epsilon(\lambda)$ is the corresponding
polarization vector;
$m(\lambda)=0,-1,0,1$ are for
$\lambda=t,-1,0,1$, respectively.
We have introduced 
the matrix element in the helicity-$LSJ$ mixed representation denoted by
$T_{MB,J^{\rm } N}(\lambda;k,q;W,Q^2)$ in which the label $MB$ is
understood to include quantum numbers such as $L,S,J$, and $I$.
The matrix element on the l.h.s. of Eq.~(\ref{eq:j-expand}) is
$\langle MB^{(-)}|J^{{\rm }}_\lambda(0)|N\rangle_{\rm hCM}$ in
Eq.~(\ref{eq:hadron-tensor-hcm}).
Thus we are now left with evaluating the hadron current matrix elements in
the helicity-$LSJ$ mixed representation,
$T_{MB,J^{\rm } N}$,
in order to calculate neutrino cross sections using the formulae
presented in Sec.~\ref{sec:xsc}.

Following the manner in the previous subsection, we divide the matrix
element into non-resonant and resonant parts as
\begin{equation}
T_{MB,J^{\rm } N}(\lambda;k,q;W,Q^2) = t_{MB,J^{\rm } N}(\lambda;k,q;W,Q^2) 
+ t^R_{MB,J^{\rm }N}(\lambda;k,q;W,Q^2) \ ,
\label{eq:tmbgn}
\end{equation}
where $t_{MB,J^{\rm } N}$ is
non-resonant amplitude, 
corresponding to the first term in Eq.~(\ref{eq:current-me}),
and is calculated by 
\begin{eqnarray}
&&
t_{MB,J^{\rm } N} (\lambda;k,q;W,Q^2) = v_{MB, J^{\rm } N} (\lambda;k,q;Q^2)
\nonumber\\
&&+
\sum_{M'B'} \int p^2 dp \, t_{MB,M'B'}(k,p;W)\,G_{M'B'}(p;W)\,
v_{M'B', J^{\rm } N} (\lambda;p,q;Q^2),
\label{eq:nonres}
\end{eqnarray}
where $v_{MB,J^{\rm } N}$ denotes a tree-level non-resonant current
matrix element,
$\langle MB|j_{\rm non-res}\cdot\epsilon(\lambda)|N\rangle$
[$j_{\rm non-res}$ from Eq.~(\ref{eq:jjj})],
projected onto the helicity-$LSJ$ mixed representation.
We will specify $v_{MB,J^{\rm } N}$ 
in the following subsection.
Also, 
$t_{MB,M'B'}$ is the non-resonant hadronic
scattering amplitude expressed with the $LSJ$ representation, 
and $G_{M'B'}$ is a meson-baryon Green's function.
The second term in Eq.~(\ref{eq:tmbgn}) 
is the resonant amplitude,
corresponding to the second term in Eq.~(\ref{eq:current-me}),
and is given by
\begin{eqnarray}
t^R_{M B,J^{\rm } N}(\lambda;k,q; W,Q^2) 
&=&
\sum_{m,n} \bar\Gamma_{MB,N^\ast_m}(k;W) [D(W)]_{m,n}
\bar\Gamma_{N^\ast_n,J^{\rm } N}(\lambda;q;W,Q^2),
\label{eq:resamp}
\\
\bar\Gamma_{N^\ast,J^{\rm } N}(\lambda;q;W,Q^2)
 &=& 
\Gamma_{N^\ast,J^{\rm } N}(\lambda;q,Q^2) 
\nonumber\\
&+& \sum_{M'B'}\int p^2 dp\, \Gamma_{N^*,M'B'}(p)\, G_{M'B'}(p,W)\, 
t_{M'B',J^{\rm } N}(\lambda;p,q;W,Q^2)
\ ,
\nonumber\\
\label{eq:nstargn}
\end{eqnarray}
where $\Gamma_{N^*,J^{\rm } N}$ denotes a bare $N^*$-excitation
current, 
$\langle N^\ast|j_{\rm res}\cdot\epsilon(\lambda)|N\rangle$
[$j_{\rm res}$ from Eq.~(\ref{eq:jjj})],
 for which explicit expressions are given in
Appendix~\ref{app:axial-nstar} (Appendix~\ref{app:vector-nstar})
for the axial-current (vector-current).
We have also used $\Gamma_{MB,N^\ast}$, $\bar\Gamma_{MB,N^\ast}$ and $D(W)$ that are
bare, dressed $N^*\to MB$ decay amplitudes 
and a dressed $N^*$ propagator, respectively.
These quantities and $t_{MB,M'B'}$ and $G_{MB}$ in Eq.~(\ref{eq:nonres})
have been defined and well discussed in Sec.~II~A of
Ref.~\cite{knls13}, and we will not repeat it here.

Before closing this subsection, we derive the hadron tensor for
neutrino-induced double-pion productions. 
Because we have specified coupled channels considered in our DCC model,
we can manipulate Eq.~(\ref{eq:W-two-pion}) to a more definite form to
be used in actual calculations.
As we mentioned, $\pi\Delta, \rho N, \sigma N$ are the doorway states
into the $\pi\pi N$ channel in the DCC model, and they contribute to the
double pion productions as in Eq.~(\ref{eq:pipin-amp}).
In addition, we also include a mechanism of
the hadron current-induced
$N \to \pi + N$ transition
followed by a perturbative
$N\to \pi N$ process
for the double-pion productions, as in 
the first term of Eq.~(\ref{eq:pipin-amp}).
In principle, different doorway state contributions
shown on the r.h.s. of Eq.~(\ref{eq:pipin-amp}) can interfere with each other.
Also, even within the $\pi\Delta$ doorway state contribution,
there is still an interference between diagrams in which the final two
pions are interchanged.
However, we can estimate these interference contributions to be
small 
for total cross sections and $d^2\sigma/dWdQ^2$
 because of the fact that the Z-diagrams in our
currently used
 DCC model give rather
small contributions to the $\pi N$ scattering~\cite{knls13}. 
We note however that the interference could be more important for
differential cross sections with respect to the pion emission angles,
which we will not calculate in this work, 
as has been seen in Ref.~\cite{msl07}.
Thus we ignore the interference to derive a simpler
hadron tensor for the double-pion production.
For example, the contribution from the $\pi\Delta$ doorway state 
to $\pi_1\pi_2 N$ production is given by
\begin{eqnarray}
W^{{\rm Y},N\to\pi\Delta\to\pi_1\pi_2 N}_{\mu\nu}  & = &  
{\cal B}_{\pi\pi}
\sum_{\{ab\}}
\int^{W-m_\pi}_{m_\pi+m_N} dM_{\pi_b N} 
\frac{1}{2\pi}
\frac{M_{\pi_b N}}{E_\Delta(k)}
\frac{ (t_{\pi_b} t^z_{\pi_b} t_N t^z_N | t_\Delta t^z_{\pi_b}+t^z_N)^2
\Gamma_{\Delta\to \pi_b N}(k;W)
}
{|W-E_{\pi_a}(k)-E_\Delta(k)-\Sigma_{\Delta\pi_a}(k;W)|^2}
\nonumber \\
&&\times
W_{\mu\nu}^{{\rm Y},N\to\pi_a\Delta}
 ,\label{eq:tcs-23body}
\end{eqnarray}
where $M_{\pi_b N}$ is
the invariant mass of the $\pi_b N$ pair from the $\Delta$-decay,
and $k$ is given by the relation,
$W = E_{\pi_a}(k) + \sqrt{M_{\pi_b N}^2+k^{2}}$;
$\Sigma_{\pi_a\Delta}(k;W)$ is the self-energy of the $\pi_a\Delta$ Green's function
as summarized in Appendix A of Ref.~\cite{knls13};
$\Gamma_{\Delta\to \pi_b N}(k;W) = -2 \mathrm{Im}[\Sigma_{\pi_a\Delta}(k;W)]$;
$(t_{\pi_b} t^z_{\pi_b} t_N t^z_N | t_\Delta t^z_{\pi_b}+t^z_N)^2 \Gamma_{\Delta\to\pi_b N}(k;W)$
gives the partial width of $\Delta$ into a given isospin state of $\pi_b N$;
$\sum_{\{ab\}}$ indicates the sum over the permutations of the two
pions, i.e., ${\{ab\}}={\{12\}},{\{21\}}$;
the symmetry factor ${\cal B}_{\pi\pi}$ has been defined below
Eq.~(\ref{eq:W-two-pion}).
The quasi two-body hadron tensor, 
$W_{\mu\nu}^{{\rm Y},N\to\pi_a\Delta}$,
is calculated using Eq.~(\ref{eq:hadron-tensor-hcm})
similarly to the other stable two-body channels. 
A difference from the stable channel case is that $k$ determined above is used instead of the
on-shell momentum.
The hadron tensors for the double-pion productions due to the other
doorway states can be derived in a similar manner.

\subsection{Matrix elements of non-resonant currents}
\label{sec:nonres}

We here specify matrix elements of non-resonant current, 
$j^\mu_{\rm non-res}$ in Eq.~(\ref{eq:jjj}),
for meson productions.
The matrix elements are projected onto the helicity-$LSJ$ state and
plugged into
$v_{MB,J N}$ in Eq.~(\ref{eq:nonres}).
As in Eqs.~(\ref{eq:J}) and (\ref{eq:Jnc}), the current consists of the
vector and axial currents.
Matrix elements of the non-resonant vector current at $Q^2=0$
have been fixed through the previous analysis of
photon-induced meson-production data, and explicit
expressions are given in Appendix~D of Ref.~\cite{knls13}.
We also need to fix the $Q^2$-dependence of the matrix elements
to study electron- and neutrino-induced reactions, and 
we use the parametrization given in Appendix~A of Ref.~\cite{nl1990}.

Regarding the axial current, 
we take advantage of the fact that most of our $\pi N\to MB$
potentials are derived from a chiral Lagrangian.
Thus, we basically follow the way how the
axial current is introduced in the chiral Lagrangian:
an external axial current ($a^\mu_{\rm ext}$) enters into the chiral Lagrangian in
combination with the pion field as 
$\partial^\mu \pi +  f_\pi a^\mu_{\rm ext}$ where $f_\pi$ is the pion decay constant.
Therefore, matrix elements of the hadronic axial current are obtained from 
those of $\pi N \rightarrow MB$ 
by a replacement 
$k^\mu\to i f_\pi \epsilon^\mu(\lambda)$, 
where 
$k^\mu$ is the four-momentum of the incoming pion and 
$\epsilon^\mu(\lambda)$ 
is the polarization vector for $a^\mu_{\rm ext}$ 
with a polarization $\lambda$.
We apply this replacement to the $\pi N \rightarrow MB$ potentials of
the DCC model presented in Appendix~C of Ref.~\cite{knls13}. 
The tree-level axial current matrix elements
constructed in this way 
are the non pion-pole part of the axial currents, 
$\langle MB |A^{i}_{\rm NP,tree}\cdot\epsilon(\lambda)| N\rangle$
($i$: isospin component),
and their expressions
are presented in Appendix~\ref{app:axial} of this paper.
By construction, $A^{i,\mu}_{\rm NP,tree}$ and the meson-baryon potential
$v$ satisfy the PCAC relation at $Q^2=-m^2_\pi$~\cite{yamagishi}:
\begin{eqnarray}
\langle MB |q\cdot A^{i}_{\rm NP,tree} |  N\rangle
= i f_\pi \langle MB | v | \pi^i N\rangle \ .
 \label{eq:pcac}
\end{eqnarray}
The axial-current matrix element $A^{i,\mu}$ in Eqs.~(\ref{eq:J}) and (\ref{eq:Jnc})
is related to the non pion-pole part, $A^{i,\mu}_{\rm NP}$, by
\begin{eqnarray}
A^{i,\mu} = A^{i,\mu}_{\rm NP} + q^{\mu} {1\over Q^2+m^2_\pi}
q\cdot A^i_{\rm NP} \ ,
 \label{eq:pion-pole}
\end{eqnarray}
where the second term is the pion-pole term.
In the DCC model of Ref.~\cite{knls13},
we needed to introduce some meson-baryon potentials that are not from a
chiral Lagrangian in order to fit a large amount of $\pi$-induced
reaction data.
Although we cannot apply the above replacement to derive the axial currents
for those potentials, 
we can still add the corresponding axial currents 
to maintain the PCAC relation, 
as given in
Eqs.~(\ref{eq:sigma-d}), (\ref{eq:sigma-nond}), (\ref{eq:s31}), and
(\ref{eq:pi-rho-contact}). 

The $Q^2$-dependence of the axial-coupling of the nucleon 
has been studied through data analyses of
quasi-elastic neutrino scattering
and single pion electroproduction near threshold.
In the analyses, the axial mass ($M_A$) of the dipole form factor 
$F_A(Q^2)=1/(1+Q^2/M^2_A)^2$,
has been determined to be $M_A$ = $1.026 \pm 0.021$~GeV~\cite{axial-mass}.
We employ this axial form factor,
and assume that all non-resonant axial current amplitudes have the same
$Q^2$-dependence.

\subsection{Matrix elements of $N^*$-excitation currents}

Here, we specify the last piece of the DCC model,
$\Gamma_{N^\ast,J N}(\lambda;q,Q^2)$ in Eq.~(\ref{eq:nstargn}).

\subsubsection{Vector current}
\label{sec:vec-nstar}

The hadronic vector current 
contributes to the neutrino-induced reactions
in the finite $Q^2$ region.
In Ref.~\cite{knls13},
we have done a combined analysis of 
$\pi N,\gamma p\to \pi N, \eta N, K\Lambda, K\Sigma$ reaction data, 
and fixed matrix elements of the vector current at $Q^2=0$ for the proton target. 
The bare $N$-$N^*$-transition matrix elements 
induced by the vector current are parametrized and presented
in Appendix~\ref{app:vector-nstar}.
What we need to do is to extend the matrix elements of the vector current of Ref.~\cite{knls13}
to the finite $Q^2$ region for application to the neutrino reactions.
More concretely, we determine
$Q^2$-dependence of 
$\tilde M_{l\pm}^{NN^*}(Q^2)$, $\tilde E_{l\pm}^{NN^*}(Q^2)$, 
$\tilde S_{l\pm}^{NN^*}(Q^2)$,
$x_{A_{3/2}}(Q^2)$ and $x_{A_{1/2}}(Q^2)$,
which we collectively denote by $F^V_{NN^*}(Q^2)$,
for the proton and the neutron.
(See Appendix~\ref{app:vector-nstar} for the definition of the symbols.)
This can be done by analyzing 
data for electron-induced reactions on the proton and the neutron,
and analysis results will be presented in Sec.~\ref{sec:electron}.
Then we separate the vector form factors for $N^*$ of $I=1/2$ ($I$: isospin)
into isovector and isoscalar parts as discussed in Appendix~\ref{app:vector-nstar}.
Regarding $N^*$ of $I=3/2$ for which only the isovector current contributes,
we can determine the vector form factors by analyzing the proton-target data.

\subsubsection{Axial current}
\label{sec:axial-nstar}

Because of rather scarce neutrino reaction data, it is difficult to
determine 
$N$-$N^*$ transition matrix elements induced by the axial-current.
This is in sharp contrast with the situation for the vector form factors
that are well determined by a large amount of electromagnetic reaction data.
Thus, we need to take a different path to fix the axial form factors. 
The conventional practice is 
to write down 
a $N$-$N^*$ transition
matrix element induced by the axial-current
in a general form with three or four form factors.
Then the PCAC relation,
$\langle N^*|q\cdot A^i_{\rm NP}|N\rangle
=if_\pi\langle N^*|\Gamma|\pi^i N\rangle$,
is invoked
to relate the presumably most important axial form factor at $Q^2=-m^2_\pi$
to the corresponding $\pi NN^*$ coupling.
The other form factors are ignored except for the pion pole term. 
We then assume 
$A^{i,\mu}_{\rm NP}(Q^2=-m^2_\pi)\sim A^{i,\mu}_{\rm NP}(Q^2=0)$.
In the present work, we consider 
the axial currents for bare $N^*$ of
the spin-parity $1/2^\pm$, $3/2^\pm$, $5/2^\pm$ and $7/2^\pm$,
and determine their axial form factors at $Q^2=0$ using the above procedure.
For more detail including explicit formulae, see Appendix~\ref{app:axial-nstar}.
It is even more difficult to determine 
the $Q^2$-dependence of 
the axial couplings to $N$-$N^*$ transitions because of the limited
amount of data.
Thus we assume that
the $Q^2$-dependence of the axial form factors is the same as that used
for the non-resonant axial-current amplitudes, i.e., the conventional dipole form
factor with $M_A$=1.02~GeV.
When necessary, we can adjust the axial mass for an axial $N$-$N^*$ coupling
to fit available data.

It is worth emphasizing that a great advantage of our approach over
the existing models is that relative phases between resonant and
non-resonant amplitudes are made under control within the DCC model.
This is possible in our approach by constructing the axial-current amplitudes and
$\pi N$ interactions consistently with the requirement of the PCAC relation.
As we will see, the resonant and non-resonant contributions to the
neutrino reactions are
sometimes comparable in magnitude and, in such a circumstance, it is essential to have a
well-controlled relative phases between them to correctly describe the
neutrino reactions.
We also note that our DCC $\pi N$ model, on which the axial-current
is based, has been extensively tested by data in Ref.~\cite{knls13}.

\section{Analysis of electron-induced reaction data}
\label{sec:electron}

In this section, we analyze data for electron-induced reactions off the proton
and neutron targets
to determine the $Q^2$ dependence of $F^V_{NN^*}(Q^2)$,
namely $\tilde M^{NN^*}_{l\pm}(Q^2)$, $\tilde E^{NN^*}_{l\pm}(Q^2)$,
$\tilde S^{NN^*}_{l\pm}(Q^2)$,
$x_{A_{3/2}}(Q^2)$, and $x_{A_{3/2}}(Q^2)$, which are our model
parameters associated with the vector transition form factors and are
defined in Appendix~\ref{app:vector-nstar}.
As explained in Sec.~\ref{sec:vec-nstar},
for the isospin $I=1/2$ nucleon
resonances, the analysis of
electron-induced reactions on 
both the proton- and neutron-targets
is required to decompose the electromagnetic transition form factors
into the isovector and isoscalar parts.
This decomposition is 
necessary for calculating the CC and NC reactions.
Regarding $I=3/2$ nucleon resonances, on the other hand,
we determine the $F^V_{NN^*}(Q^2)$ 
by analyzing only proton-target reactions data.
The data we analyze span the kinematical region of 
$W\le$ 2~GeV and $Q^2\le$ 3~(GeV/$c$)$^2$
that is also shared by neutrino reactions for $E_\nu\le$ 2~GeV.
Meanwhile, it is a challenge for the DCC model to predict 
cross sections of various final hadron states
in the finite $Q^2$ region
by adjusting only the 'bare' $N$-$N^*$ transition form factors.
Thus the analysis of the electron-induced reaction data also serves as a
testing ground for the soundness of the DCC model.

\subsection{Electron-proton reactions}
\label{sec:electron-p}

Among data for electron-proton reactions in the resonance region, 
those for the single pion electroproductions
are the most abundant
over a wide range of $W$ and $Q^2$.
Therefore, these
are the most useful to determine
the $Q^2$ dependence of the $p$-$N^*$ transition form factors.
The cross sections for $p(e,e'\pi^0)p$  and $p(e,e'\pi^+)n$ have
different sensitivities to resonances of different isospin state
($1/2$ or $3/2$).
The angular distribution of the pion is useful to disentangle the
spin-parity of the resonances.
Based on the one-photon exchange approximation,
a standard formula of the angular distribution for
the single pion electroproduction
can be expressed in terms of virtual photon cross sections
$\displaystyle {d\sigma_{\beta}(Q^2,W,\cos\theta_\pi^*)}/{d\Omega_\pi^*}$ ($\beta=T,L,LT,TT,LT'$)
for the $\gamma^* N\to\pi N$ process in the hCM as,
\begin{eqnarray}
\frac{d^5\sigma_{ep\rightarrow e'\pi N}}{d E_{e'} d\Omega_{e'} d\Omega_\pi^*}
&=&
\Gamma_\gamma
\left[ \frac{d\sigma_T}{d\Omega_\pi^*} + \epsilon \frac{d\sigma_L}{d\Omega_\pi^*}
 +\sqrt{2\epsilon(1+\epsilon)}\frac{d\sigma_{LT}}{d\Omega_\pi^*} \cos \phi_\pi^\ast
\right.
\nonumber \\
& &
\left.
+ \epsilon \frac{d\sigma_{TT}}{d\Omega_\pi^*} \cos 2\phi_\pi^\ast
+ h_e\sqrt{2\epsilon(1-\epsilon)}
 \frac{d\sigma_{LT^\prime}}{d\Omega_\pi^*}\sin \phi_\pi^\ast
\right] \ ,
\label{eq:dcrst-em}
\end{eqnarray}
where
\begin{eqnarray}
\displaystyle \Gamma_\gamma=\frac{\alpha}{2\pi^2 Q^2}
\frac{E_{e'}}{E_e}\frac{q_\gamma}{1-\epsilon}\ ,
\end{eqnarray}
and 
$q_\gamma =  (W^2 - m_N^2) / 2m_N$,
$\epsilon  =  [1 + 2 ( q_\gamma^2/Q^2)\tan^2 (\theta_{e'}/2)]^{-1}$, 
the scattering angle of electron $\theta_{e'}$,
the magnitude of the virtual photon three momentum $q_\gamma$,
and the incident (outgoing) electron energy $E_e$ ($E_{e'}$)
in the laboratory frame;
$h_e$ is the helicity of the incoming electron;
$\phi_\pi^\ast$ ($\phi_M$ in Fig.~\ref{fig:kinem})
is the angle between
the $\pi$-$N$ plane and the plane of the incoming and outgoing electrons.
The formulae for calculating $d\sigma_{\beta}/d\Omega_\pi^*$
from the amplitudes defined by Eq.~(\ref{eq:tmbgn}) are given in Ref.~\cite{sl09}.

\begin{figure}
\includegraphics[width=0.5\textheight]{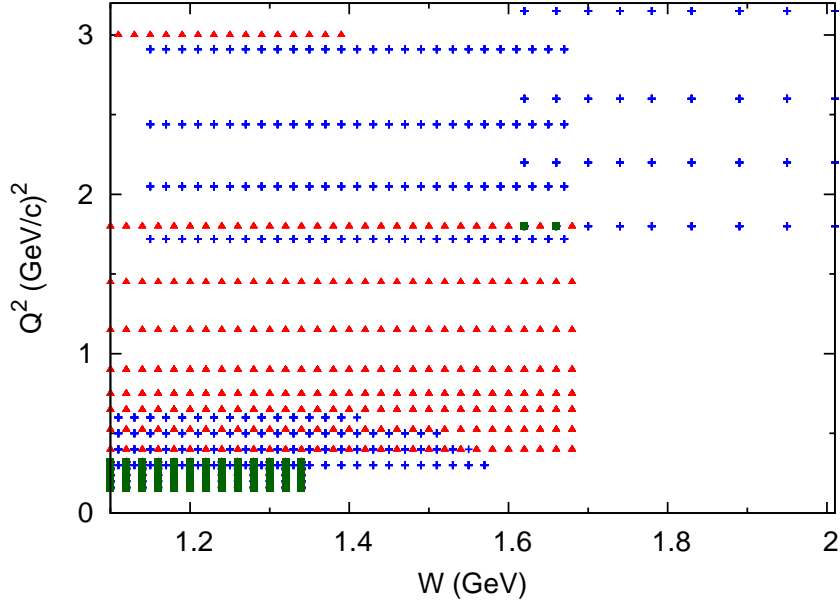}
\caption{(Color online)
Kinematical region covered by available single pion electroproduction
 data 
from the CLAS Collaboration~\cite{eepi-joo-prl,eepi-joo-prc,eepi-egiyan-prc,clas-park,clas-park2,clas-ungaro,clas-smith,clas-Aznauryan}.
The red triangle [blue cross] points indicate the kinematical points where data for
$p(e,e'\pi^0)p$ [$p(e,e'\pi^+)n$] are available. 
At the green square points, both
$p(e,e'\pi^0)p$ and $p(e,e'\pi^+)n$ data are available. }
\label{fig:ep-data}
\end{figure}

The CLAS Collaboration has collected data~\cite{eepi-joo-prl,eepi-joo-prc,eepi-egiyan-prc,clas-park,clas-park2,clas-ungaro,clas-smith,clas-Aznauryan}
for the single pion electroproduction off the proton
in the kinematical region of our interest, as shown in 
Fig.~\ref{fig:ep-data}.
Then they have extracted from the data
the virtual photon cross sections introduced above~\cite{joo-smith}, i.e.,
$d\sigma_T/d\Omega_\pi^* + \epsilon d\sigma_L/d\Omega_\pi^*$,
$d\sigma_{LT}/d\Omega_\pi^*$, $d\sigma_{TT}/d\Omega_\pi^*$, and
$d\sigma_{LT'}/d\Omega_\pi^*$.
We fit these virtual photon cross sections
to determine the $Q^2$ dependence of
the $p$-$N^*$ transition form factors.
As seen in Fig. \ref{fig:ep-data},
the single pion electroproduction data occupy a substantial portion of the relevant
kinematical region of $W$ and $Q^2$.
In some kinematical region, however, 
we still need more data to fix the vector form factors.
In particular, data are missing for the $W\gtap 1.4$~GeV and low-$Q^2$ region,
and the $W\gtap 1.7$~GeV and $Q^2\ltap 2$ (GeV/$c$)$^2$.
In those kinematical region, 
we fit the inclusive structure functions from an empirical
model due to Christy and Bosted~\cite{christy}.
The inclusive cross section is defined as 
\begin{eqnarray}
\frac{d^3\sigma_{ep\rightarrow e'X}}{d E_{e'} d\Omega_{e'}}
&=&
\Gamma_\gamma
\left[ \sigma_T(W,Q^2) + \epsilon \sigma_L(W,Q^2)
\right] \ ,
\label{eq:incl-str}
\end{eqnarray}
and the proton structure functions 
$W_i^{\rm em}$ [cf. Eqs.~(\ref{eq:W1}) and (\ref{eq:W2})]
are related 
to the transverse and longitudinal cross sections as
\begin{eqnarray}
W_1^{\rm em} & = & \frac{q_\gamma}{4\pi^2 \alpha}\sigma_T(W,Q^2)\ , \nonumber \\
W_2^{\rm em} & = & \frac{q_\gamma}{4\pi^2 \alpha}\frac{Q^2}{Q^2 + \omega^2}[\sigma_T(W,Q^2)+\sigma_L(W,Q^2)]
\ .
\end{eqnarray}

We remark that
an analysis of electron reaction data is also interesting in the context
of studying the structure of nucleon resonances, and
we are conducting a fuller and more dedicated analysis of
electroproduction data to be reported elsewhere~\cite{knls-e}.

\begin{figure}[t]
\includegraphics[width=78mm]{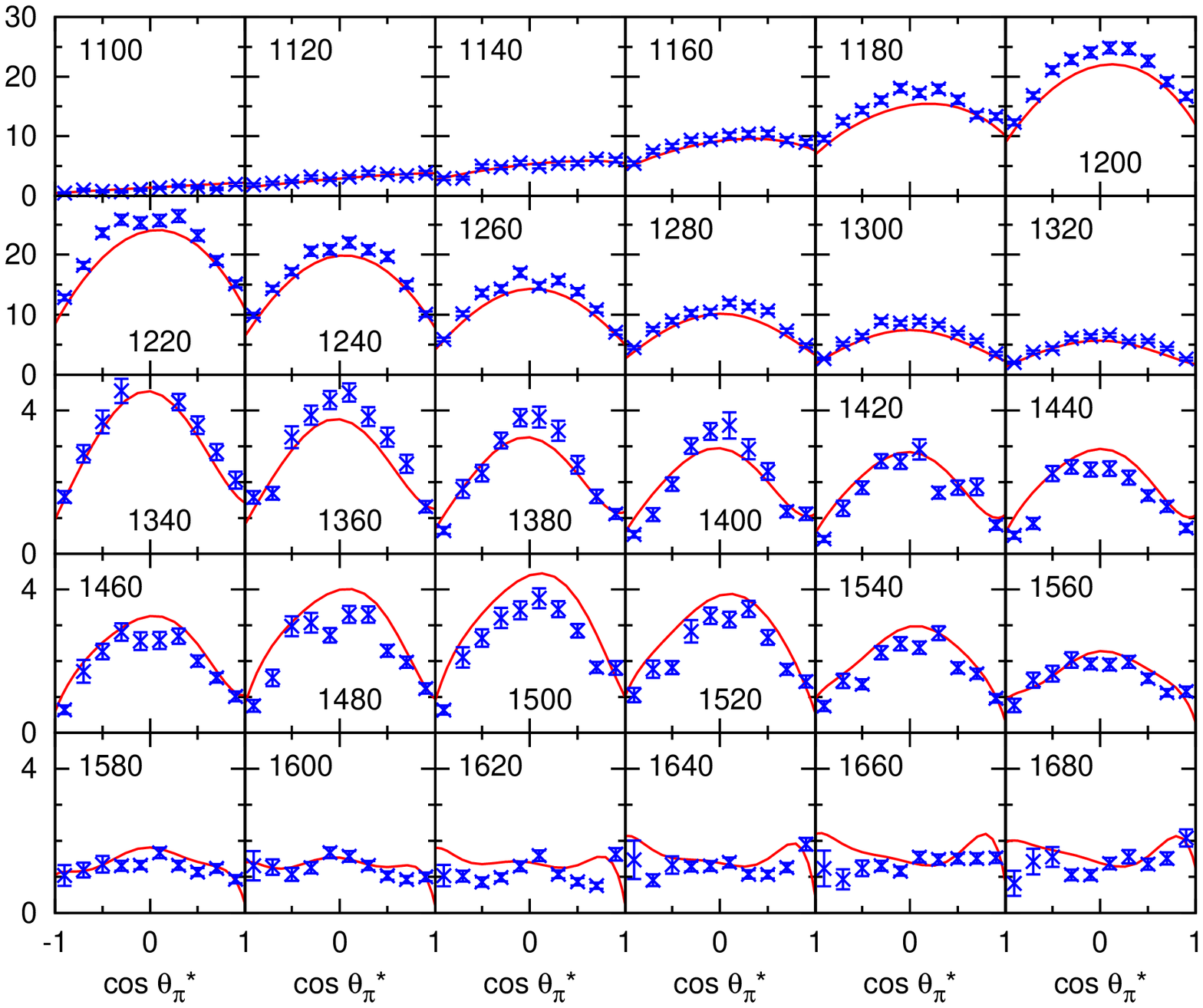}
\hspace{5mm}
\includegraphics[width=78mm]{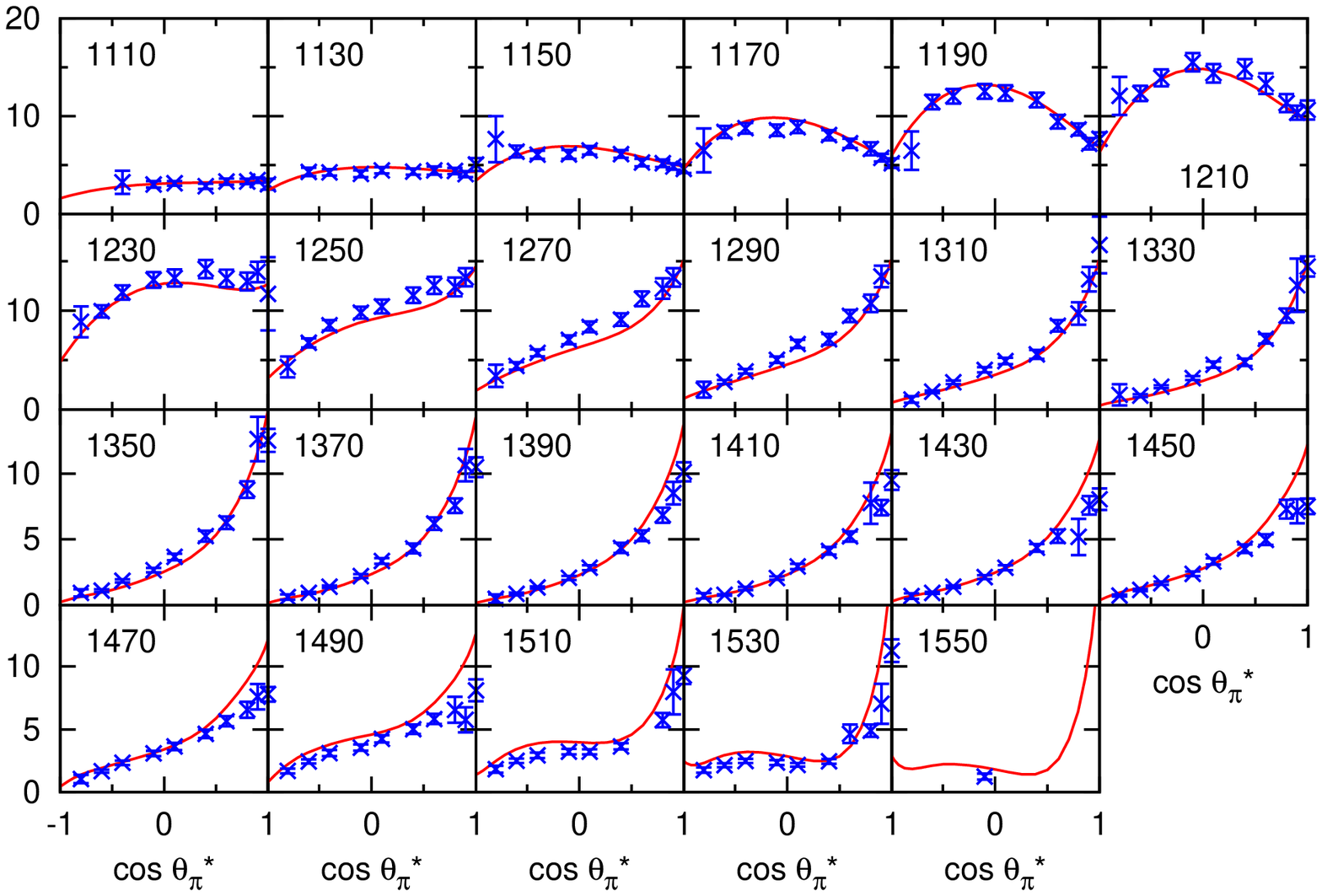}
\caption{(Color online)
The virtual photon cross section 
$d\sigma_T/d\Omega^*_\pi+\epsilon\, d\sigma_L/d\Omega^*_\pi$ ($\mu$b/sr)
at $Q^2$=0.40 (GeV/$c$)$^2$ for
$p(e,e'\pi^0)p$ (left) and $p(e,e'\pi^+)n$ (right) from the DCC model.
The number in each panel indicates $W$ (MeV).
The data are from Ref.~\cite{eepi-joo-prl} for $p(e,e'\pi^0)p$,
and Ref.~\cite{eepi-egiyan-prc} for $p(e,e'\pi^+)n$.
}
\label{fig:eepi-0.40}
\end{figure}
\begin{figure}[t]
\includegraphics[width=78mm]{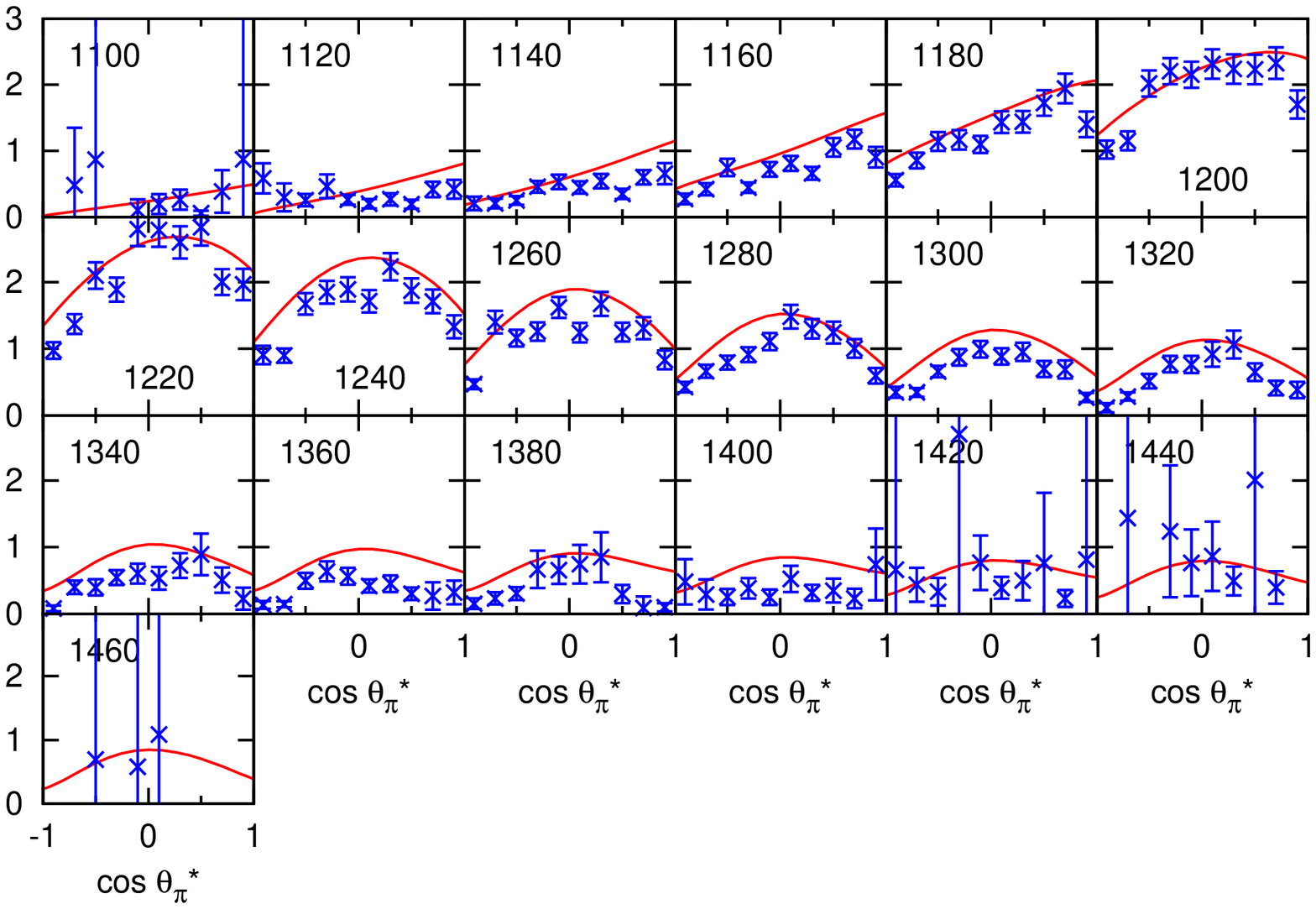}
\hspace{5mm}
\includegraphics[width=78mm]{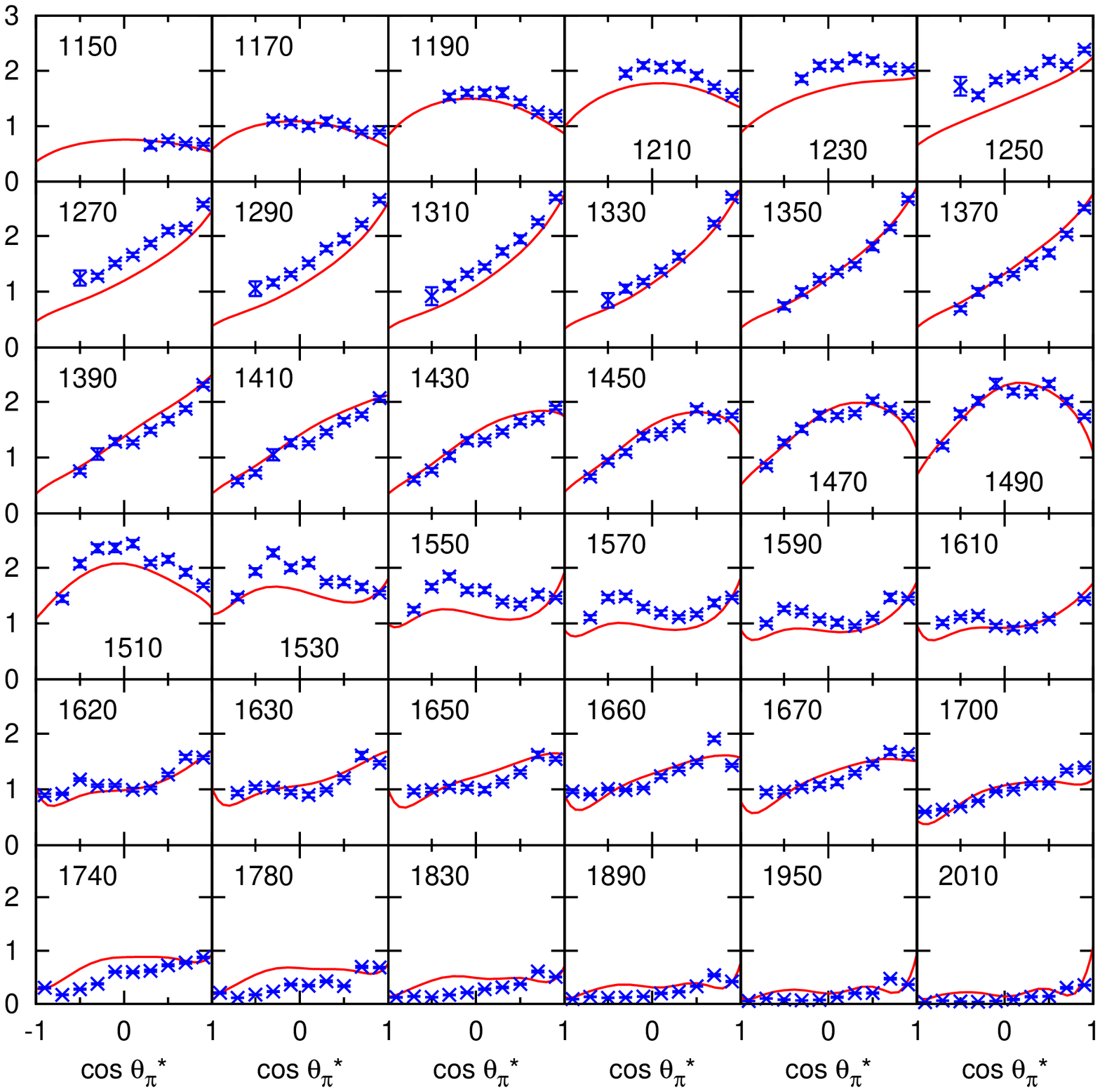}
\caption{(Color online)
The virtual photon cross section 
$d\sigma_T/d\Omega^*_\pi+\epsilon\, d\sigma_L/d\Omega^*_\pi$ ($\mu$b/sr)
at $Q^2$=1.76 (GeV/$c$)$^2$.
The data are in the range,
$1.72\le Q^2\le$1.80 (GeV/$c$)$^2$, 
and are from
Ref.~\cite{eepi-joo-prl} for $p(e,e'\pi^0)p$,
and Refs.~\cite{clas-park,clas-park2} for $p(e,e'\pi^+)n$.
The other features are the same as those in 
Fig.~\ref{fig:eepi-0.40}.
}
\label{fig:eepi-1.76}
\end{figure}
\begin{figure}[t]
\includegraphics[width=78mm]{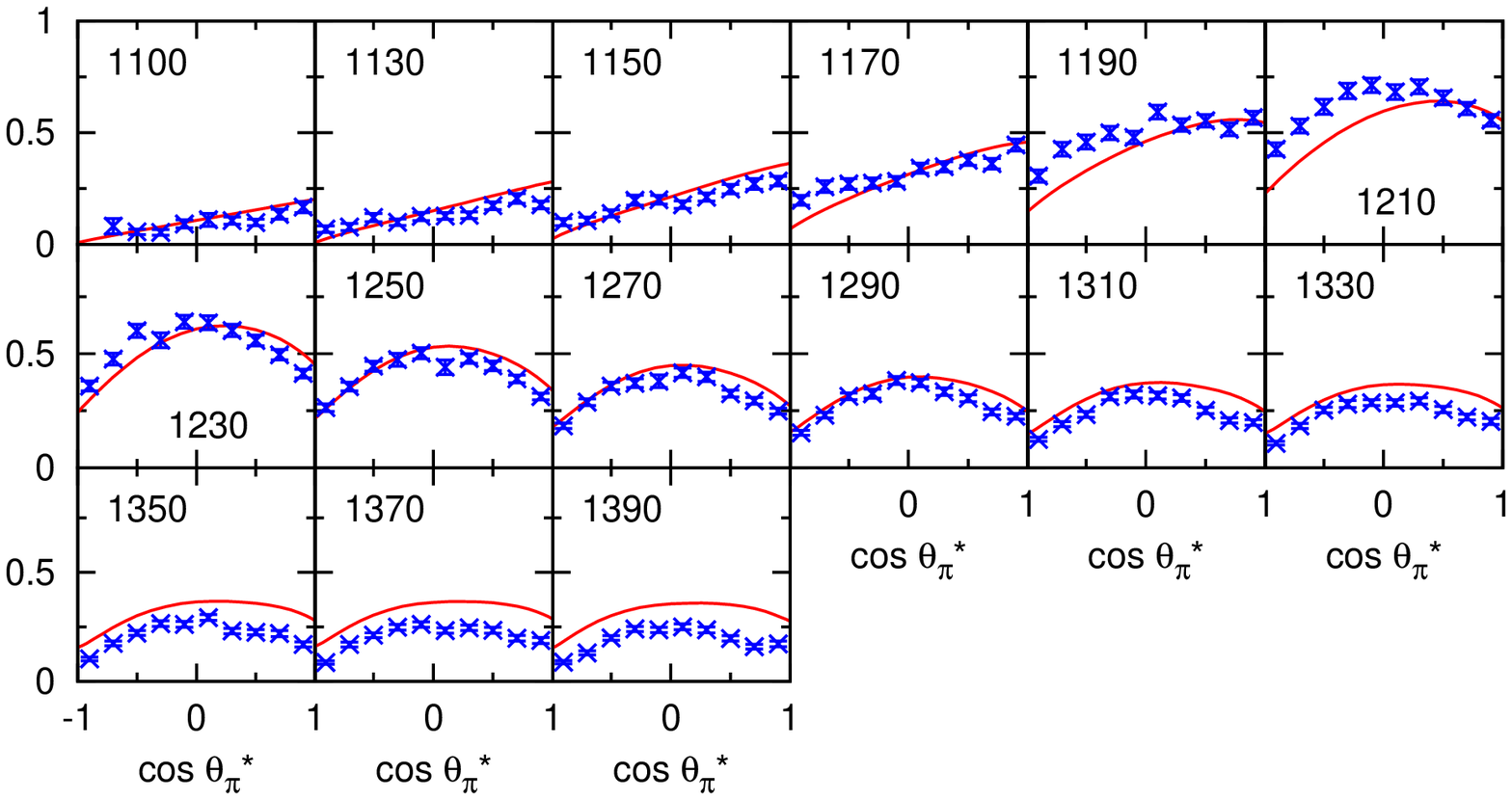}
\hspace{5mm}
\includegraphics[width=78mm]{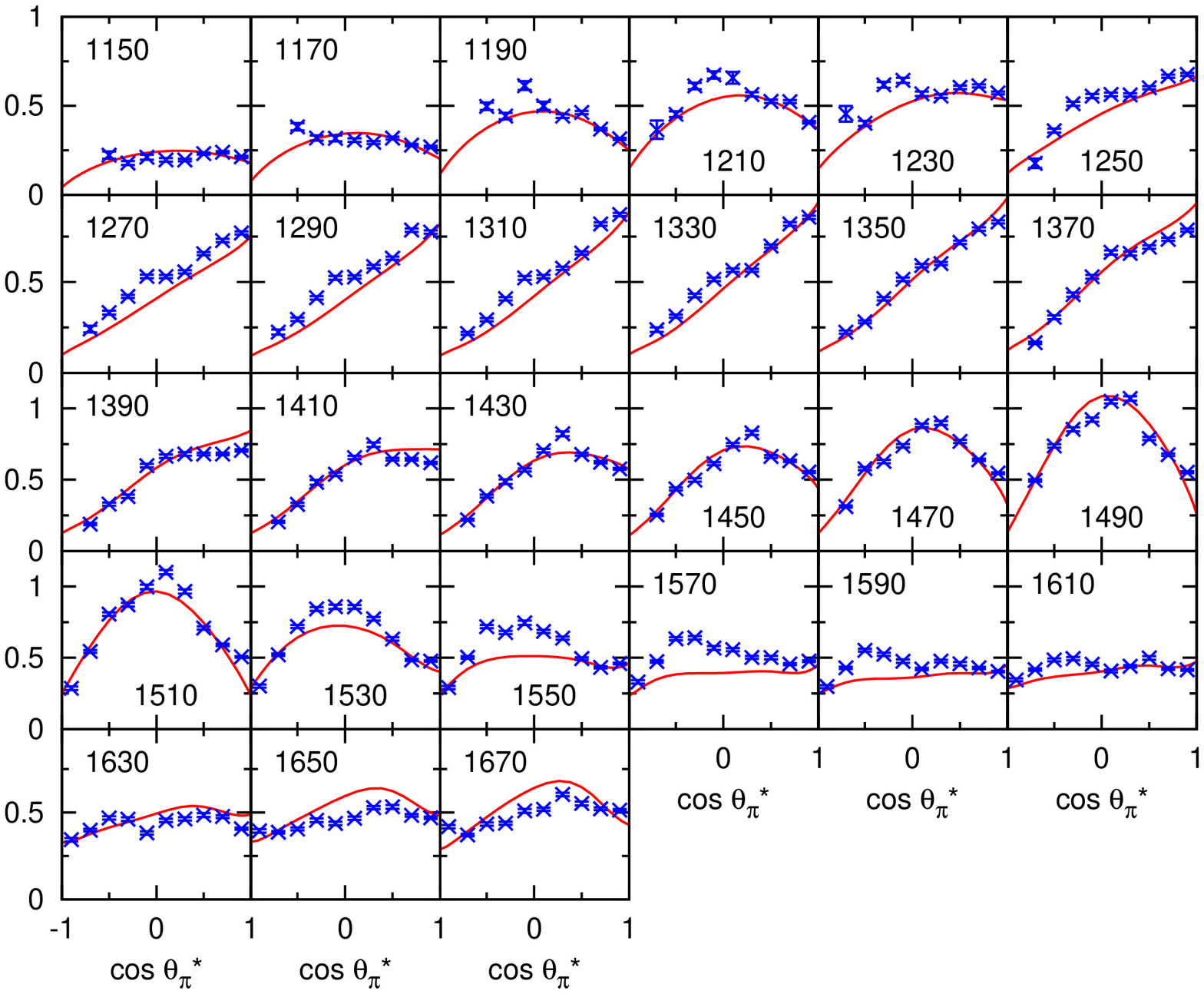}
\caption{(Color online)
The virtual photon cross section 
$d\sigma_T/d\Omega^*_\pi+\epsilon\, d\sigma_L/d\Omega^*_\pi$ ($\mu$b/sr)
at $Q^2$=2.95 (GeV/$c$)$^2$.
The data are in the range,
$2.91\le Q^2\le$3.00 (GeV/$c$)$^2$,
and are from Ref.~\cite{clas-ungaro} for $p(e,e'\pi^0)p$
and Refs.~\cite{clas-park,clas-park2} for $p(e,e'\pi^+)n$.
The other features are the same as those in 
Fig.~\ref{fig:eepi-0.40}.
}
\label{fig:eepi-2.95}
\end{figure}

\begin{figure}[t]
\includegraphics[width=0.34\textwidth]{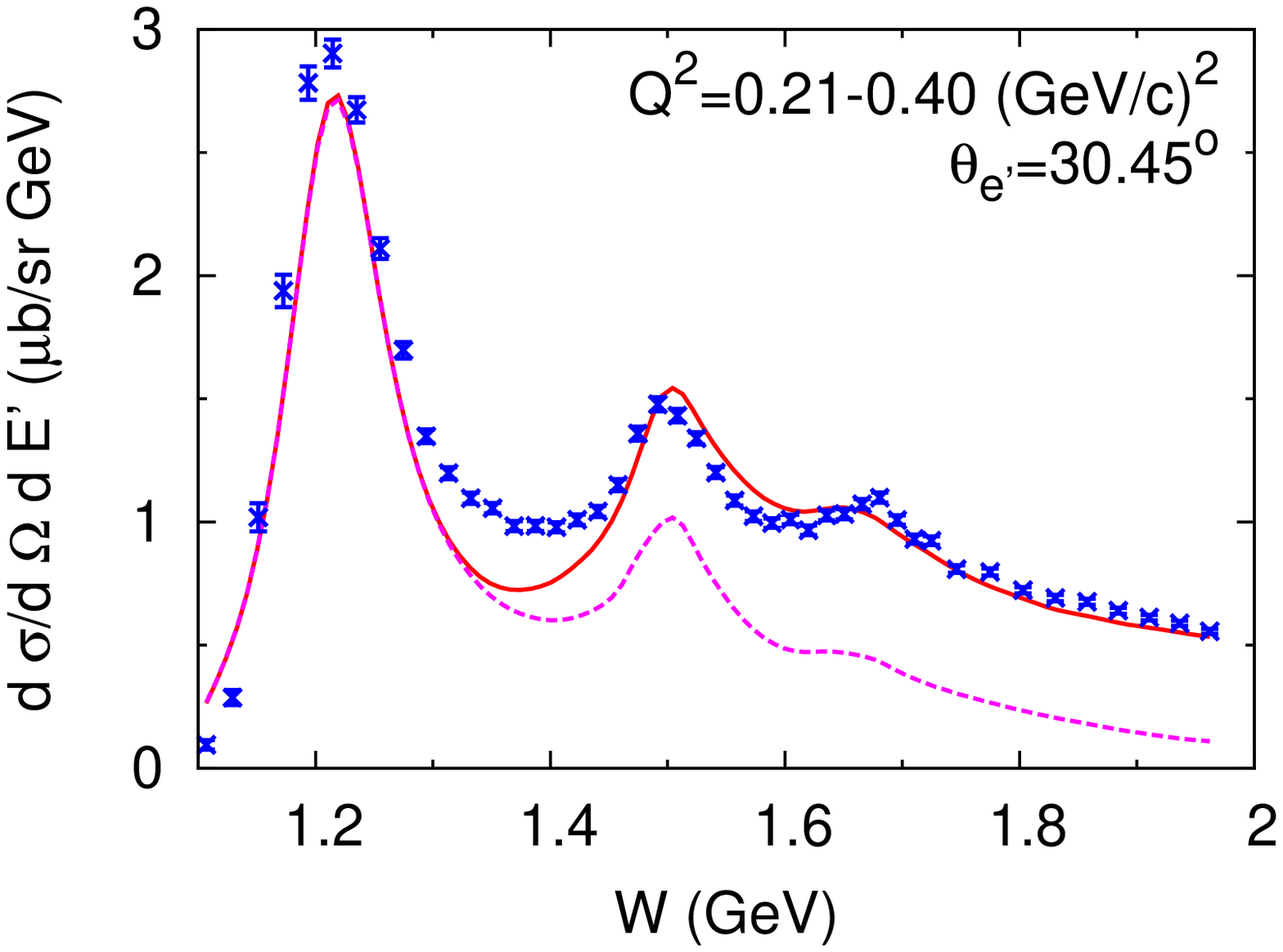}
\hspace{-5mm}
\includegraphics[width=0.34\textwidth]{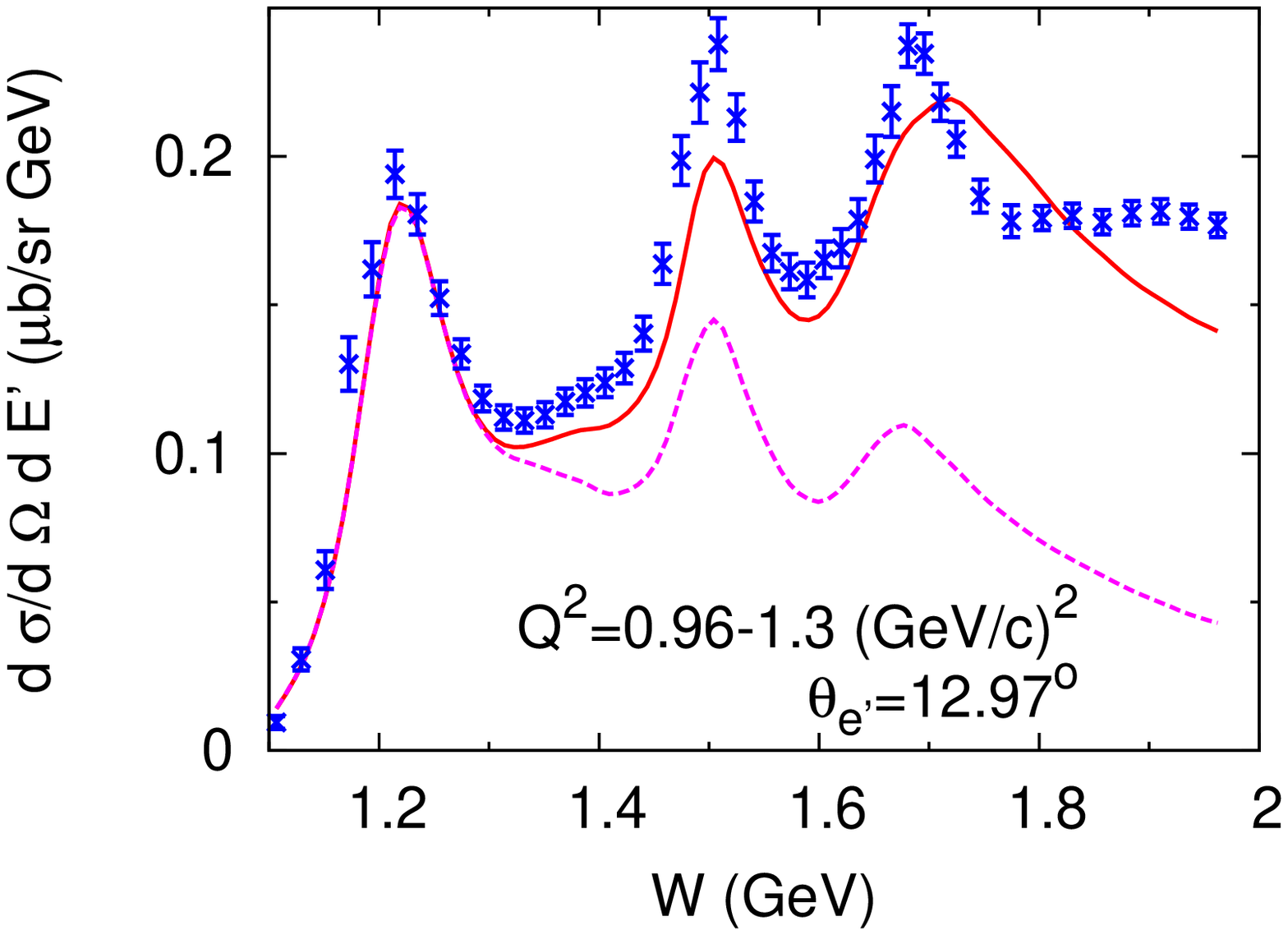}
\hspace{-5mm}
\includegraphics[width=0.34\textwidth]{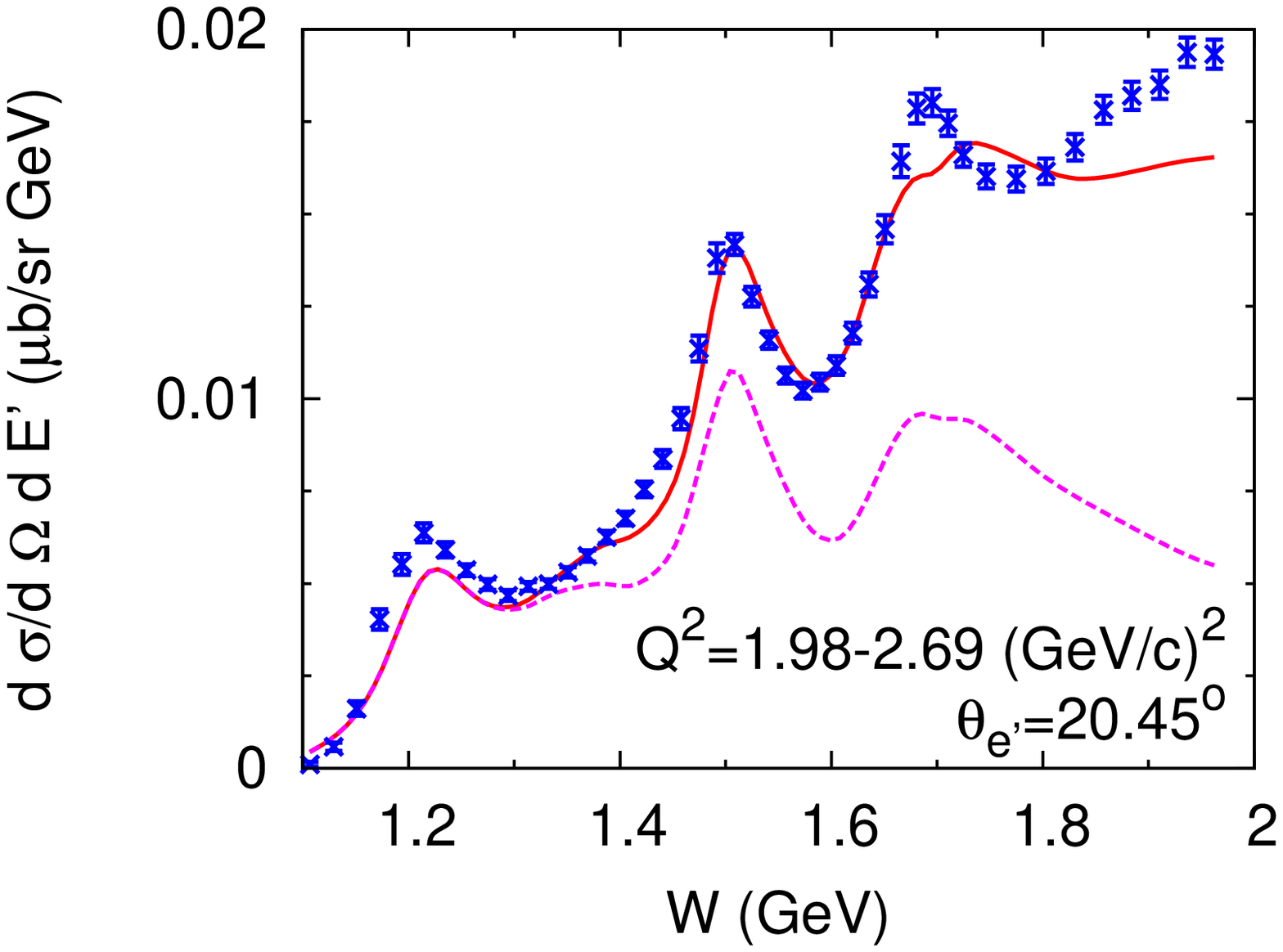}
\caption{(Color online)
Comparison of DCC-based calculation with data for inclusive electron-proton
 scattering at $E_e$=5.498~GeV.
The red solid curves are for inclusive cross sections while the magenta
dashed-curves 
are for contributions from the $\pi N$ final states.
The range of $Q^2$ and the electron scattering angle
 ($\theta_{e'}$) are indicated in each panel.
The data are from Ref.~\cite{E00-002}.
}
\label{fig:eepi-incl}
\end{figure}

We have fitted $F^V_{pN^*}(Q^2)$ to the data at several $Q^2$ values where
the data are available. All the other parameters in the DCC model,  
406 (115) parameters for hadronic ($\gamma$-$p$) interactions,
are fixed as those determined 
by the combined analysis~\cite{knls13} of 
$\pi N, \gamma p\to \pi N, \eta N, K\Lambda, K\Sigma$ data consisting of
22,348 data points.
We have successfully tested the DCC-based vector current model with the
data covering the 
whole kinematical region relevant to neutrino reactions of
$E_\nu\le 2$~GeV. 
Before presenting numerical results of the analysis, we take one more step as follows.
In the course of the analysis,
we have determined the $p$-$N^*$ vector form factors
$F^V_{pN^*}(Q_i^2)$ at particular $Q_i^2$ values where the data are available:
$Q_i^2$=0,
0.16,
0.20,
0.24,
0.28,
0.30,
0.32,
0.40,
0.50,
0.525,
0.60,
0.65,
0.75,
0.90,
1.15,
1.45,
1.72,
1.80,
2.05,
2.20,
2.44,
2.60,
2.91,
3.00~(GeV/$c$)$^2$.
When the model is applied to calculating
neutrino cross sections, we need
the form factors at arbitrary values of $Q^2$.
Therefore, it is convenient to parametrize 
$F^V_{pN^*}(Q_i^2)$ with a simple polynomial function of $Q^2$.
We approximate the form factors using the following 
parametrization:
\begin{eqnarray}
F^{V}_{NN^*}(Q^2)
\sim \sum_{n=0}^{\cal N} c^N_n (Q^2)^n \ ,
\label{eq:q2-fit}
\end{eqnarray}
where $c^N_n$ are constants. 
We set ${\cal N}=5$ in Eq.~(\ref{eq:q2-fit})
and require that $c^p_0$ for the transverse form factors are
exactly the same as the values determined by the analysis of photo-reactions,
i.e., 
$c^p_0=F^V_{pN^*}(Q_i^2=0)$.
Then the other $c^p_n$, totally 440 parameters,
are determined to fit the obtained form factors $F^V_{pN^*}(Q_i^2)$.
We present numerical values for $c^p_n$ in Table~\ref{tab:param-cn} of Appendix~\ref{app:cn}.
We will use the approximate polynomial parametrization of $F^V_{pN^*}(Q^2)$ 
in calculations presented hereafter.

Here we present some selected results of the analysis 
of electron-proton reactions.
Among the five differential virtual photon cross sections in Eq.~(\ref{eq:dcrst-em}),
only $d\sigma_T/d\Omega^*_\pi$ and $d\sigma_L/d\Omega^*_\pi$ survive after
integrating over the hadronic final states. Therefore, they are the most important
in view of applications to the neutrino reactions.
Thus we show a combination of the virtual photon cross sections,
$d\sigma_T/d\Omega^*_\pi+\epsilon\, d\sigma_L/d\Omega^*_\pi$,
at $Q^2$=0.40, 1.76 and 2.95 (GeV/$c$)$^2$ 
for $p(e,e'\pi^0)p$ and $p(e,e'\pi^+)n$ from the DCC model
in Figs.~\ref{fig:eepi-0.40}-\ref{fig:eepi-2.95}.
In the same figures, the corresponding data are also shown for
comparison. 
The DCC model fits the data reasonably well for both $\pi^0$ and $\pi^+$
channels.
We also show in Fig.~\ref{fig:eepi-incl}
our DCC-based calculation of differential
cross sections of the inclusive electron-proton scattering 
in comparison with data;
the single pion electroproduction cross sections from the DCC
model are also presented.
In each of the panels, the range of $Q^2$ is indicated, and 
$Q^2$ monotonically decreases as $W$ increases.
The figures show a reasonable agreement between our calculation with the
data, and also show the increasing importance of the multi-pion production processes 
above the $\Delta(1232)$ resonance region.
However,
we observe a discrepancy between the model and data in $W=1.3\sim 1.45$~GeV
in the left panel of Fig.~\ref{fig:eepi-incl}.
Because our DCC model gives a reasonable fit to the single pion
electroproduction data in this kinematical region as shown in
Fig.~\ref{fig:eepi-0.40}, the discrepancy points to a problem in our DCC
model in describing double-pion electroproduction in this kinematics.
By simply adjusting the vector form factors, $F^V_{pN^*}(Q_i^2)$, we were not
able to fit the single-pion and inclusive
data at the same time in this kinematics.
A resolution to the discrepancy in the inclusive cross sections may
require a combined analysis including double-pion production data.
Also, as $Q^2$ increases, the DCC model starts to underestimate 
the inclusive cross section towards $W\sim$~2~GeV 
where the kinematical
region is entering the DIS 
and multi-meson production region.
Currently available data of neutrino cross sections in the resonance region
are not very precise compared to the high precision data of the electron-induced reactions.
Therefore, for our present purpose of constructing a model of neutrino
reaction comparable to the current neutrino data, it is sufficient to
fit the electron-induced reactions data at the level as presented 
in Figs.~\ref{fig:eepi-0.40}-\ref{fig:neutron-pi0},
and we refrain from showing the $\chi^2$ values for the fits.

\subsection{Photon-neutron and electron-neutron reactions}

Because a free neutron target is not available, 
``neutron''-target data are extracted from deuteron-target data.
Effects of the final state interaction and the Fermi motion
on the $\gamma d \rightarrow \pi^-  p  p$ reaction has been studied 
in Ref.~\cite{taras11}.
Here we analyze the data of pion photoproduction on ``neutron'' available in literature.
We analyze unpolarized differential cross sections for
$\gamma n\to \pi N$ from $\pi N$ threshold to $W=2$~GeV,
and determine $F^V_{nN^*}(Q^2=0)$ 
and the cutoffs $\Lambda_{N^*}^{\rm e.m.}$ 
[Eqs.~(\ref{eq:M-para})-(\ref{eq:long-amp})]
for $I$=1/2 $N^*$ states
($F_{nN^*}^V(Q^2) \equiv F^V_{pN^*}(Q^2)$
for $I$=3/2 $N^*$ states).
A formula to calculate differential cross sections from the amplitudes
of Eq.~(\ref{eq:tmbgn}) can be found in Ref.~\cite{msl07}, and we will
not repeat it here. 
In the finite $Q^2$ region,
we use empirical inclusive structure functions from Ref.~\cite{christy2} as data
to determine the transition vector form factors $F^V_{nN^*}(Q^2)$.
Bosted et al.~\cite{bosted} fitted inclusive electron-deuteron reaction data to
obtain their model for the inclusive deuteron structure functions, 
and the inclusive ``neutron'' structure functions are
obtained from that by subtracting the proton structure function of
Ref.~\cite{christy}.
We use an improved version~\cite{christy2} of this 
``neutron'' structure functions.
After determining $F^V_{nN^*}(Q_i^2)$ at $Q_i^2$=0, 0.20, ..., 3.00
(GeV/$c$)$^2$ at every 0.20 (GeV/$c$)$^2$,
we parametrize them using Eq.~(\ref{eq:q2-fit}), as we did for 
the $p$-$N^*$ vector form factors.
We present numerical values for $c^n_n$ (258 parameters) 
and those for 
the cutoffs $\Lambda_{N^*}^{\rm e.m.}$ 
(16 parameters)
in Tables~\ref{tab:param-cn-n} and \ref{tab:param-cutoff} of
 Appendix~\ref{app:cn}.
The following results are obtained with this approximate polynomial parametrization.

\begin{figure}[t]
\includegraphics[width=1\textwidth]{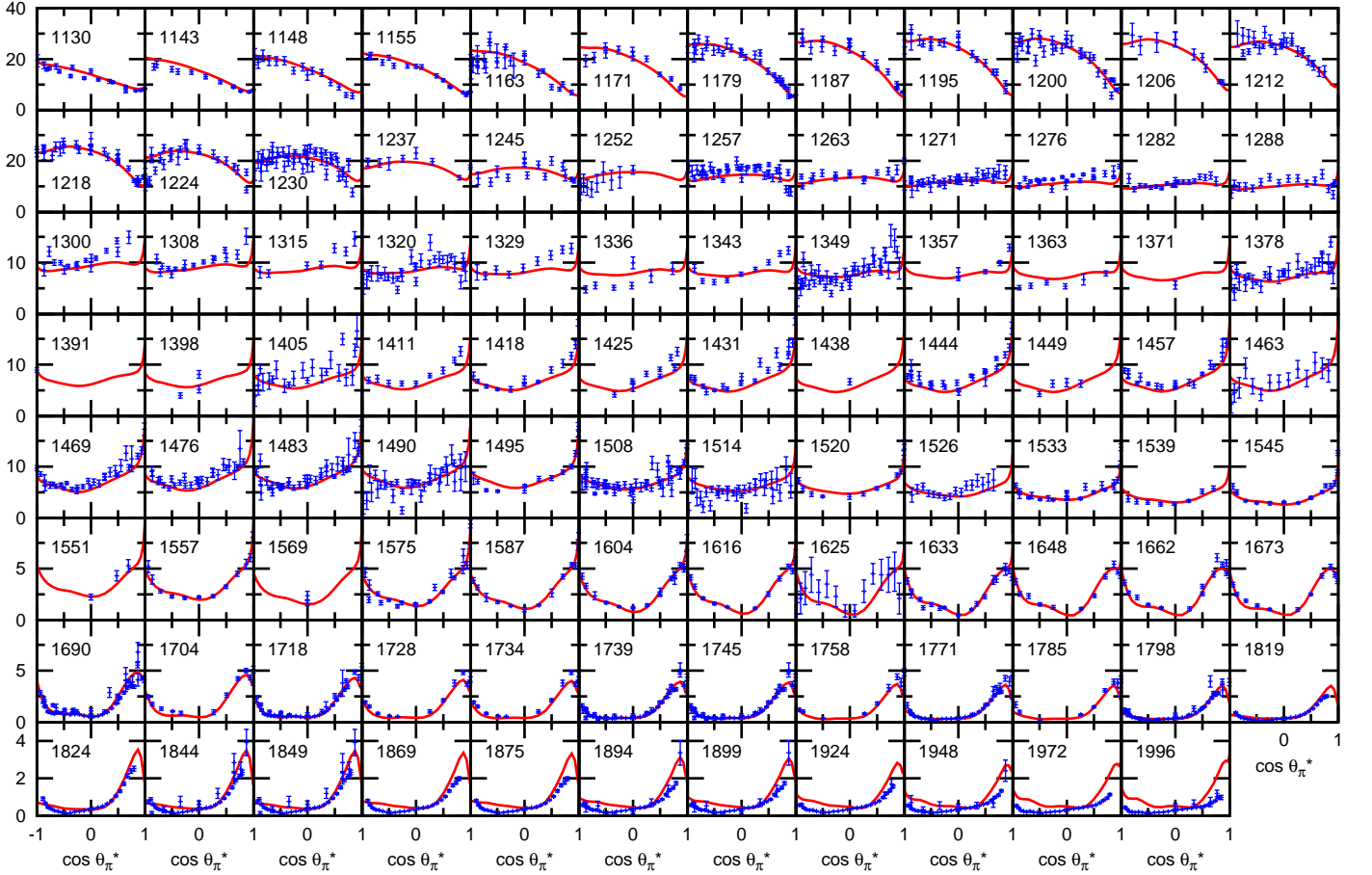}
\caption{(Color online)
Unpolarized differential cross sections,
$d\sigma/d\Omega^*_\pi$ ($\mu$b/sr),
for $\gamma n\to \pi^- p$.
The data are from 
Refs.~\cite{gn-pmp-1,gn-pmp-2,gn-pmp-3,gn-pmp-4,gn-pmp-5,gn-pmp-6,gn-pmp-7,gn-pmp-8,gn-pmp-9,gn-pmp-10,gn-pmp-11,gn-pmp-12,gn-pmp-13,gn-pmp-14,gn-pmp-15,gn-pmp-16,gn-pmp-17,gn-pmp-18,gn-pmp-19,gn-pmp-20,gn-pmp-21,gn-pmp-22,gn-pmp-23,gn-pmp-24}.
\label{fig:neutron-pim}}
\end{figure}
\begin{figure}[t]
\includegraphics[width=1\textwidth]{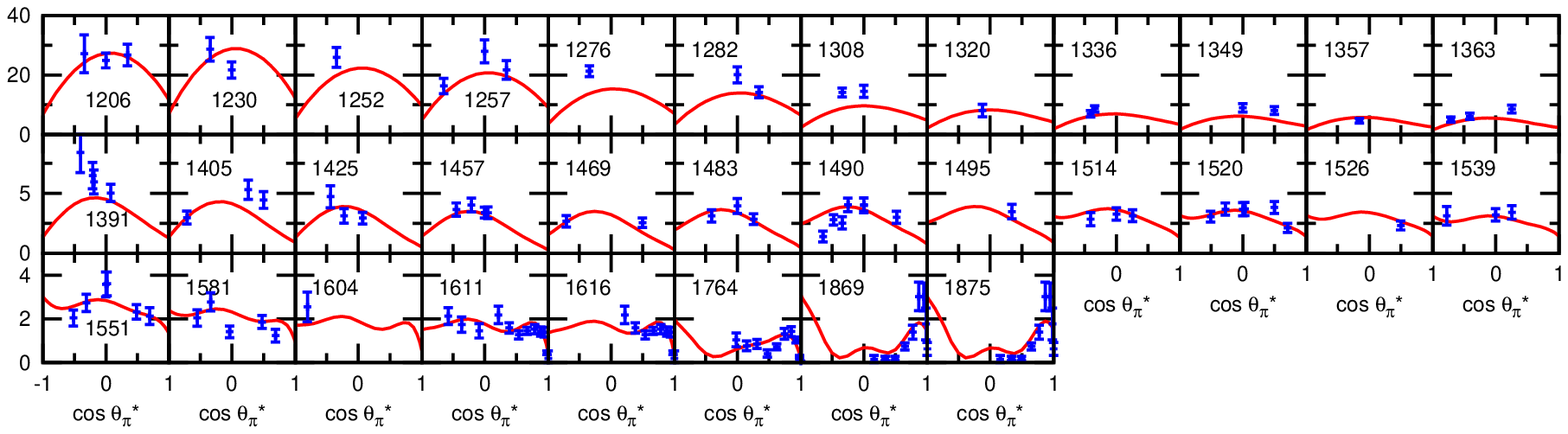}
\caption{(Color online)
Unpolarized differential cross sections
$d\sigma/d\Omega^*_\pi$ ($\mu$b/sr),
for $\gamma n \to \pi^0 n$.
The data are from Refs.~\cite{gn-p0n-1,gn-p0n-2,gn-p0n-3,gn-p0n-4}.
}
\label{fig:neutron-pi0}
\end{figure}
\begin{figure}[h]
\includegraphics[width=0.34\textwidth]{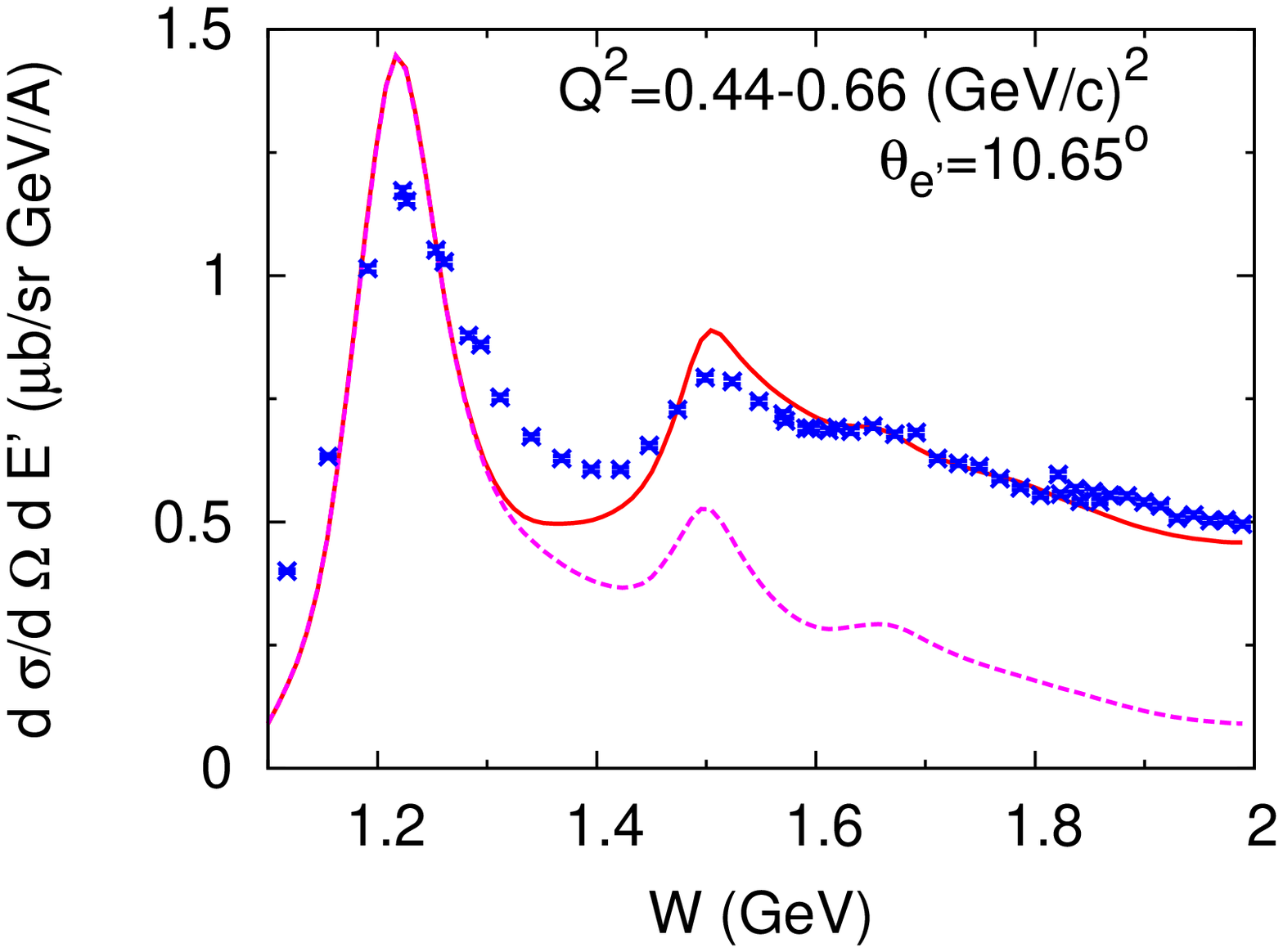}
\hspace{-5mm}
\includegraphics[width=0.34\textwidth]{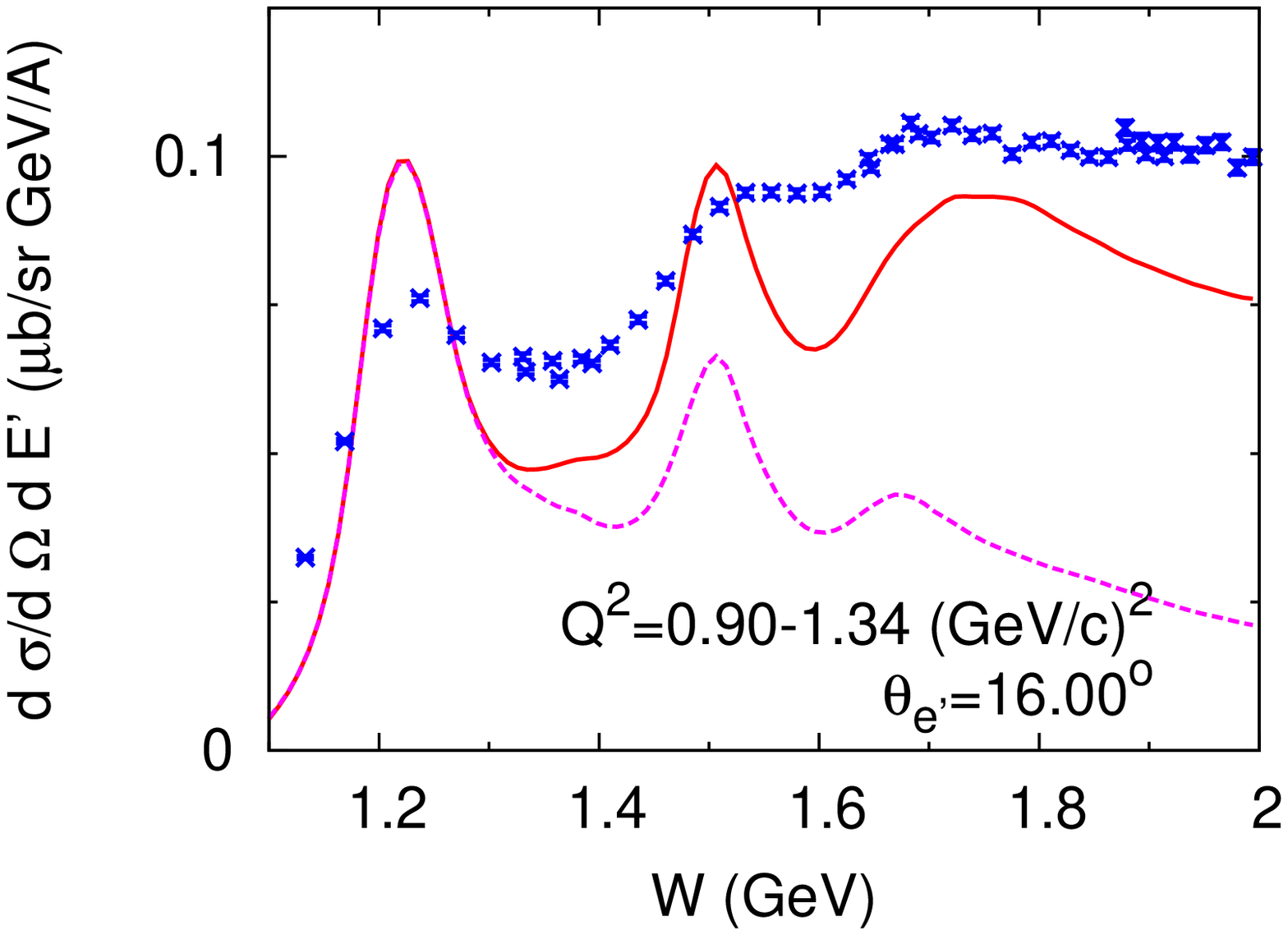}
\hspace{-5mm}
\includegraphics[width=0.34\textwidth]{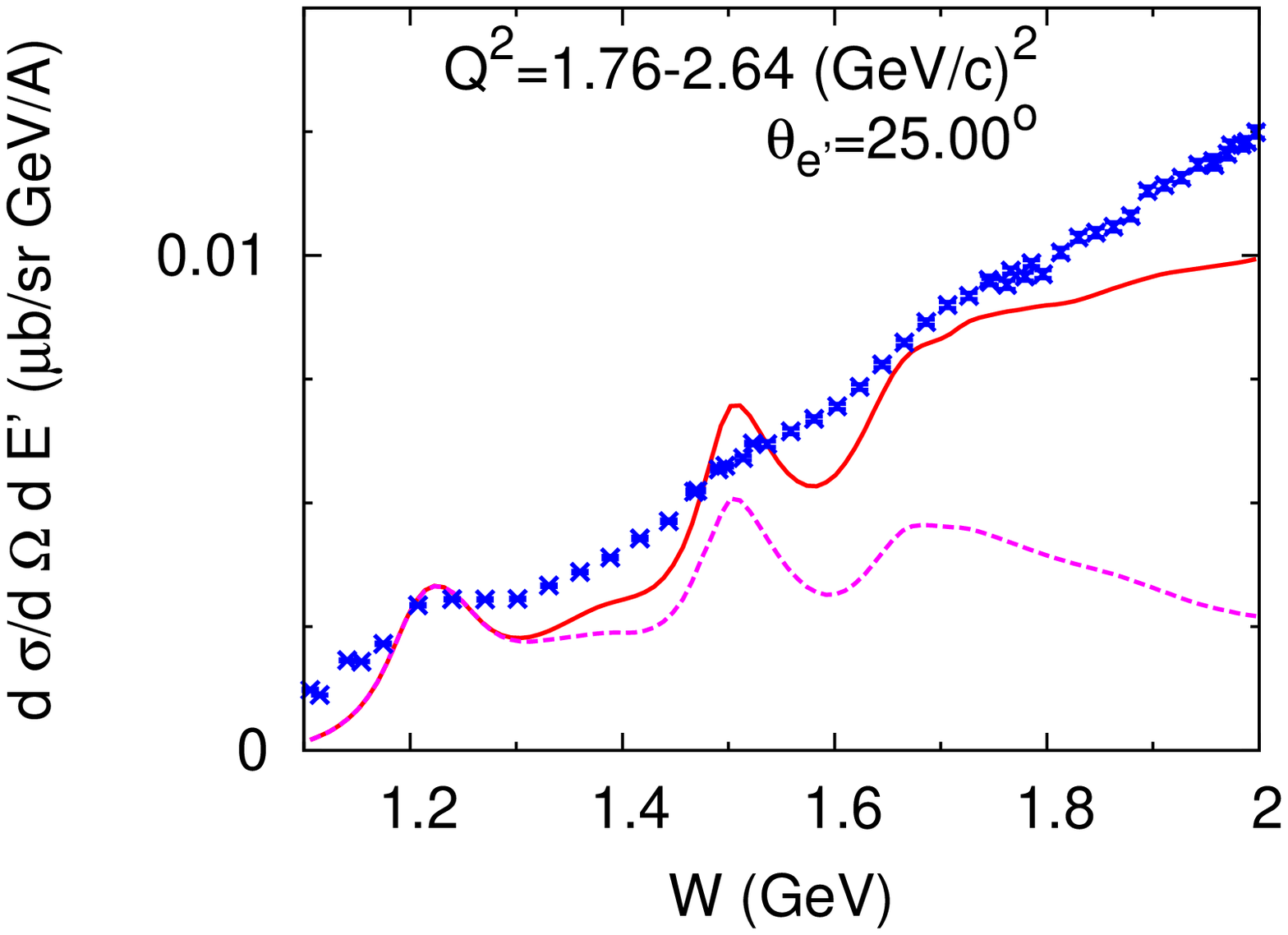}
\caption{\label{fig:neutron-str}
(Color online)
Comparison of DCC-based calculation with data for 
differential cross sections per nucleon of
inclusive electron-deuteron
 scattering at $E_e$=4.628~GeV.
The red solid curves are for inclusive cross sections while the magenta
dashed-curves are for contributions from the $\pi N$ final states.
The range of $Q^2$ and the electron scattering angle ($\theta_{e'}$)
are indicated in each panel.
The data are from Ref.~\cite{en-incl}.
}
\end{figure}

We show unpolarized differential cross sections for $\gamma n\to \pi N$
calculated with the DCC model in comparison with data in 
Figs.~\ref{fig:neutron-pim} and \ref{fig:neutron-pi0},
and find a reasonable agreement.
We also show a comparison of the DCC-based calculation with data of
differential cross sections per nucleon for the inclusive
electron-deuteron scattering in Fig.~\ref{fig:neutron-str}.
In this calculation, we simply take an average of electron-proton and
electron-neutron differential cross sections in the free space.
As seen in the figures, the calculated resonance peaks are sharper than
the data, indicating that the smearing due to the Fermi motion 
and final state interaction needs to
be taken into account to obtain a good agreement with the data~\cite{hirata}.
Finally we note that
a more comprehensive analysis including $\gamma n\to \pi N$ data is
currently underway, which will be reported elsewhere~\cite{neutron-dcc}.

\section{Results for neutrino reactions}
\label{sec:results}

\begin{figure}[t]
\includegraphics[width=0.45\textwidth]{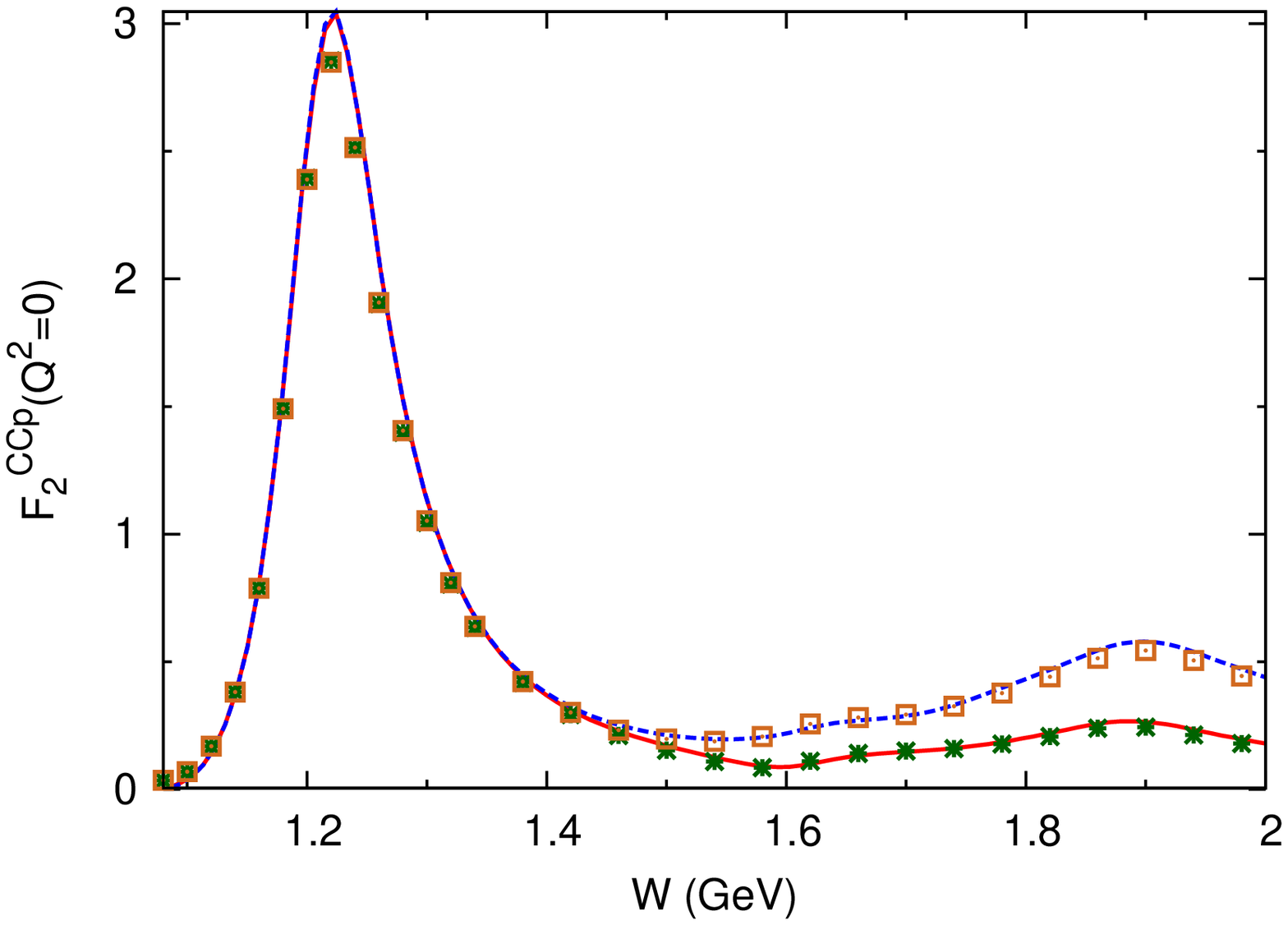}
\includegraphics[width=0.45\textwidth]{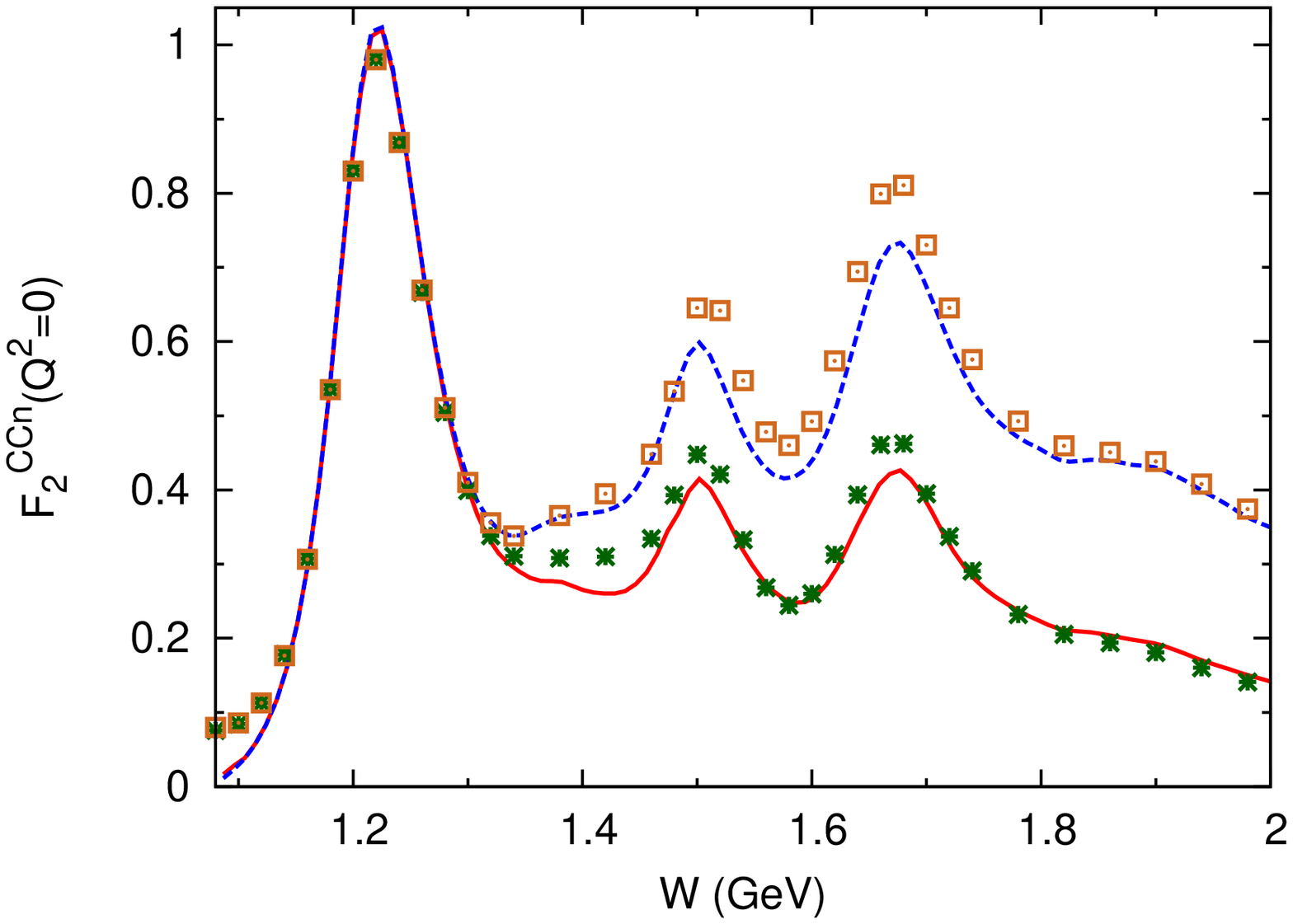}
\caption{(Color online)
The structure function $F^{\rm CC}_2$ at $Q^2=0$.
$F^{\rm CC}_2$ calculated with 
the axial current amplitude (lines) are compared with
those from $\pi N$ cross sections via the PCAC relation (points).
The upper (lower) line and point are for inclusive ($\pi N$) final state(s).
The left (right) panel is for CC $\nu$-proton/CC $\bar\nu$-neutron
(CC $\nu$-neutron/CC $\bar\nu$-proton) process.
}
\label{fig:neutrino-f2}
\end{figure}

Before discussing neutrino reaction cross sections, 
first we examine the inclusive structure function $F^{\rm CC}_2$ defined in
Eq.~(\ref{eq:F}) at $Q^2$=0.
At this particular kinematics, the neutrino cross sections are solely
determined by $F^{\rm CC}_2$.
$F^{\rm CC}_2(Q^2=0)$ has been often calculated with a PCAC model in which 
$F^{\rm CC}_2(Q^2=0)$ is related to (experimental) $\pi N$ total cross
section by applying the PCAC hypothesis~\cite{knls12,paschos-pcac}.
Within our DCC model, we can calculate $F^{\rm CC}_2(Q^2=0)$
either with the $\pi N$ amplitudes
or with the hadronic axial-current amplitudes.
In Fig.~\ref{fig:neutrino-f2},
we compare $F^{\rm CC}_2(Q^2=0)$ calculated by these two ways, and 
find a good agreement. Some comments are in order.
For the CC $\nu$-proton/CC $\bar\nu$-neutron process 
(left panel of Fig.~\ref{fig:neutrino-f2}), 
only $I$=3/2 resonances contribute, and the $\Delta(1232)$ dominates
$F^{\rm CC}_2$ in low energies.
On the other hand, for 
CC $\nu$-neutron/CC $\bar\nu$-proton process 
(right panel of Fig.~\ref{fig:neutrino-f2}), 
both  $I$=1/2 and 3/2 resonances contribute.
Thus not only the $\Delta(1232)$ but also 
$N(1535)~1/2^-$ and $N(1520)~3/2^-$ resonances in the second resonance region,
$N(1675)~5/2^-$ and $N(1680)~5/2^+$ resonances in the third resonance
region create characteristic energy dependence of
$F^{\rm CC}_2$.
Although we found the two calculations agree well by construction of the
model,
we still notice that the $\Delta(1232)$ peak from the axial-current amplitude somewhat
overshoots that from the $\pi N$ model. 
Also, there are slight differences in the second and third resonance region
in the right panel of Fig.~\ref{fig:neutrino-f2}.
These differences originate from 
the fact that the spatial momentum transfer is fixed either at
$Q^2$=0 or at $Q^2=-m^2_\pi$.
For a more meaningful comparison, we could have compared 
the two $F^{\rm CC}_2$ calculated at the same $Q^2$.
Indeed, we have confirmed a significantly better agreement between $F^{\rm CC}_2(Q^2=-m^2_\pi)$ from
the axial current amplitudes and those from the $\pi N$ cross sections,
which should be the case by definition of the model.
Here, we showed
$F^{\rm CC}_2$ that is actually used in calculating the neutrino cross
sections, and how much it can deviate from those obtained with the 
$\pi N$ cross sections.
Also, we remark that even though most existing neutrino-nucleon reaction
models in the resonance region have axial-currents based on the PCAC
relation, they do not necessarily give 
$F^{\rm CC}_2(Q^2=0)$ in agreement with those from the $\pi N$ cross
sections; we will come back to this point later in this section.

\begin{figure}[t]
\includegraphics[height=0.33\textwidth]{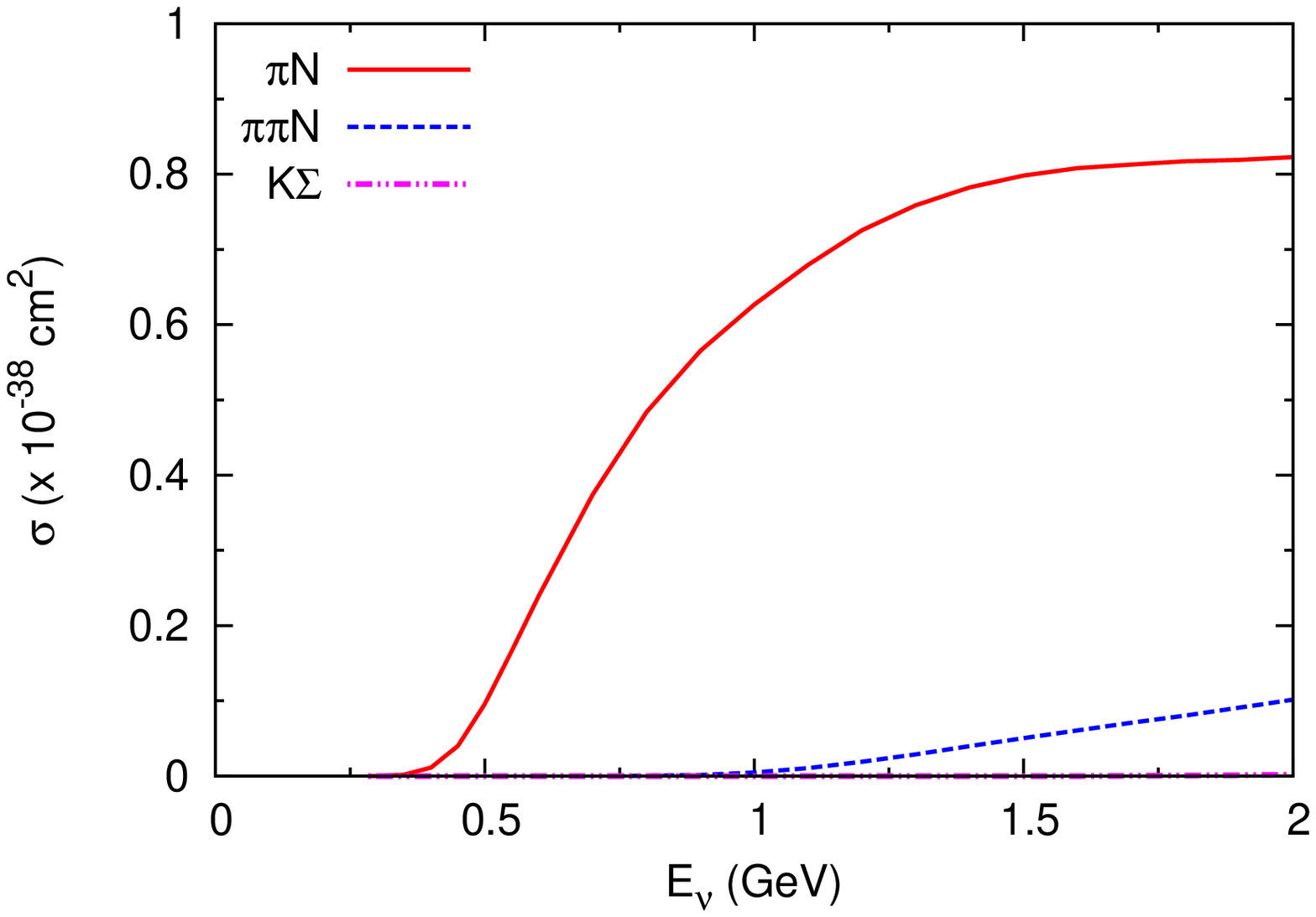}
\includegraphics[height=0.33\textwidth]{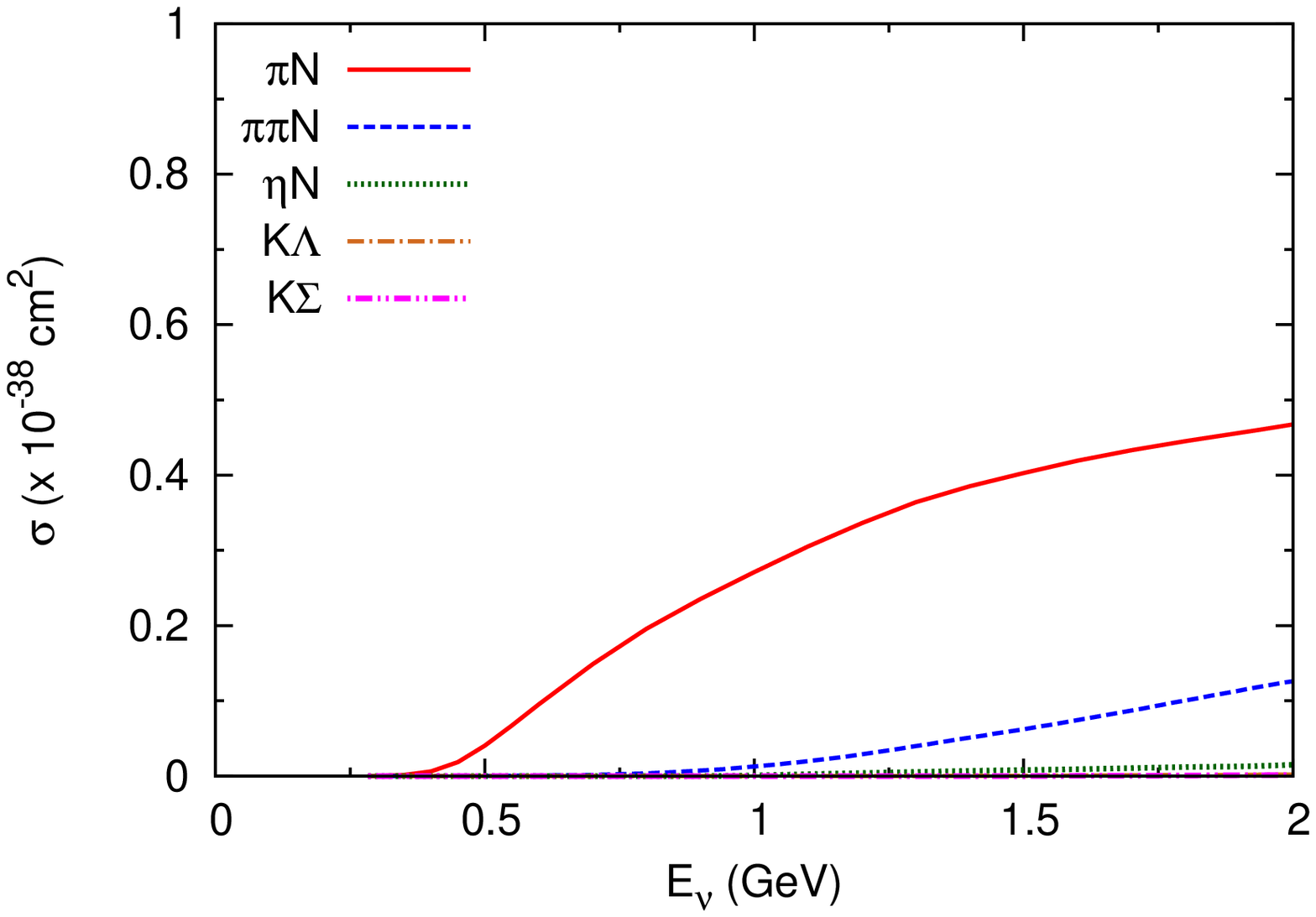}
\caption{(Color online)
Total cross sections for the CC $\nu_\mu\, p$ (left)
and $\nu_\mu n$ (right) reactions.
}
\label{fig:neutrino-tot}
\end{figure}
\begin{figure}[t]
\includegraphics[height=0.33\textwidth]{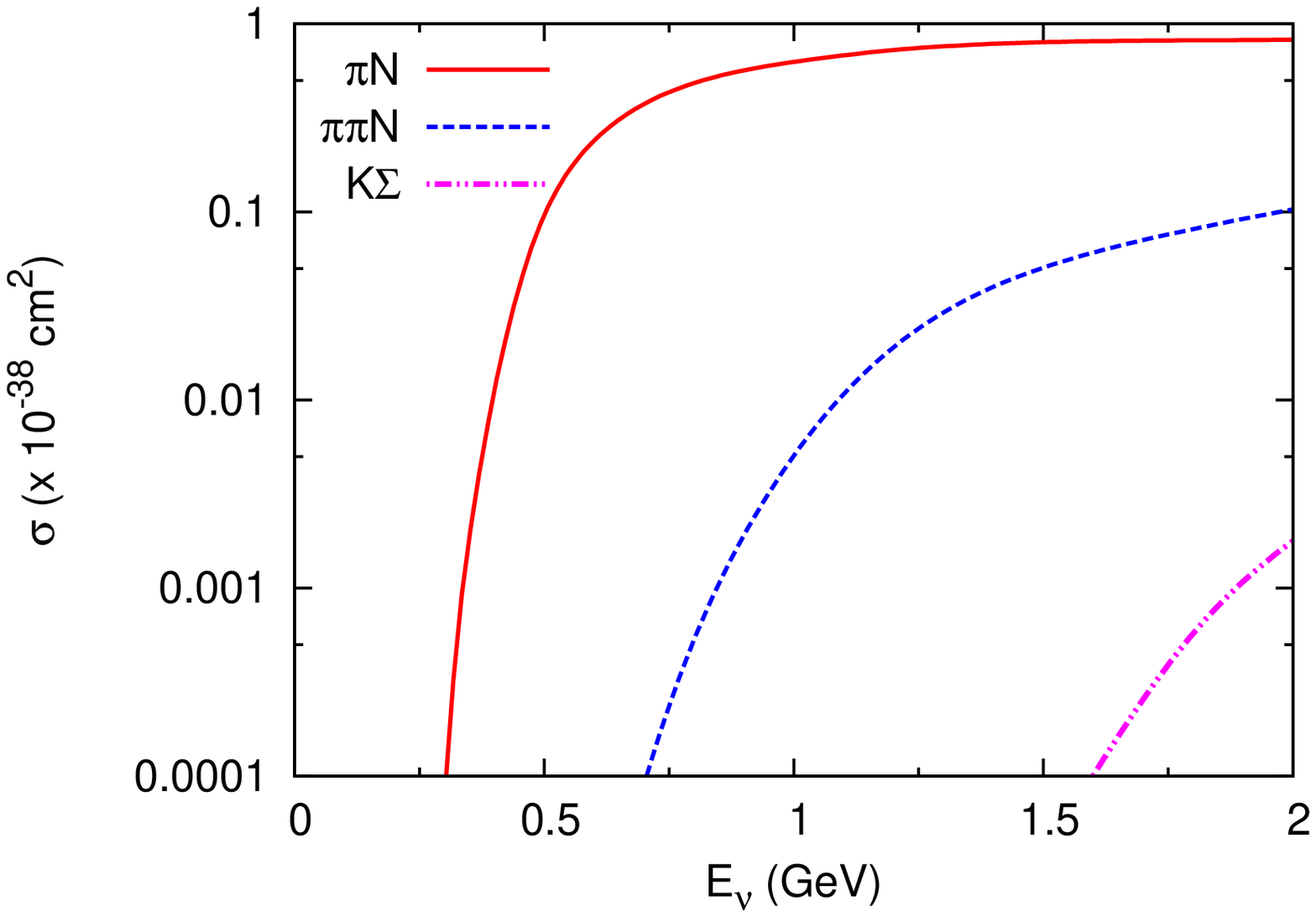}
\includegraphics[height=0.33\textwidth]{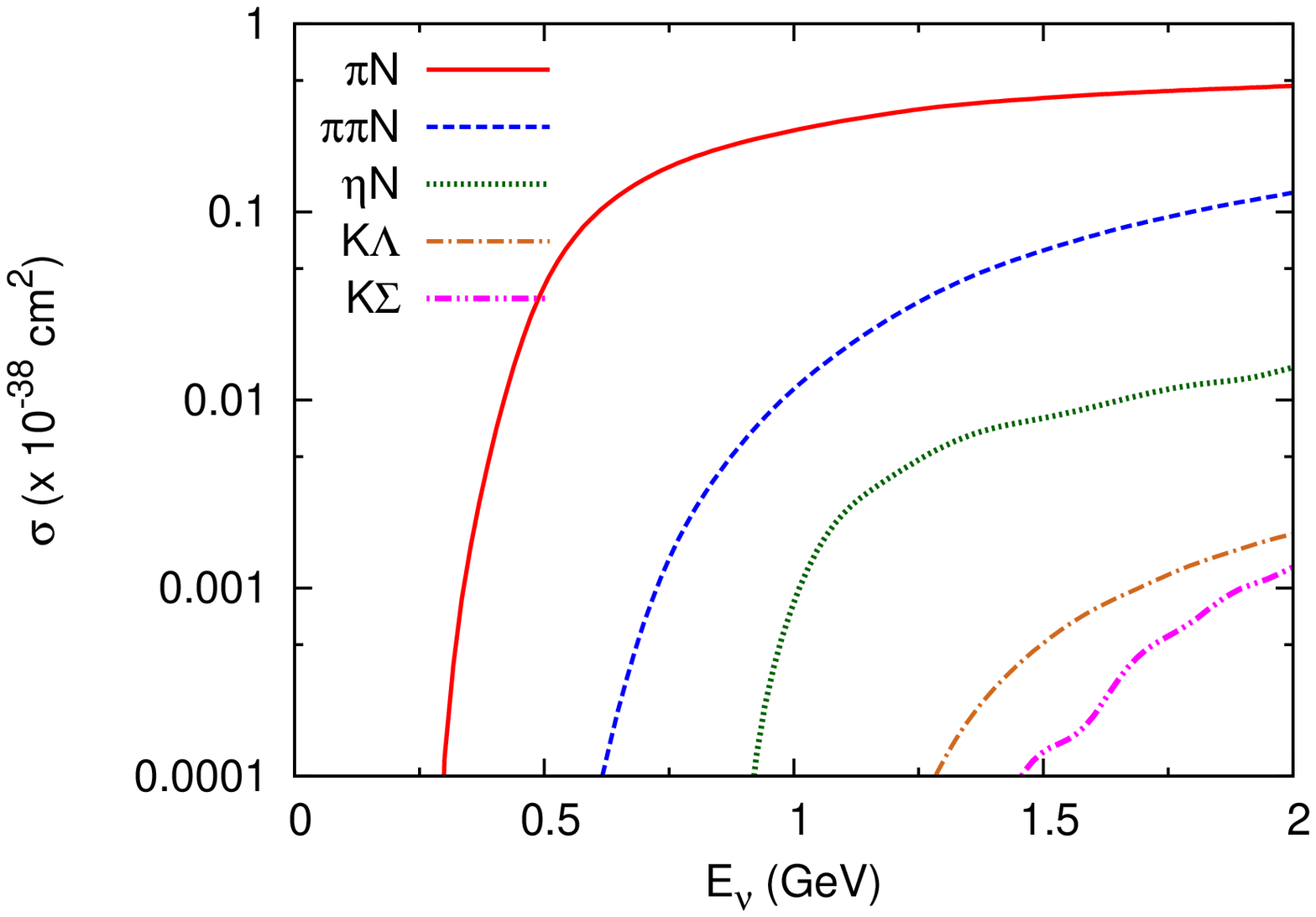}
\caption{(Color online)
The same as Fig.~\ref{fig:neutrino-tot} but in log scale.
}
\label{fig:neutrino-tot-log}
\end{figure}
Now we present cross sections for the
$\nu_\mu\, N$ reactions.
With the DCC model, we can predict contributions from all the final states
included in our model.
Also, the DCC model provides all possible differential cross sections
for each channel.
Here, we present total cross sections for the CC
$\nu_\mu\, N$ reactions up to $E_\nu=2$~GeV
in Fig.~\ref{fig:neutrino-tot};
we also show them in Fig.~\ref{fig:neutrino-tot-log} in log scale to see 
contributions from all of the final states.
For the proton-target, the single pion production dominates in the
considered energy region.
For the neutron-target, the single pion production is still the largest,
but double-pion production becomes relatively more important towards $E_\nu=2$~GeV.
The $\eta N$ and $KY$ production cross sections are ${\cal O}(10^{-1}$-$10^{-2})$ smaller.

\begin{figure}[t]
\includegraphics[height=0.33\textwidth]{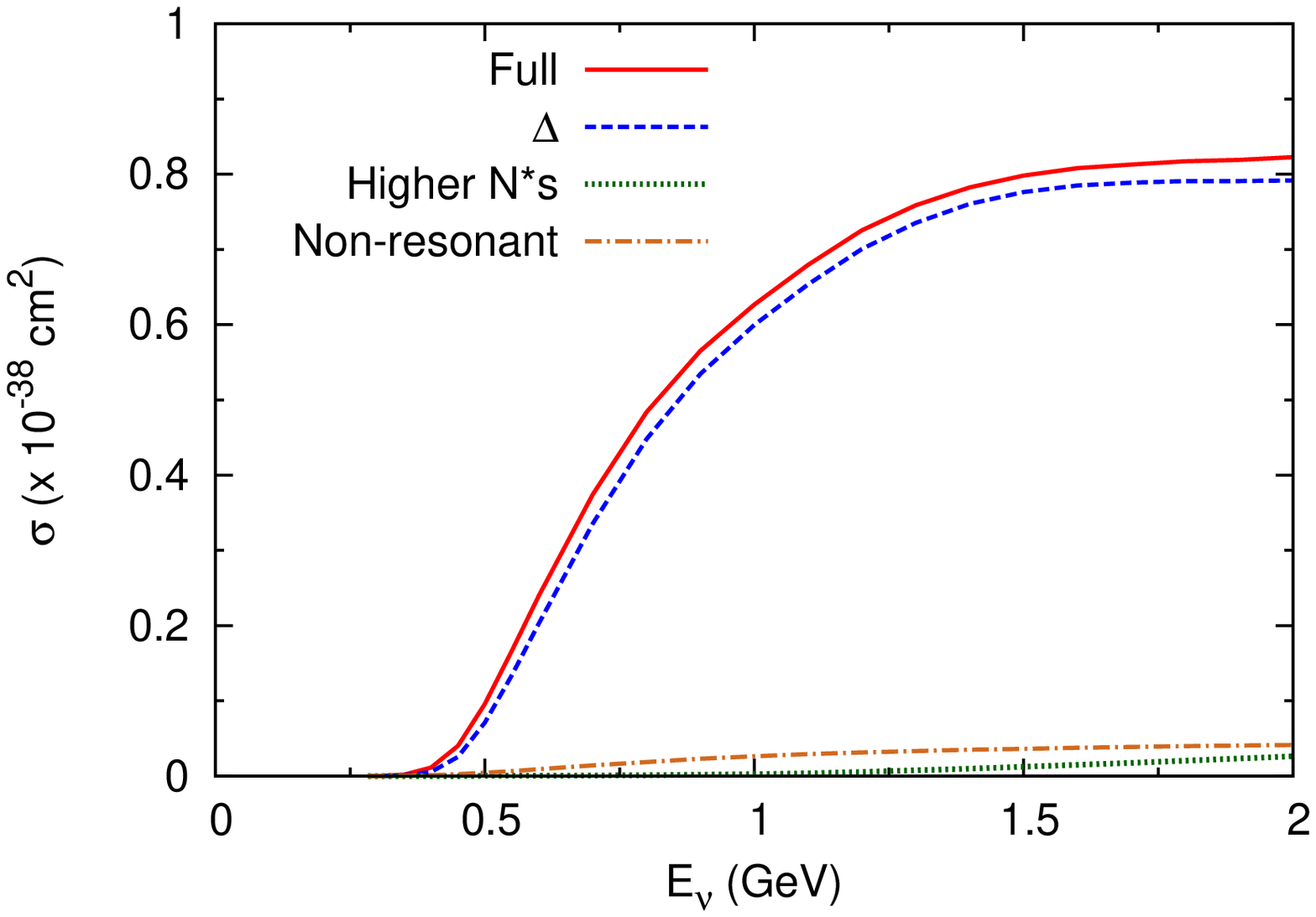}
\includegraphics[height=0.33\textwidth]{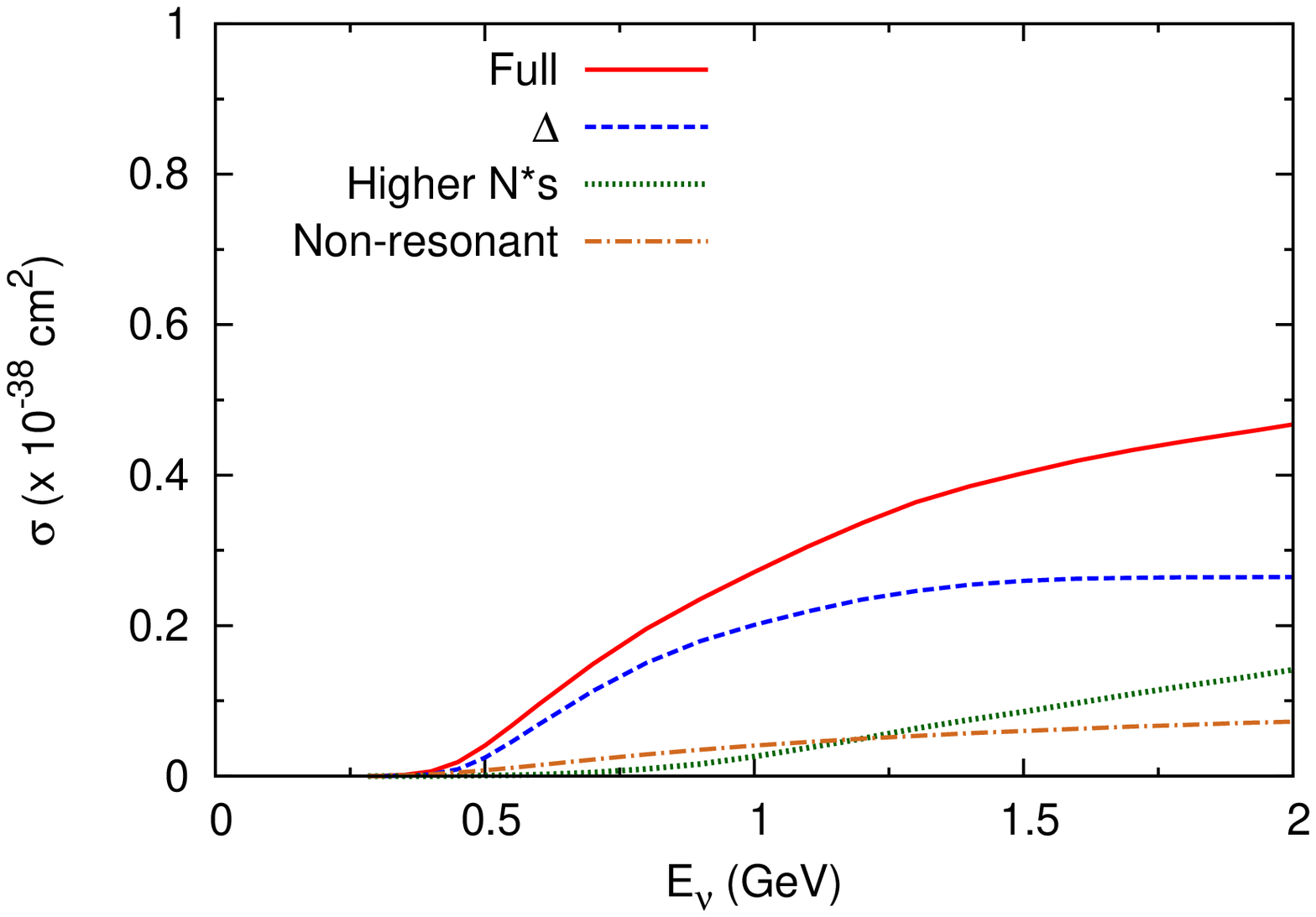}
\caption{(Color online)
Various mechanisms contributing to
$\nu_\mu\, p\to \mu^- \pi^+ p$ (left)
and $\nu_\mu n\to \mu^- \pi N$ (right).
}
\label{fig:neutrino-tot-comp}
\end{figure}
Next we examine reaction mechanisms of the $\nu_\mu\, N$ scattering.
In Fig.~\ref{fig:neutrino-tot-comp}, we break down the single-pion
production cross sections into several contributions each of which contains a
set of certain mechanisms.
For the proton-target process, 
the contribution from the $\Delta$(1232) resonance dominates,
while the higher $N^*$ contribution is very small.
The $\Delta$ contribution here is the neutrino cross section calculated
with the resonant amplitude, Eq.~(\ref{eq:resamp}), 
of the $P_{33}$ partial wave only,
while the higher $N^*$ contribution is from 
the resonant amplitude including all partial waves other than
$P_{33}$.
The non-resonant cross sections
calculated from the non-resonant amplitude of Eq.~(\ref{eq:nonres})
is small for the proton-target process.
In contrast, the situation is more complex in the neutron-target
process where the $\Delta$ gives a smaller contribution and 
both $I=$1/2 and 3/2 resonances contribute.
As can be seen in the right panel of 
Fig.~\ref{fig:neutrino-tot-comp},
the $\Delta$ dominates for $E_\nu\ltap 1$~GeV, 
and higher resonances and non-resonant mechanisms give comparable
contributions towards $E_\nu\sim 2$~GeV.
This shows an importance of including both resonant and non-resonant
contributions with the interferences among them under control, as has been
stressed in Sec.~\ref{sec:axial-nstar}.
\begin{figure}[t]
\includegraphics[height=0.33\textwidth]{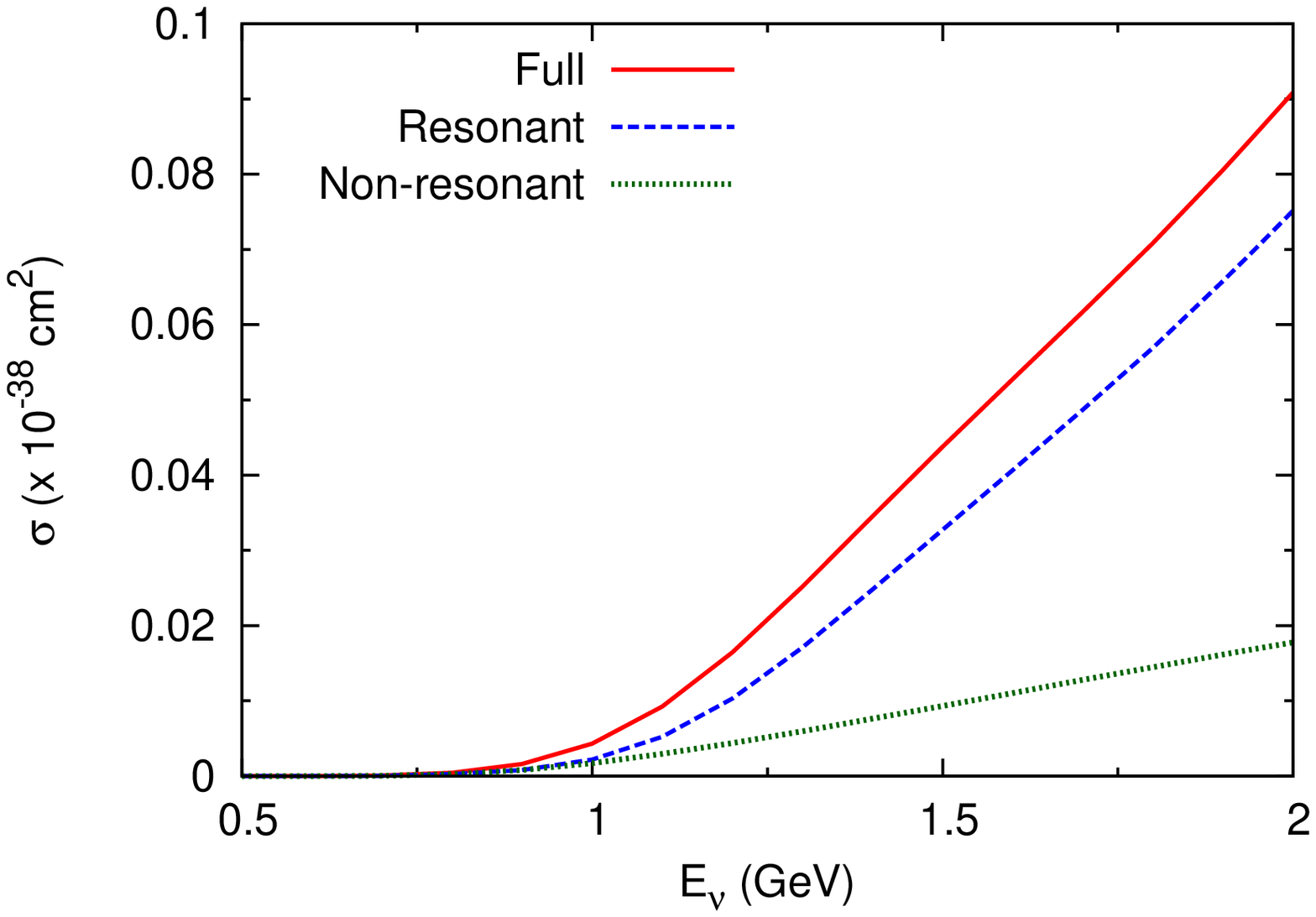}
\includegraphics[height=0.33\textwidth]{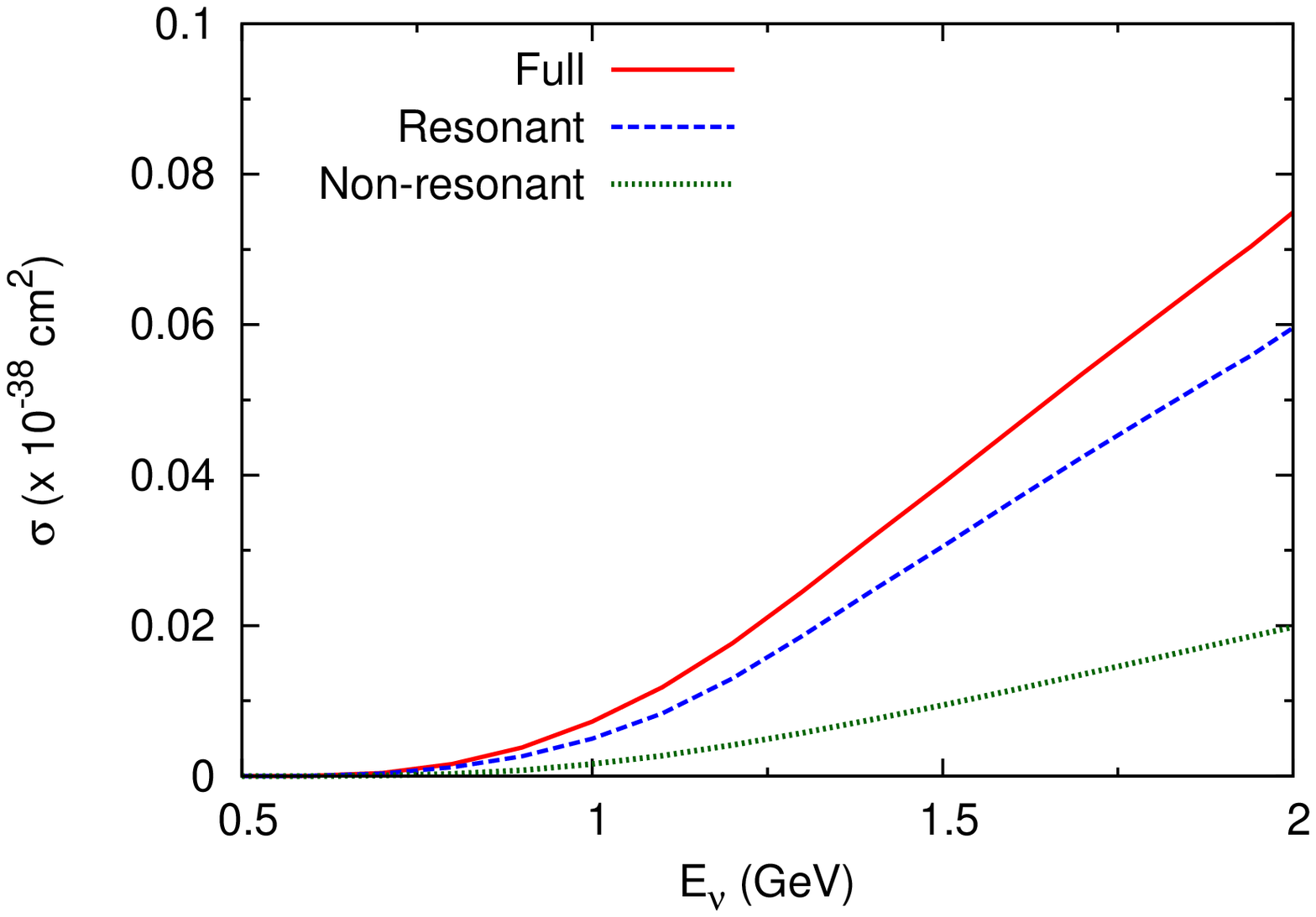}
\caption{(Color online)
Various mechanisms contributing to
$\nu_\mu\, p\to \mu^- \pi^+\pi^0 p$ (left)
and $\nu_\mu n\to \mu^- \pi^+\pi^- p$ (right).
}
\label{fig:neutrino-tot-pipin-comp}
\end{figure}
Similarly, 
the contribution of resonant and non-resonant amplitudes are
shown in 
Fig.~\ref{fig:neutrino-tot-pipin-comp}
for the two-pion production reaction.
Because $\Delta(1232)$ mainly contributes below the $\pi\pi N$
production threshold and thus gives a small contribution here,
the resonant and non-resonant contributions are more comparable.
Still, the figures show that the resonance-excitations are the main
mechanism for the double-pion production in this energy region.

\begin{figure}[t]
\includegraphics[height=0.24\textwidth]{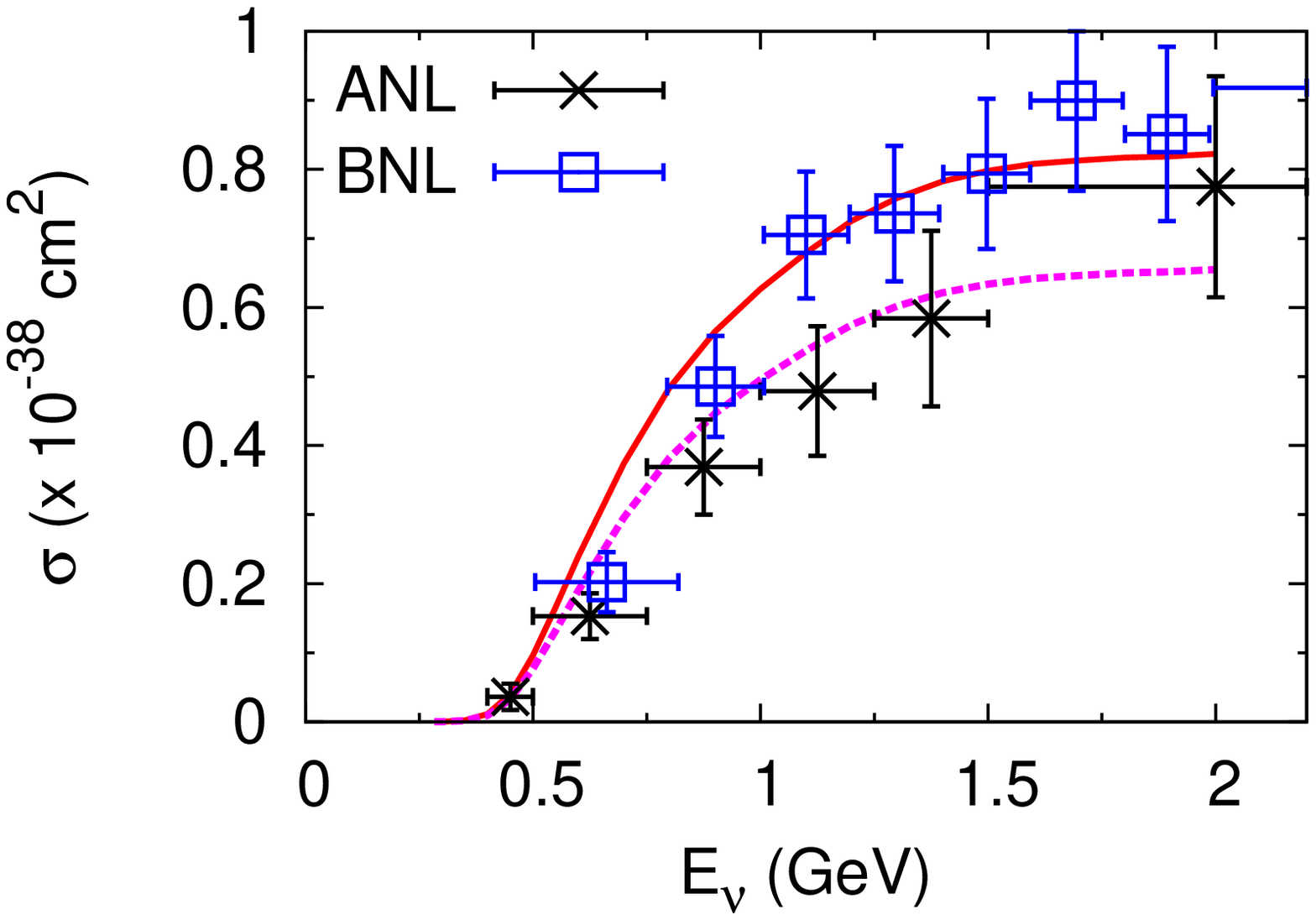}
\hspace{-5mm}
\includegraphics[height=0.24\textwidth]{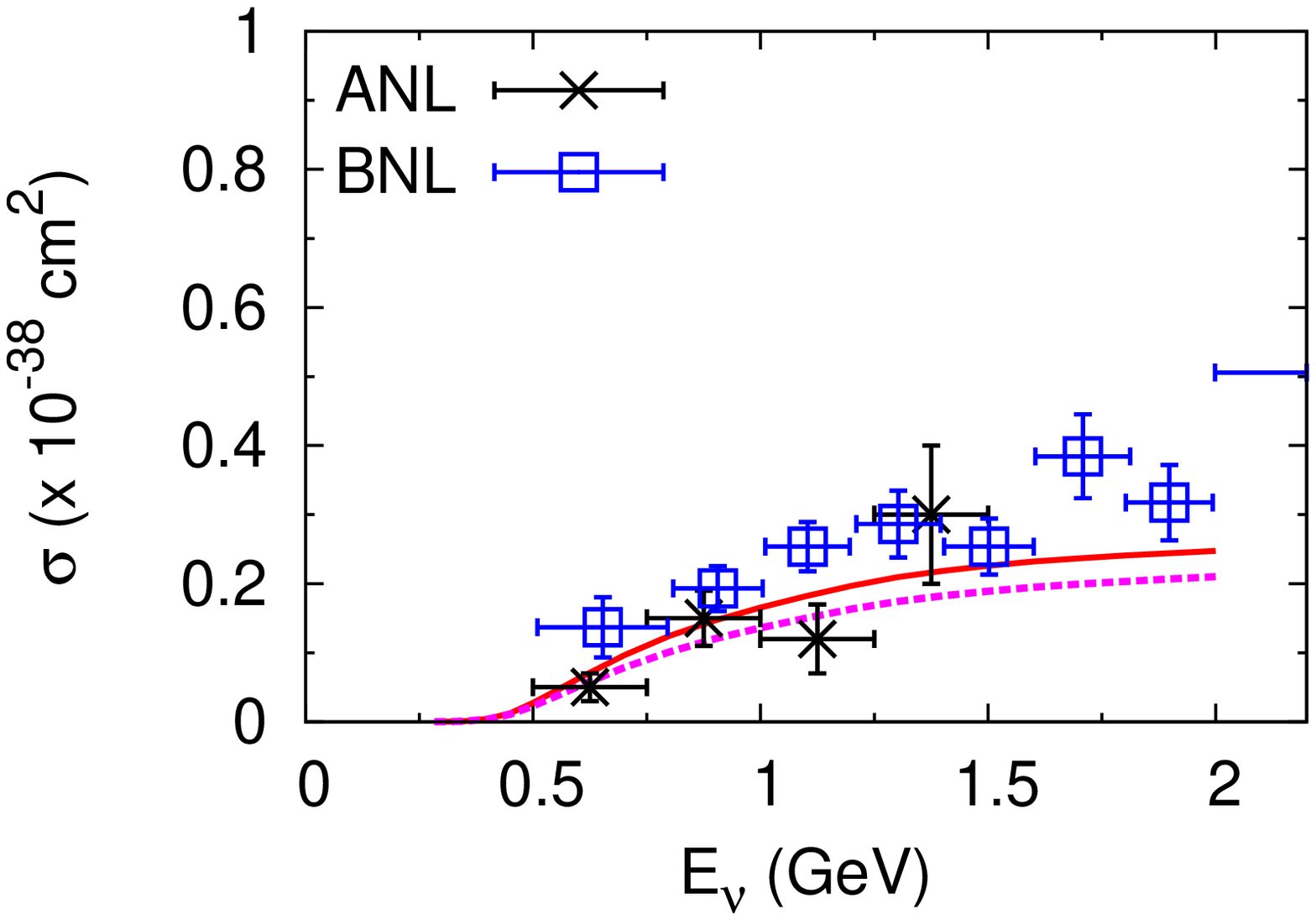}
\hspace{-5mm}
\includegraphics[height=0.24\textwidth]{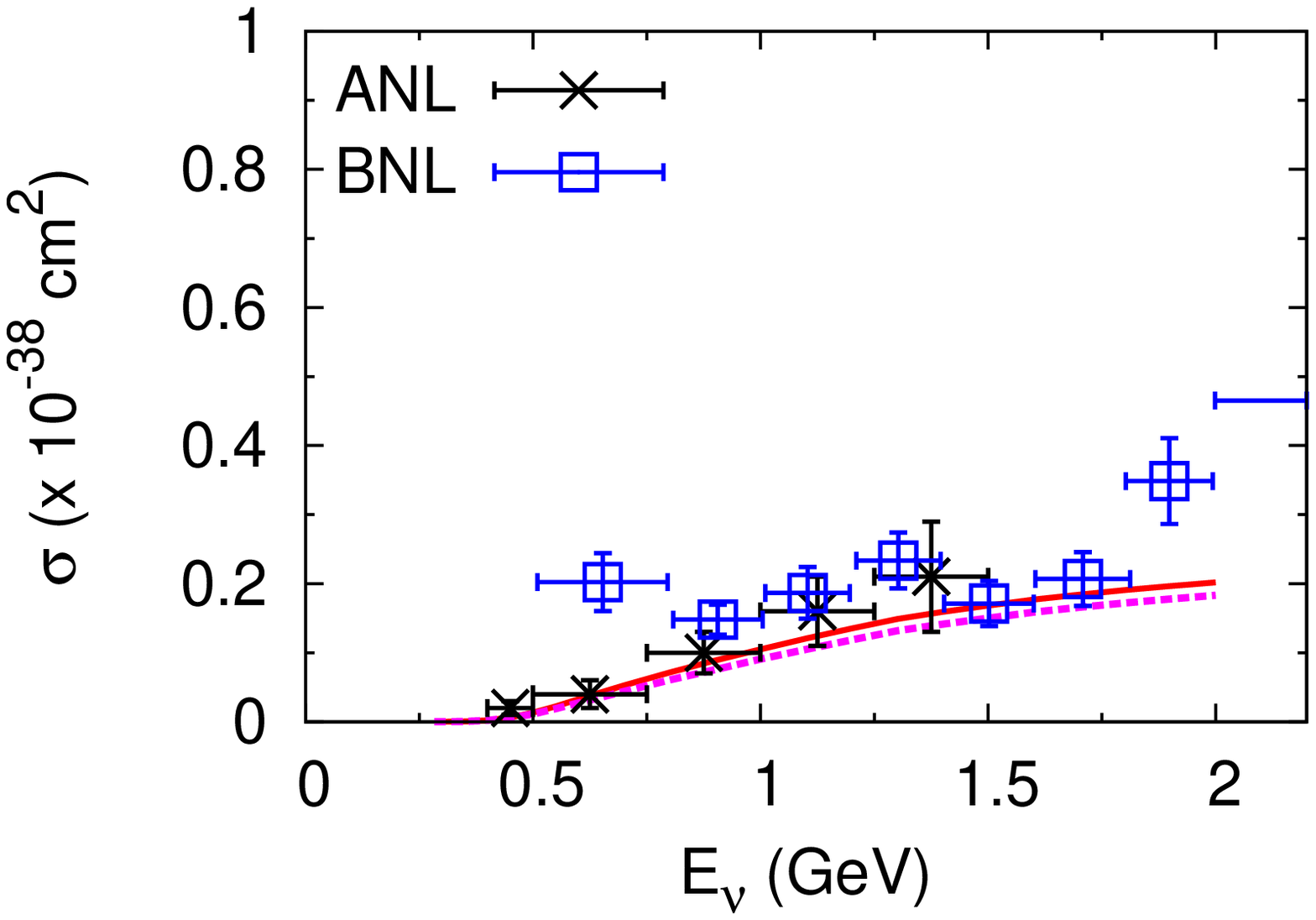}
\caption{(Color online)
Comparison of the DCC-based calculation (red solid curves)
with data for 
$\nu_\mu\, p\to \mu^- \pi^+ p$ (left),
$\nu_\mu n\to \mu^- \pi^0 p$ (middle)
and $\nu_\mu n\to \mu^- \pi^+ n$ (right).
The DCC calculation with $0.8\times g_{AN\Delta(1232)}^{\rm PCAC}$
is also shown (magenta dashed curve).
ANL (BNL) data are from Ref.~\cite{anl} (\cite{bnl}).
}
\label{fig:neutrino-tot-data}
\end{figure}
Next we compare the CC neutrino-induced single pion production cross
sections from the DCC model with
available data from Refs.~\cite{anl,bnl} in Fig.~\ref{fig:neutrino-tot-data}.
The left panel shows the total cross sections for
$\nu_\mu\, p\to \mu^- \pi^+ p$ for which $\Delta(1232)$ dominates as we
have seen in Fig.~\ref{fig:neutrino-tot-comp}.
If the $\Delta(1232)$-dominance persists in the neutron-target processes
shown in the middle and right panels of
Fig.~\ref{fig:neutrino-tot-data},
the isospin Clebsch-Gordan coefficients determine the relative strength
as
$\sigma(\nu_\mu n\to \mu^- \pi^0 p)/\sigma(\nu_\mu p\to \mu^- \pi^+ p) = 2/9 \sim 0.22$, and
$\sigma(\nu_\mu n\to \mu^- \pi^+ n)/\sigma(\nu_\mu p\to \mu^- \pi^+ p) = 1/9 \sim 0.11$.
The actual ratios from the DCC model are
$\sigma(\nu_\mu n\to \mu^- \pi^0 p)/\sigma(\nu_\mu p\to \mu^- \pi^+ p) =$
0.28, 0.27, 0.29, and 
$\sigma(\nu_\mu n\to \mu^- \pi^+ n)/\sigma(\nu_\mu p\to \mu^- \pi^+ p) =$
0.13, 0.17, 0.21
at $E_\nu$=0.5, 1, 1.5~GeV, respectively.
The deviations from the naive isospin analysis are due to the
the non-resonant and higher-resonances 
contributions mostly in the neutron-target processes,
as we have seen in Fig.~\ref{fig:neutrino-tot-comp}.
The two datasets from BNL and ANL 
for $\nu_\mu p\to \mu^- \pi^+ p$
shown in the left panel of
Fig.~\ref{fig:neutrino-tot-data} are not consistent
as has been well known,
and our result is closer to the BNL data~\cite{anl}.
For the other channels, our result is fairly consistent with both of the
BNL and ANL data.
It seems that 
the bare axial $N$-$\Delta(1232)$ coupling constants
determined by the PCAC relation are too large to 
reproduce the ANL data.
Because 
axial $N$-$N^*$ coupling constants should be better determined by analyzing
neutrino-reaction data,
it is tempting to 
multiply the bare axial
$N$-$\Delta(1232)$ coupling constants,
$g_{AN\Delta(1232)}^{\rm PCAC}$,
defined in
Eq.~(\ref{eq:ANN*(P33)}) by 0.8, so that 
the DCC model better fits the ANL data.
The resulting cross sections are shown by the dashed curves in 
Fig.~\ref{fig:neutrino-tot-data}.
We find that
$\sigma(\nu_\mu p\to \mu^- \pi^+ p)$ is reduced due to the
dominance of the $\Delta(1232)$ resonance in this channel, while 
$\sigma(\nu_\mu n\to \mu^- \pi N)$ is only slightly reduced.
As mentioned in the introduction,
the original data of these two experimental data have been
reanalyzed recently~\cite{reanalysis}, and it is
pointed out that the discrepancy between the two datasets is resolved. 
The resulting cross sections are closer to the previous ANL data.
However, the number of data is still very limited, and 
a new measurement of neutrino cross sections on the hydrogen and
deuterium is highly desirable.
We also note that the data shown in Fig.~\ref{fig:neutrino-tot-data} were taken from
experiments using the deuterium target.
Thus one should analyze the data considering the nuclear effects such as
the initial two-nucleon correlation and the final state interactions.
Recently, the authors of Ref.~\cite{wsl} have taken a first step towards 
such an analysis.
They developed a model that consists of
elementary amplitudes for neutrino-induced single pion production off the
nucleon~\cite{sul}, pion-nucleon rescattering amplitudes, and 
the deuteron and final $NN$ scattering wave functions.
Although they did not analyze the ANL and BNL data with their model, 
they examined how much the cross sections at certain kinematics
can be changed by considering the nuclear effects. 
They found that the cross sections can be reduced as much as 30\% 
for $\nu_\mu d\to \mu^- \pi^+ pn$ due to the $NN$ rescattering.
Meanwhile, the cross sections 
for $\nu_\mu d\to \mu^- \pi^0 pp$ are hardly changed by the final state
interaction. 
It will be important to analyze the ANL and BNL data with this kind of model to
determine the axial nucleon current, particularly the axial
$N$-$\Delta$(1232) transition strength.

\begin{figure}[t]
\includegraphics[height=0.24\textwidth]{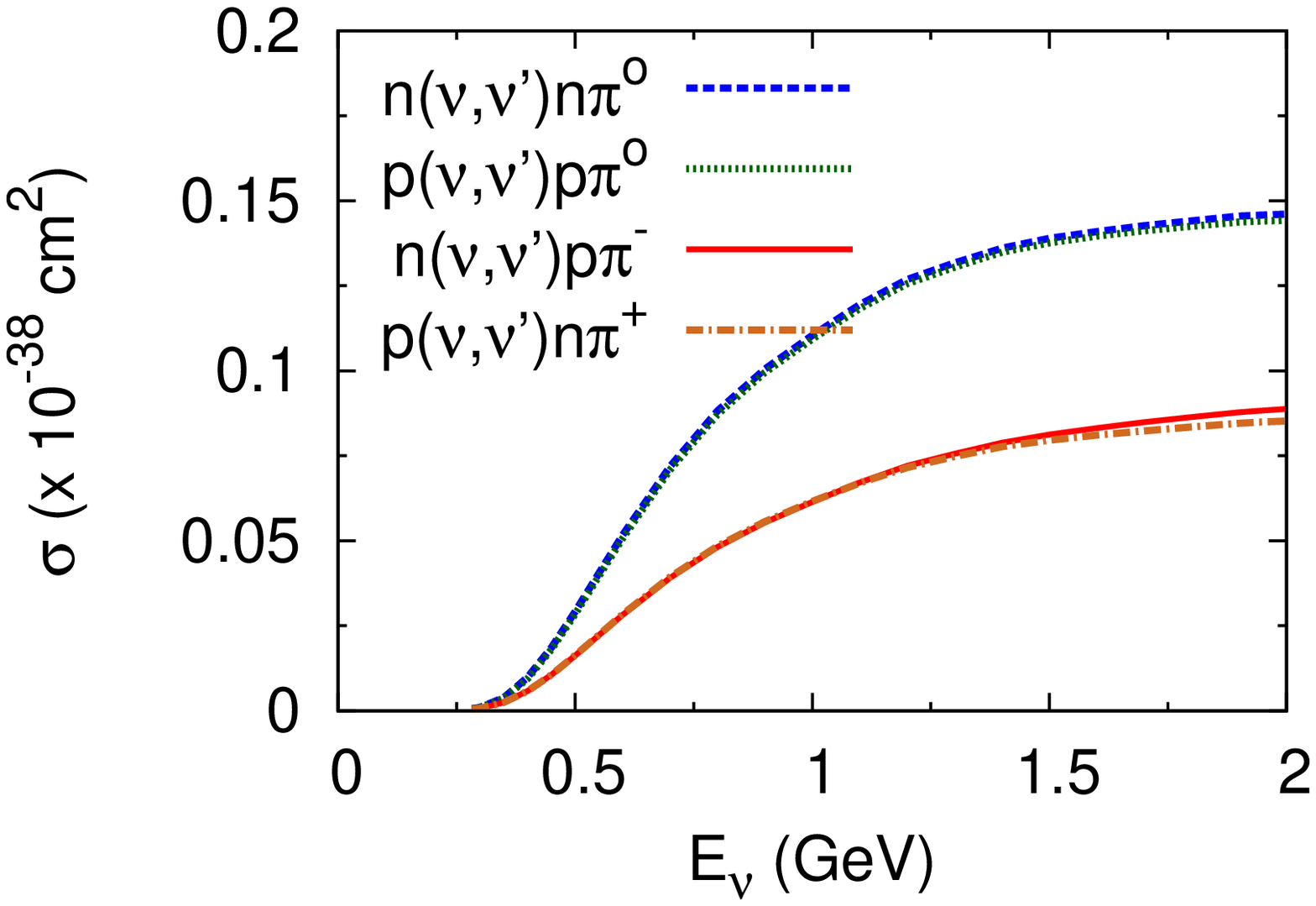}
\hspace{-5mm}
\includegraphics[height=0.24\textwidth]{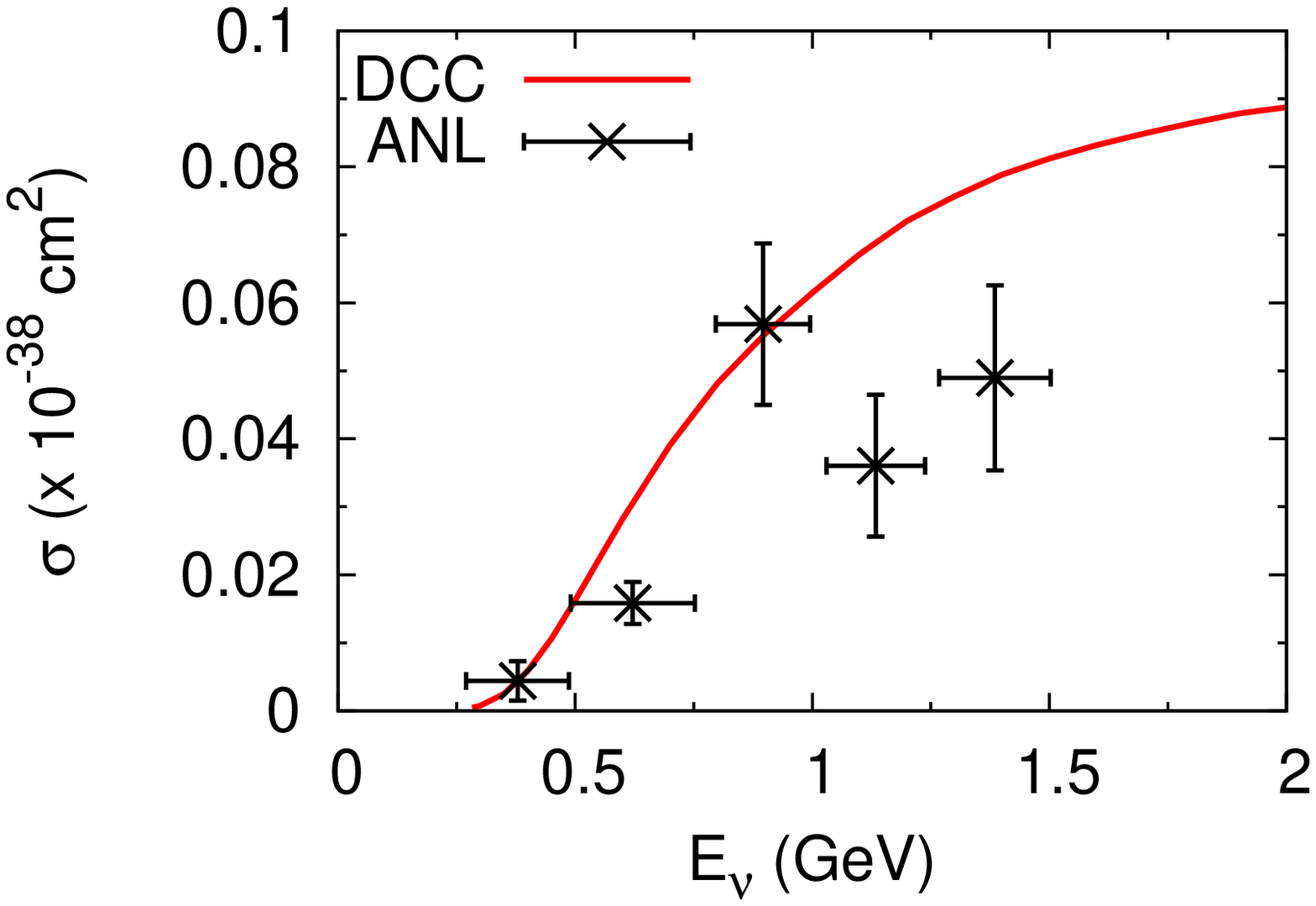}
\hspace{-5mm}
\includegraphics[height=0.24\textwidth]{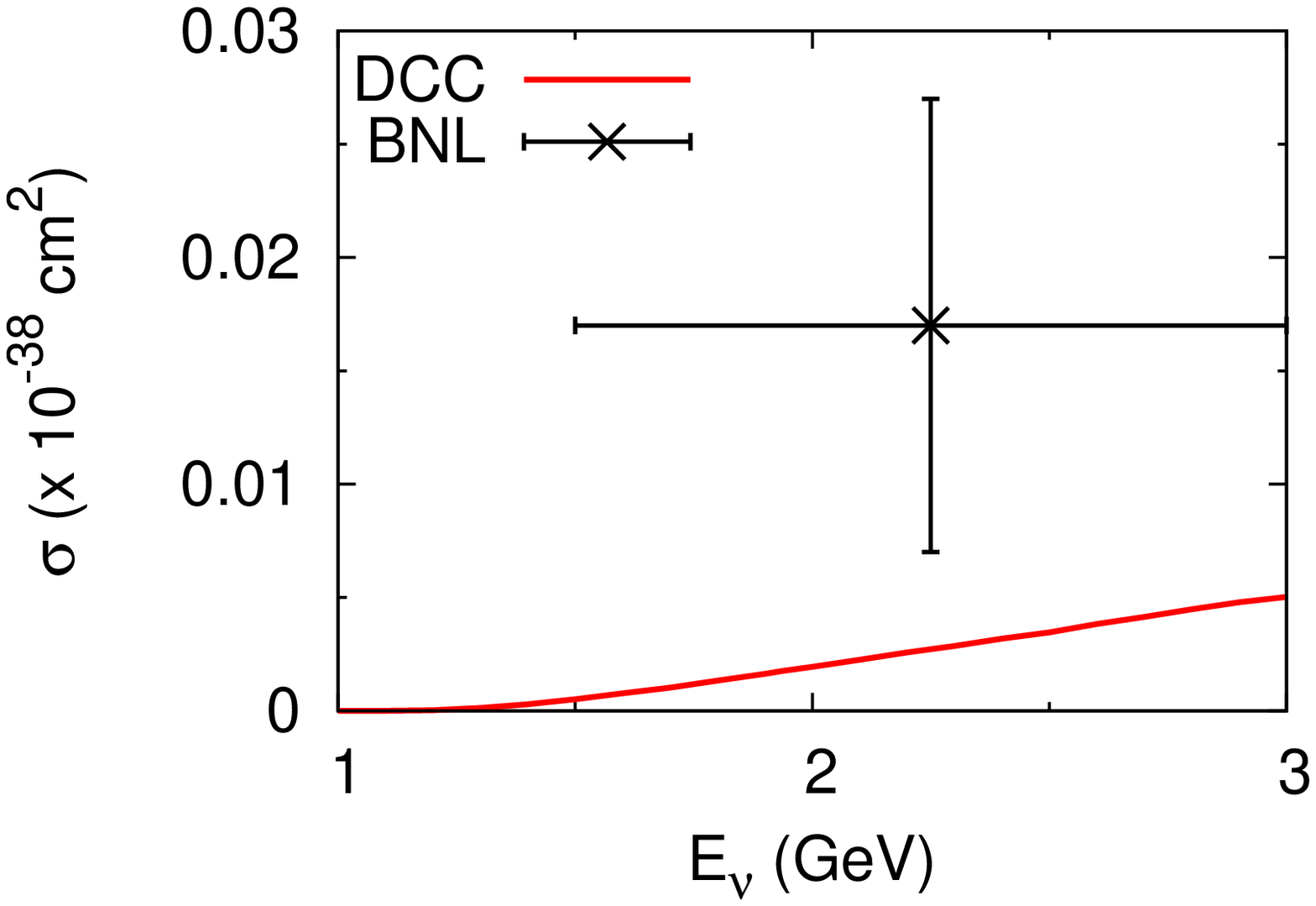}
\caption{(Color online)
(left) DCC-based calculation for NC neutrino-induced single pion
 productions.
(middle) Comparison of the DCC model and ANL data~\cite{NC-data}
 for $\nu n \to \nu p \pi^-$.
(right)  Comparison of the DCC model and BNL data~\cite{KL-data}
 for $\nu n \to \mu^- K^+ \Lambda$.
}
\label{fig:NC}
\end{figure}
Regarding the NC single pion production, we show results in
Fig.~\ref{fig:NC}.
In the left panel, we show the cross sections for all final charge
states.
The ratios $\sigma(\nu p\to \nu p \pi^0)/\sigma(\nu p\to \nu n \pi^+)
\sim \sigma(\nu n\to \nu n \pi^0)/\sigma(\nu n\to \nu p \pi^-)\sim 2$
can be mostly understood from the isospin Clebsch-Gordan coefficient
accompanied by the $\Delta\to\pi N$ vertex. 
A slight difference between 
$\sigma(\nu p\to \nu p \pi^0)$ and $\sigma(\nu n\to \nu n \pi^0)$
[also between $\sigma(\nu p\to \nu n \pi^+)$ and $\sigma(\nu n\to \nu p \pi^-)$]
is mostly from different interference patterns between the isovector
and isoscalar currents.
In the middle panel of Fig.~\ref{fig:NC},
we compare the NC $\nu n\to \nu p \pi^-$ cross sections from the DCC
model with ANL data~\cite{NC-data}, and find a fair consistency.
In closing this paragraph, we compare our DCC-based result for another
single meson production, $\nu n\to\mu^- K^+\Lambda$, with data in the right
panel of Fig.~\ref{fig:NC}.
Although our result undershoots the data, 
the data are based on statistically very limited number of events
(3 events) and we still cannot say something conclusive.

\begin{figure}[t]
\includegraphics[height=0.35\textwidth]{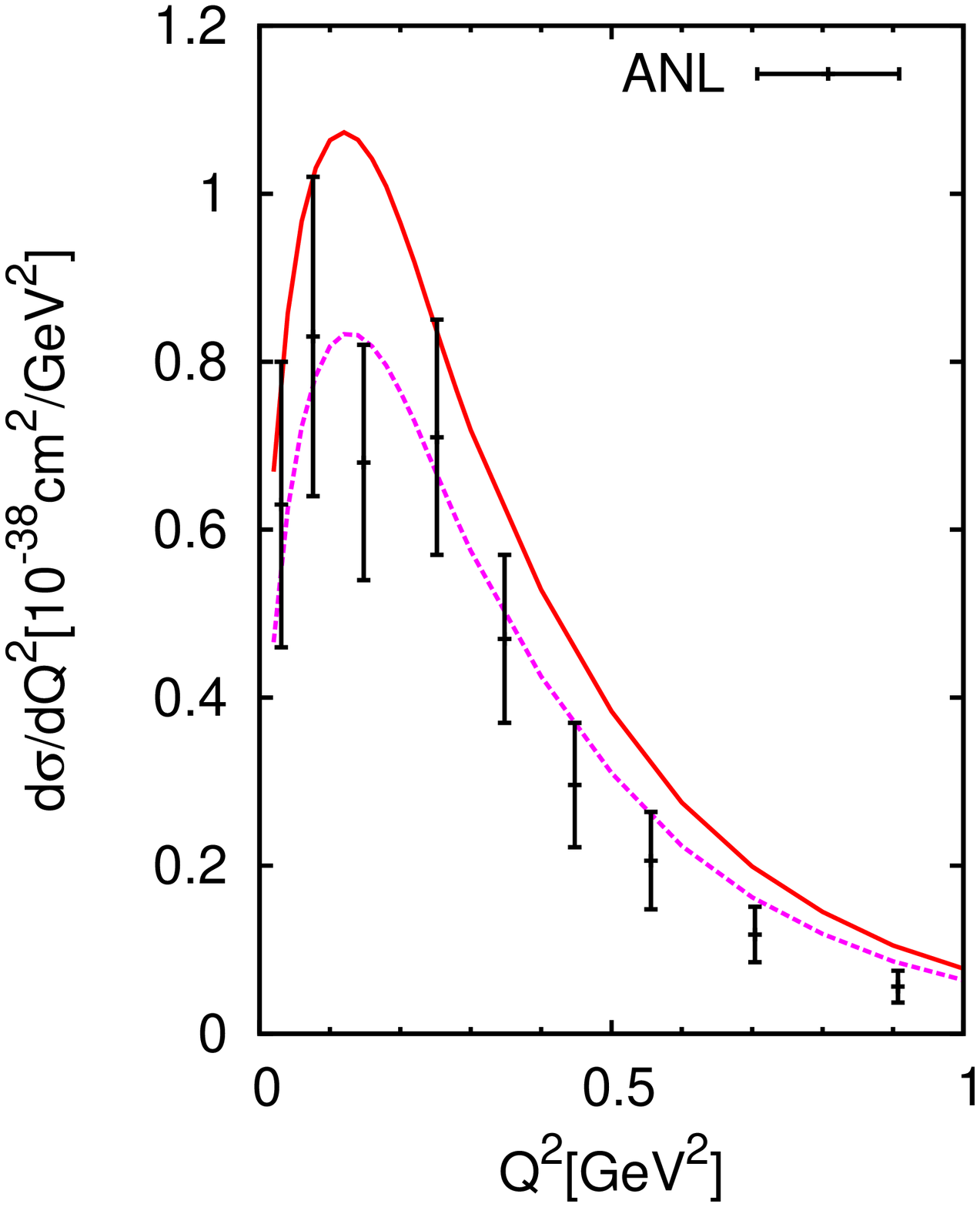}
\includegraphics[height=0.35\textwidth]{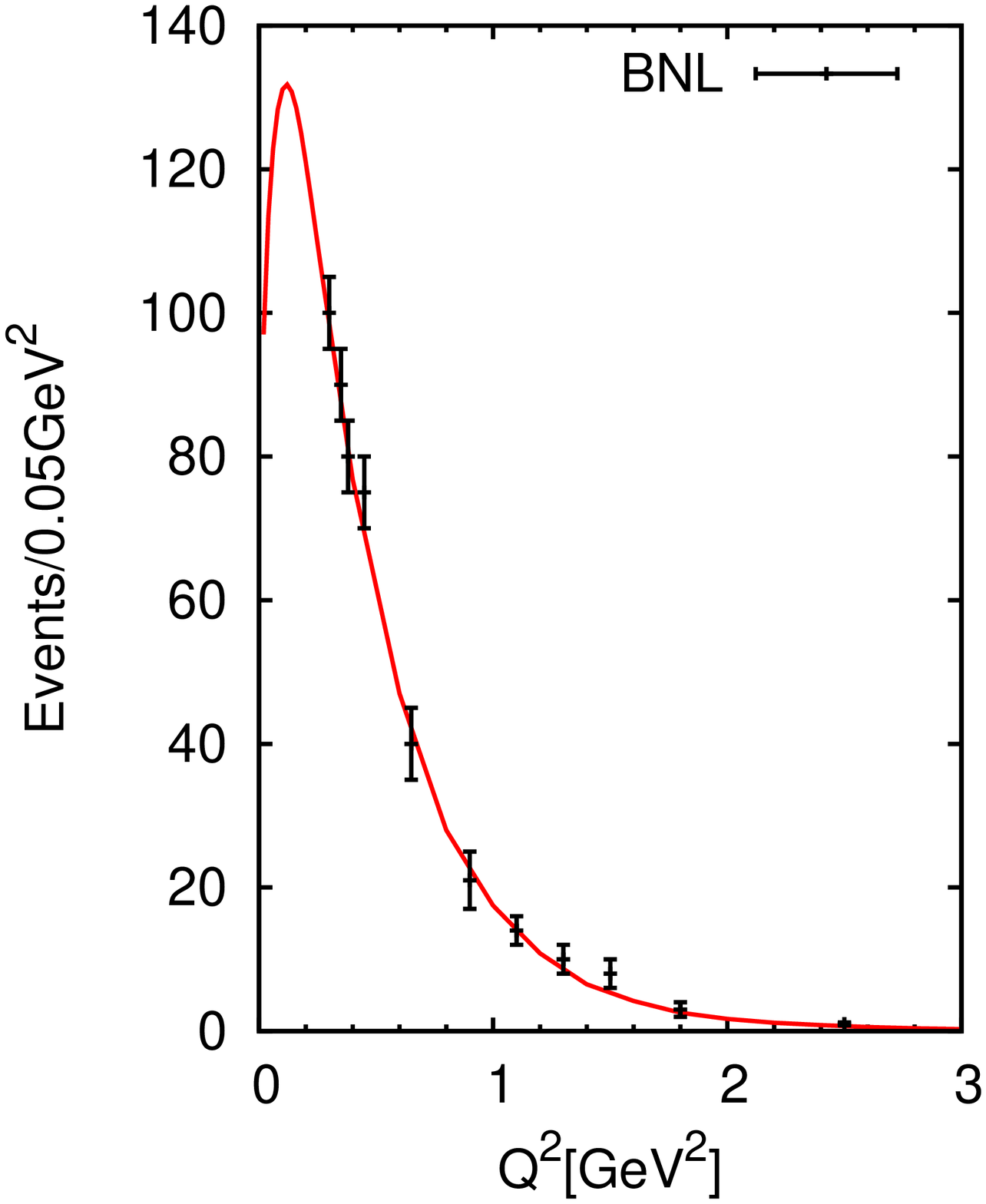}
\caption{(Color online)
The flux-averaged (0.5~GeV~$\le E_\nu\le$ 6~GeV)
$d\sigma/dQ^2$ for $\nu_\mu\, p\to \mu^- \pi^+ p$.
(Left)
The DCC-based calculation (red solid curve) is
compared with the ANL data~\cite{anl}.
The DCC calculation with $0.8\times g_{AN\Delta(1232)}^{\rm PCAC}$
is also shown (magenta dashed curve).
Contributions from high $W$ 
($>$1.4~GeV) are cut in the calculation in order
to be consistent with the data.
(Right)
The DCC-based calculation (red solid curve) is
compared with the BNL data~\cite{bnl}.
}
\label{fig:Q2-dep}
\end{figure}
We also compare our calculation of the single pion production
with data for differential cross sections with respect to
$Q^2$ ($d\sigma/dQ^2$).
To make contact with the data in Refs.~\cite{anl,data90}, 
we calculate the following flux-averaged cross sections:
\begin{eqnarray}
\frac{d\bar{\sigma}}{dQ^2}=
{\displaystyle \int_{E_{\rm min}}^{E_{\rm max}} d E_\nu 
\frac{N(E_\nu)}{\sigma_{\rm model}(E_\nu)}\frac{d\sigma_{\rm model}}{dQ^2}(E_\nu)
\over \displaystyle
\int_{E_{min}}^{E_{max}} d E_\nu \frac{N(E_\nu)}{\sigma_{\rm model}(E_\nu)}}
\ ,
\end{eqnarray}
where $N(E_\nu)$ is the number of events at neutrino
energy $E_\nu$, and 
is given in Fig.~6 of Ref.~\cite{anl} and Fig.~4 of Ref.~\cite{data90}.
The DCC model gives cross sections denoted as $\sigma_{\rm model}(E_\nu)$.
In the left panel of Fig.~\ref{fig:Q2-dep}, we compare the DCC-based
calculation with the ANL data~\cite{anl}.
We find here again that the DCC model overshoots the ANL data for 
$\nu_\mu\, p\to \mu^- \pi^+ p$.
This tempts us 
to plot $d\sigma/dQ^2$ obtained with
the bare axial
$N$-$\Delta(1232)$ coupling constants, 
$g_{AN\Delta(1232)}({\rm PCAC})$, multiplied by 0.8,
as we did in Fig.~\ref{fig:neutrino-tot-data}(left).
This result is in better agreement with the ANL data
as seen in Fig.~\ref{fig:Q2-dep} (left).
The assumed dipole form factor ($M_A=1.02$~GeV)
for the bare axial $N$-$\Delta(1232)$ vertex
seems fairly consistent with the data. 
In the right panel of Fig.~\ref{fig:Q2-dep}, we compare the DCC-based
calculation with the BNL data~\cite{data90}.
The $Q^2$-dependence of the BNL data with the arbitrary scale
is well explained with our DCC model.

\begin{figure}[t]
\includegraphics[height=0.24\textwidth]{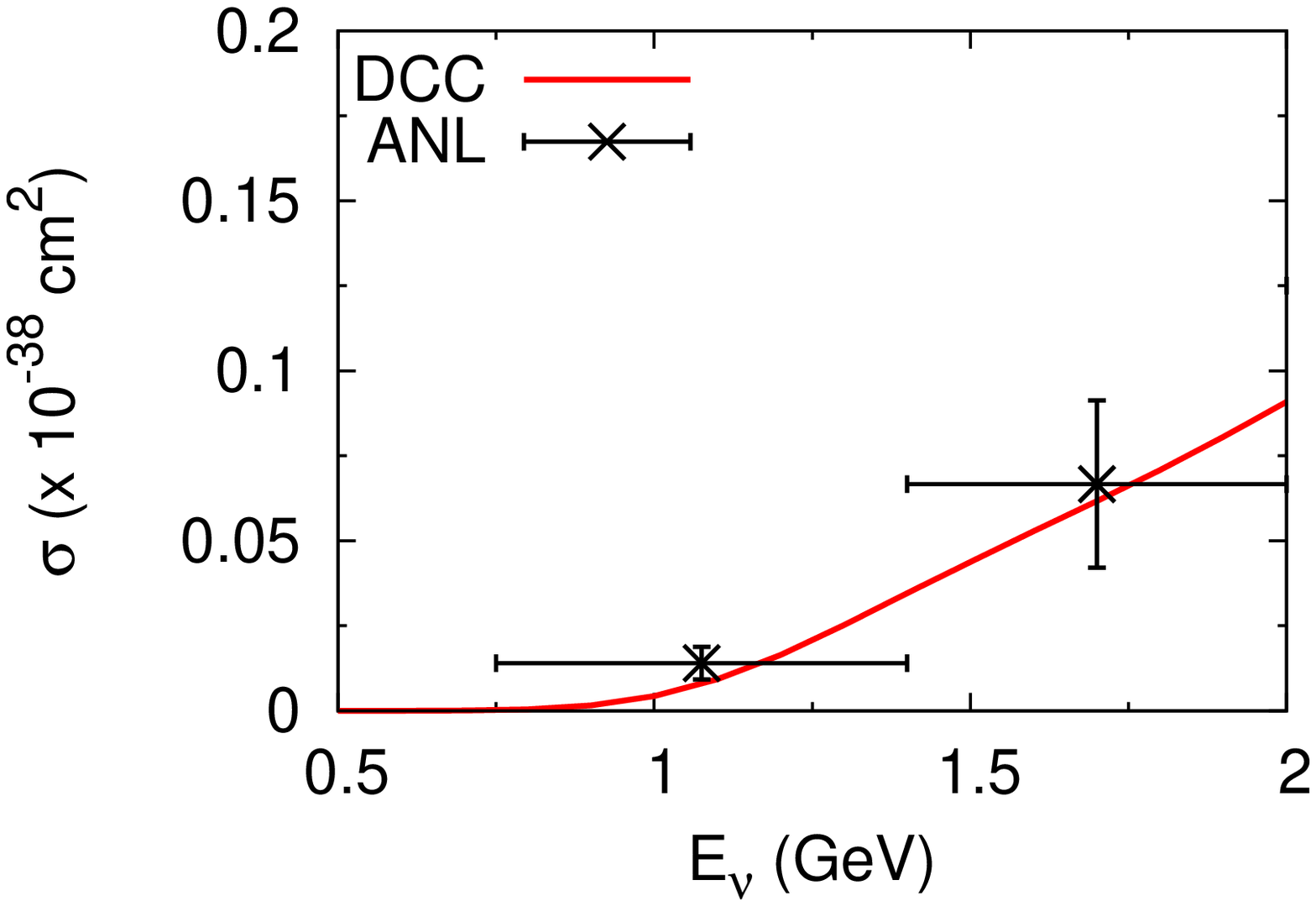}
\hspace{-5mm}
\includegraphics[height=0.24\textwidth]{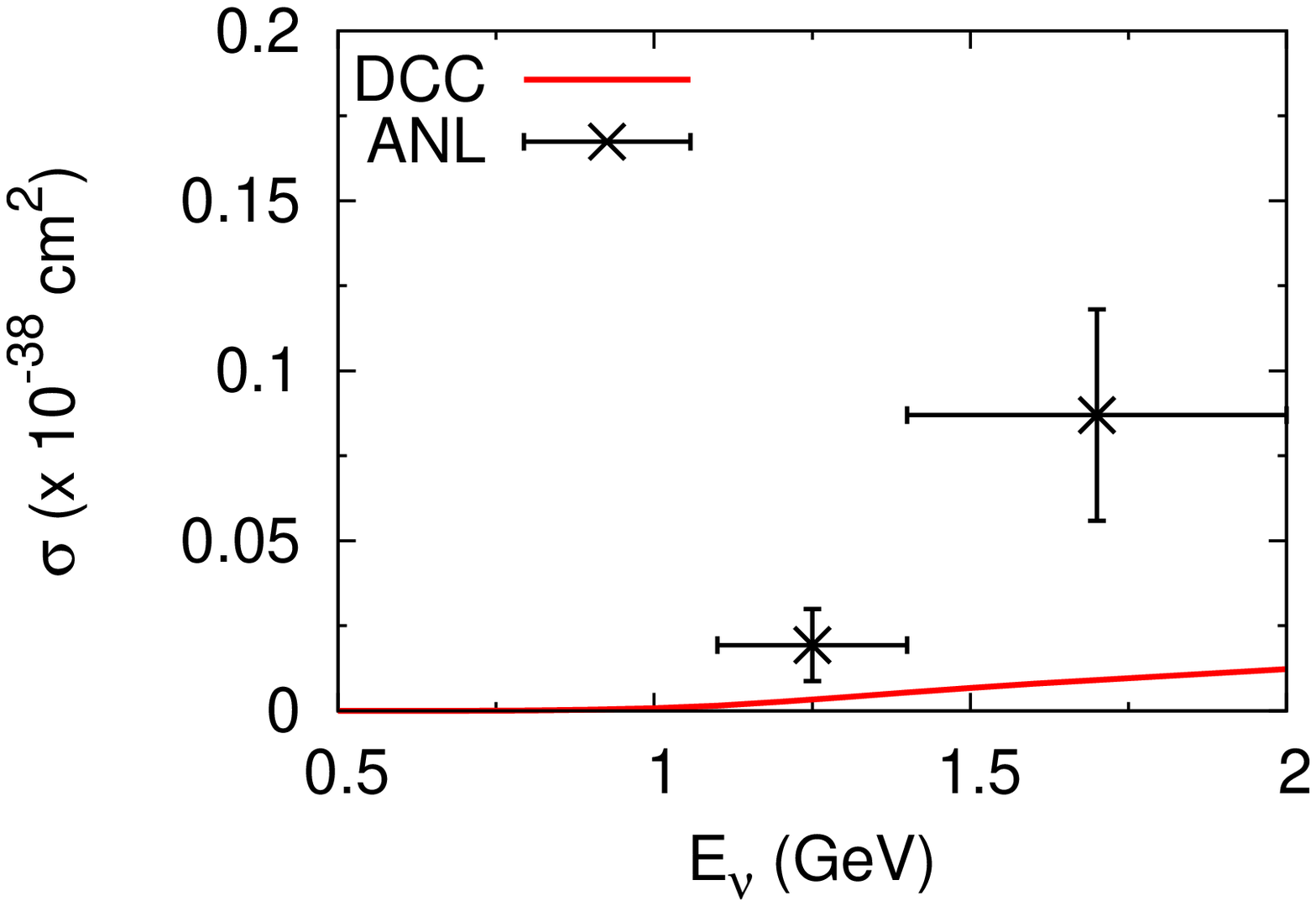}
\hspace{-5mm}
\includegraphics[height=0.24\textwidth]{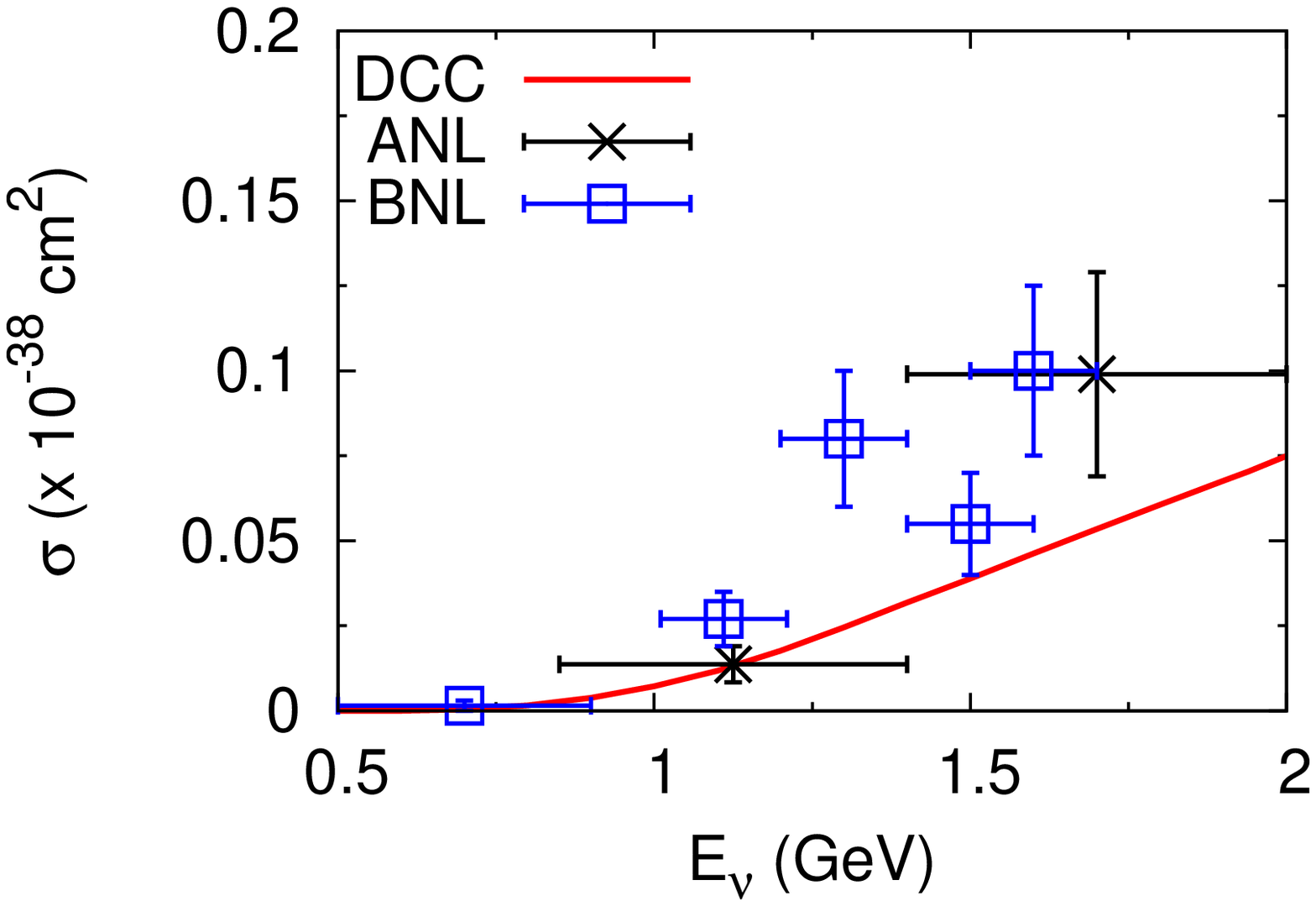}
\caption{(Color online)
Comparison of the DCC-based calculation with data for 
$\nu_\mu\, p\to \mu^- \pi^+\pi^0 p$ (left),
$\nu_\mu p\to \mu^- \pi^+\pi^+ n$ (middle)
and $\nu_\mu n\to \mu^- \pi^+\pi^- p$ (right).
ANL (BNL) data are from Ref.~\cite{anl2} (\cite{bnl}).
}
\label{fig:neutrino-pipin-data}
\end{figure}
We finally compare our results for double-pion productions with existing
data in Fig.~\ref{fig:neutrino-pipin-data}.
Although there exist a few theoretical works on the neutrino-induced double-pion
production near threshold~\cite{biswas,adjei,spain-pipin},
our calculation for the first time takes account of relevant resonance
contributions for this process.
The DCC-based prediction is fairly consistent with the data in the order
of the magnitude.
Particularly,
the cross sections for $\nu_\mu\, p\to \mu^- \pi^+\pi^0 p$ from the DCC
model are in agreement with data.
However, the DCC prediction underestimates 
the $\nu_\mu\, p\to \mu^- \pi^+\pi^+ n$ data.
The rather small ratio of
$\sigma(\nu_\mu\, p\to \mu^- \pi^+\pi^+ n) /
\sigma (\nu_\mu\, p\to \mu^- \pi^+\pi^0 p)\sim 13\%$
at $E_\nu$=2~GeV from our calculation
can be understood as follows.
Within the present DCC-based calculation, $\pi\pi N$ final states are 
from decays of 
the $\pi N$ and
of the $\pi\Delta$, $\rho N$, $\sigma N$ quasi two-body states.
For a neutrino CC process on the proton for which hadronic states have
$I=3/2$, the $\pi N$, $\pi\Delta$, $\rho N$ channels can contribute.
Within the current DCC model, we found that the $\pi\Delta$ channel
gives a dominant contribution to the double pion productions.
Then, retaining only the $\pi\Delta$ contribution,
the ratio is given by
the isospin Clebsch-Gordan coefficients as,
$\sigma(\nu_\mu\, p\to \mu^- \pi^+\pi^+ n) /
\sigma (\nu_\mu\, p\to \mu^- \pi^+\pi^0 p)=2/13\sim 15\%$,
in good agreement with the ratio from the full calculation.
With a very limited dataset, we do not further pursue the origin
of the difference between our calculation and the data.
If the double-pion data are further confirmed, 
then the model needs to incorporate some other mechanisms
and/or adjust model parameters of the DCC model to explain the data.

\begin{figure}[t]
\includegraphics[height=0.33\textwidth]{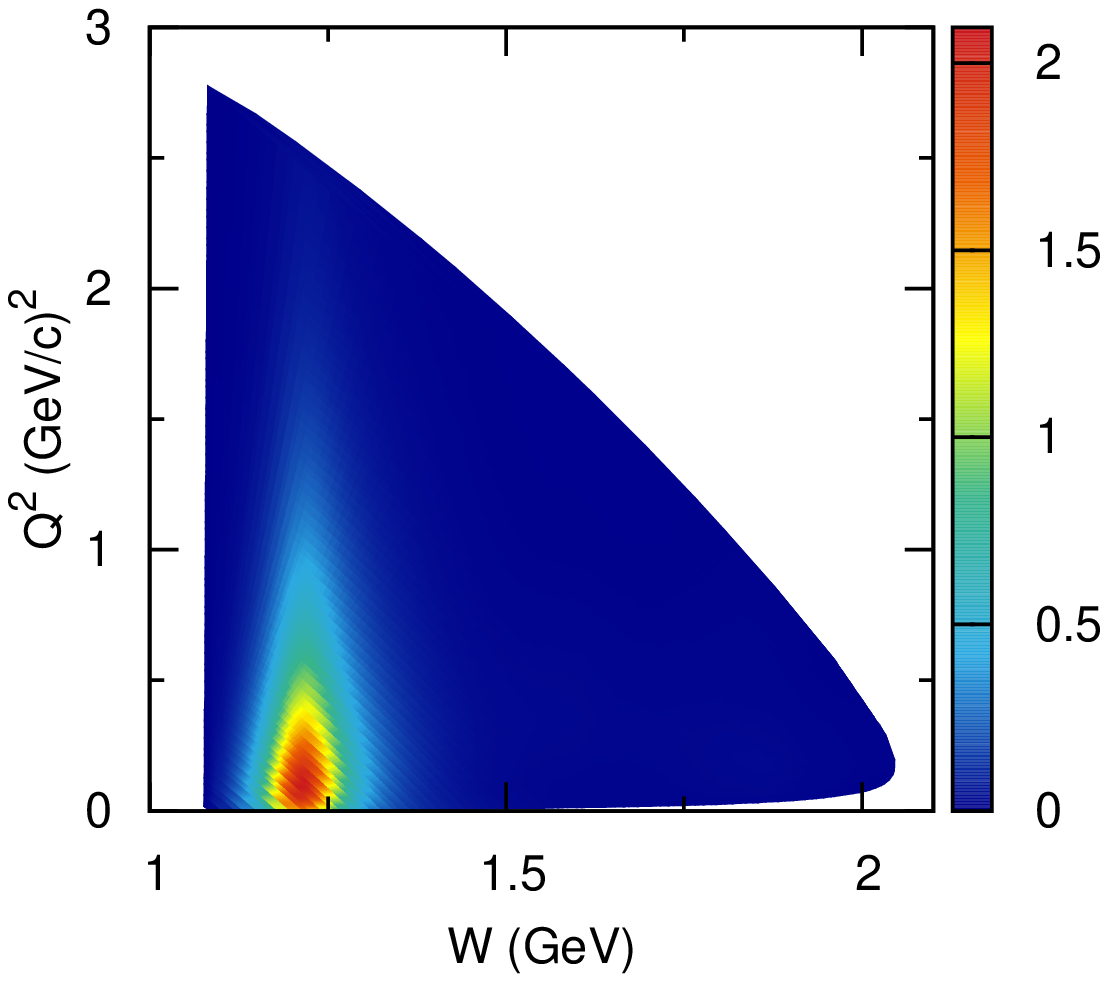}
\hspace{10mm}
\includegraphics[height=0.33\textwidth]{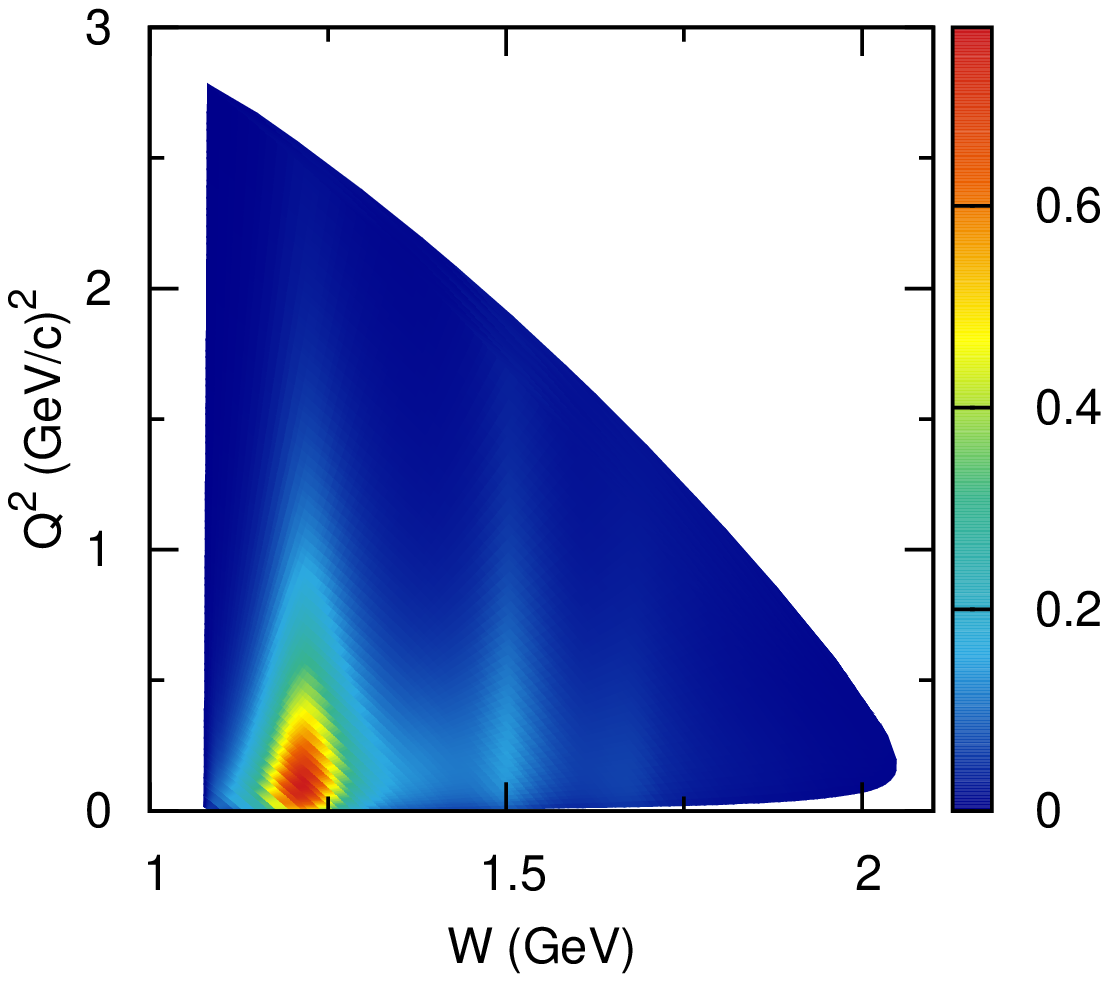}
\caption{(Color online)
Contour plots of $d^2\sigma/dWdQ^2$ for
$\nu_\mu\, p\to \mu^- \pi^+ p$ (left)
and $\nu_\mu n\to \mu^- \pi N$ (right)
at $E_\nu=2$~GeV.
}
\label{fig:ds_dw_dq2_pin}
\end{figure}
\begin{figure}[t]
\includegraphics[height=0.33\textwidth]{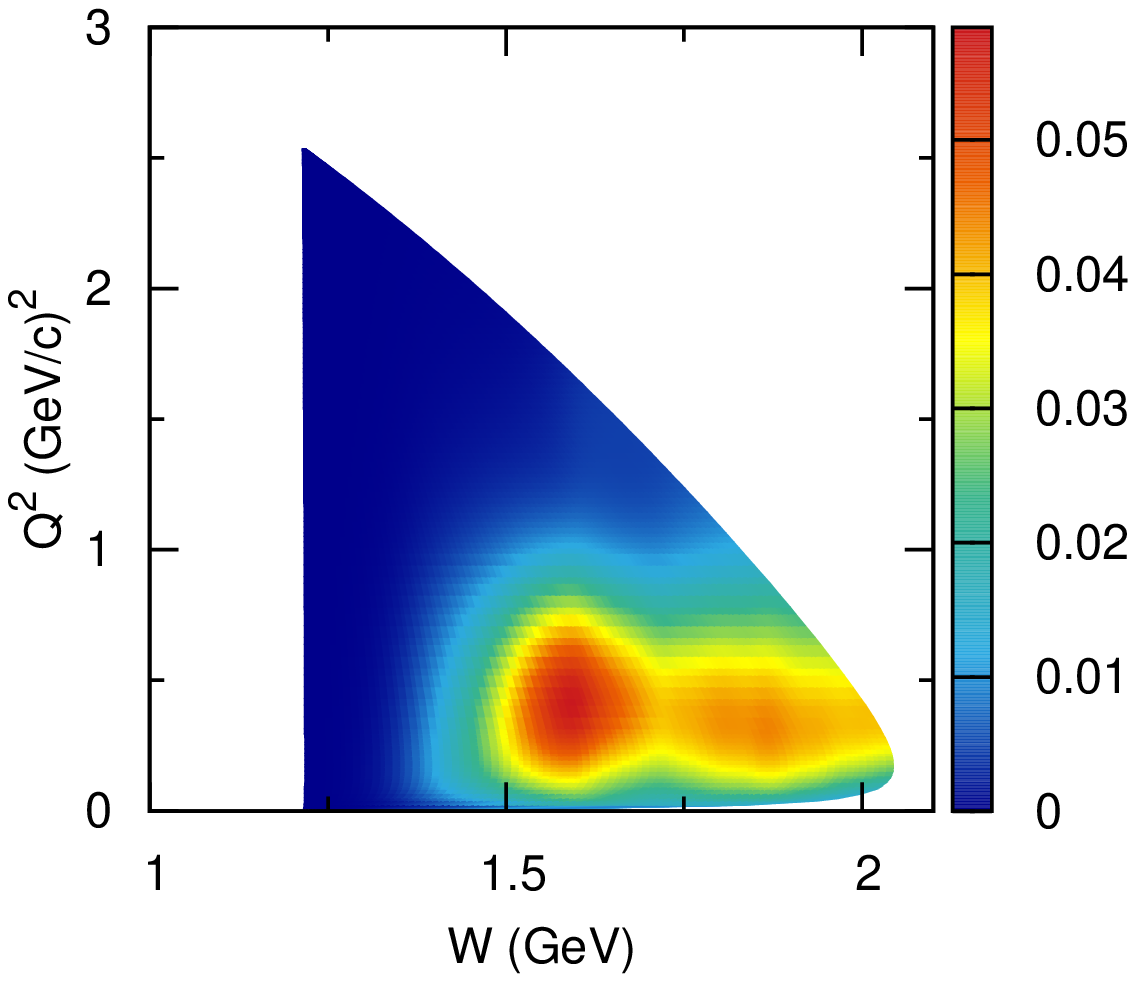}
\hspace{10mm}
\includegraphics[height=0.33\textwidth]{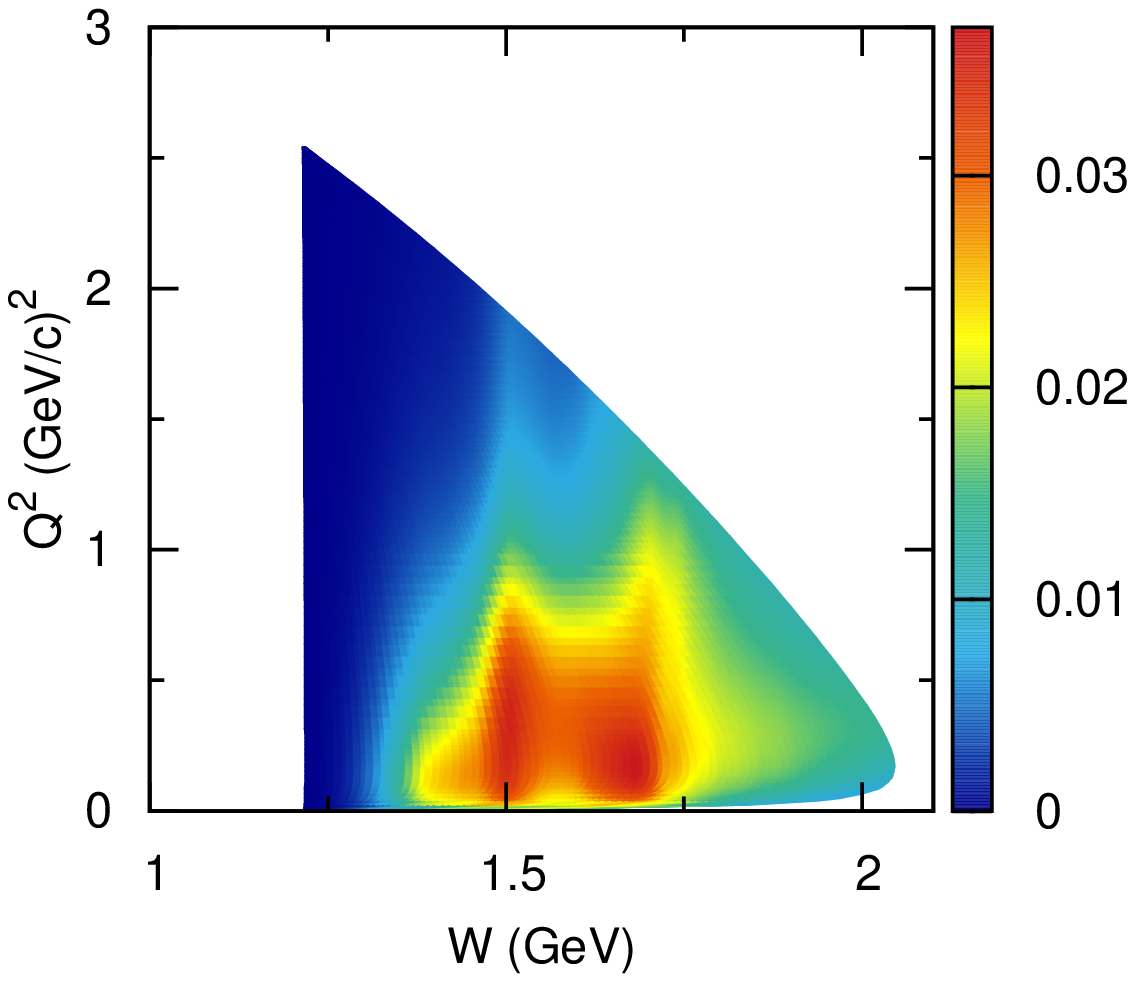}
\caption{(Color online)
Contour plots of $d^2\sigma/dWdQ^2$ for
$\nu_\mu\, p\to \mu^- \pi^+ \pi^0 p$ (left)
and $\nu_\mu n\to \mu^- \pi^+\pi^- p$ (right)
at $E_\nu=2$~GeV.
}
\label{fig:ds_dw_dq2_pipin}
\end{figure}
Important and characteristic hadronic dynamics changes as
$W$ and $Q^2$ change.
Thus, it would be interesting to see 
double-differential cross sections, $d^2\sigma/dWdQ^2$, 
as shown in Fig.~\ref{fig:ds_dw_dq2_pin} 
for the single-pion productions.
The prominent peak due to $\Delta(1232)$ has a long tail toward higher
$Q^2$ region. 
For the neutron-target, the resonant behavior in the second resonance
region is also seen.
Similar contour plots are also shown for double-pion productions in 
Fig.~\ref{fig:ds_dw_dq2_pipin}.
Here, the situation is very different from the single pion case, and
the main contributors are resonances in the second and third
resonance regions. 

\begin{figure}[t]
\includegraphics[height=0.33\textwidth]{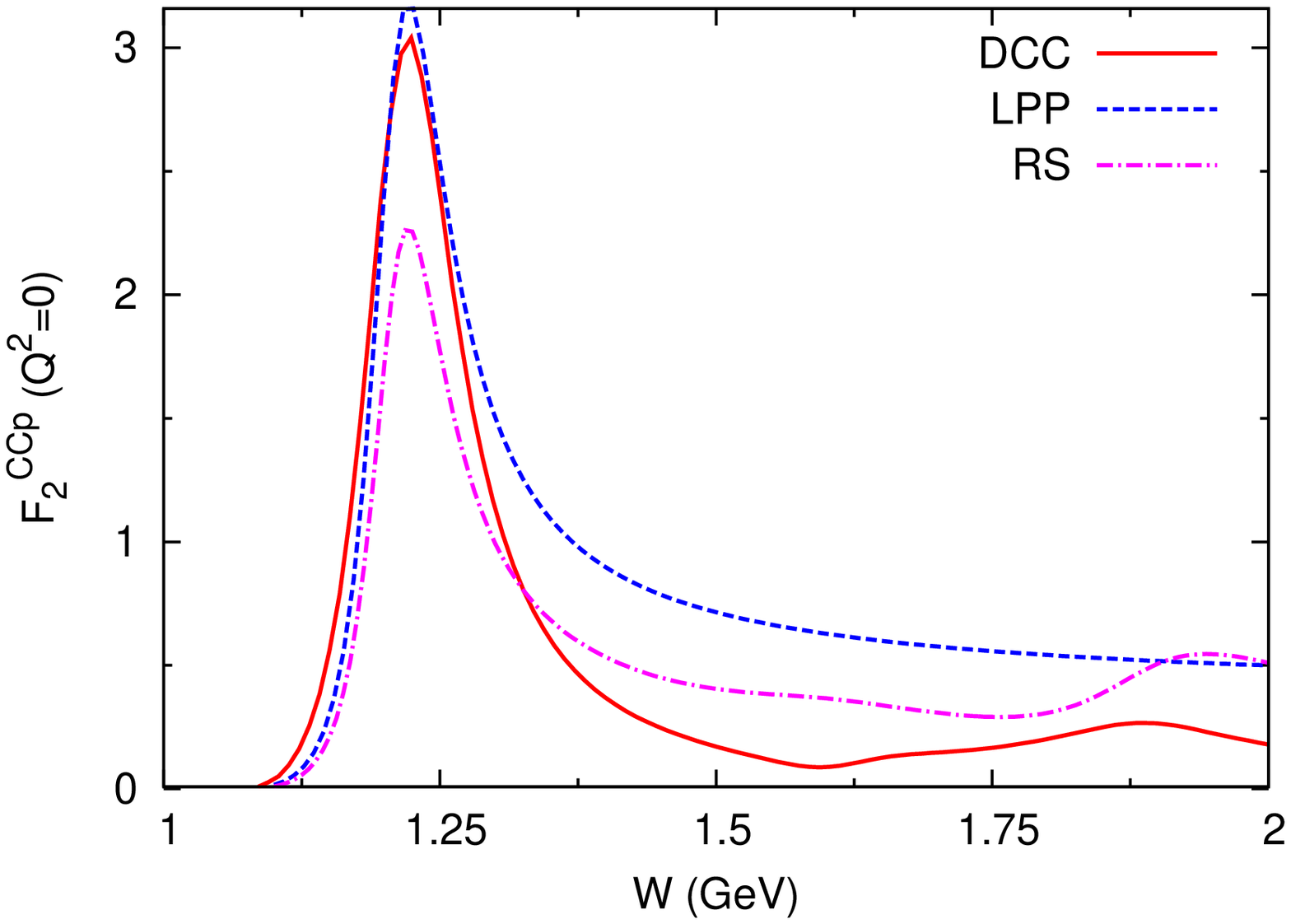}
\includegraphics[height=0.33\textwidth]{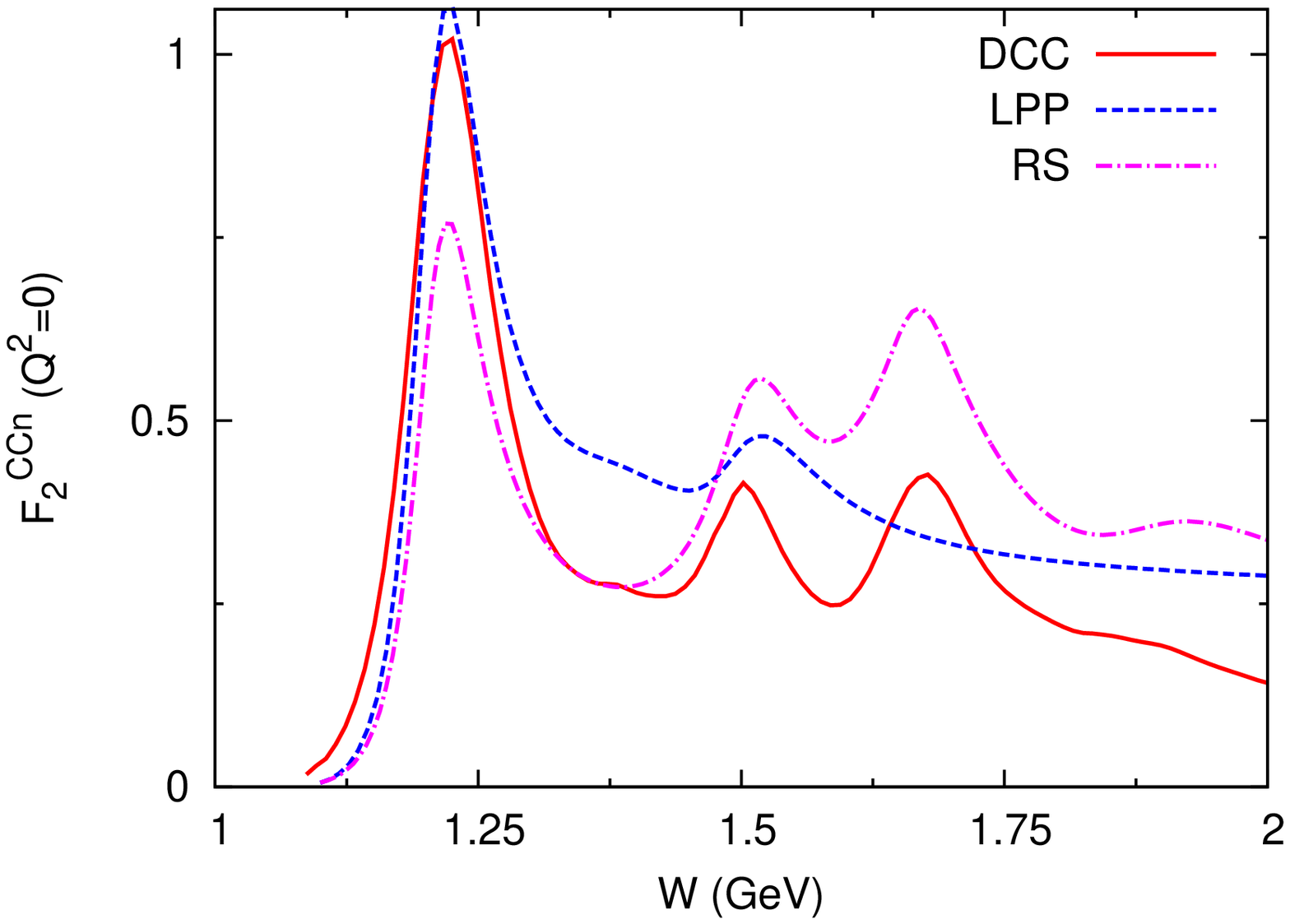}
\caption{(Color online)
$F^{\rm CC}_2$ at $Q^2=0$ 
from contribution of the single pion production only. 
The DCC model is compared with
the LPP model due to Lalakulich et al.~\cite{lalakulich}
and the Rein-Sehgal (RS) model~\cite{RS,RS2}.
The left (right) panel is for the CC
$\nu_\mu\, p$ ($\nu_\mu\, n$) reaction.
}
\label{fig:LP}
\end{figure}

\begin{figure}[t]
\includegraphics[height=0.33\textwidth]{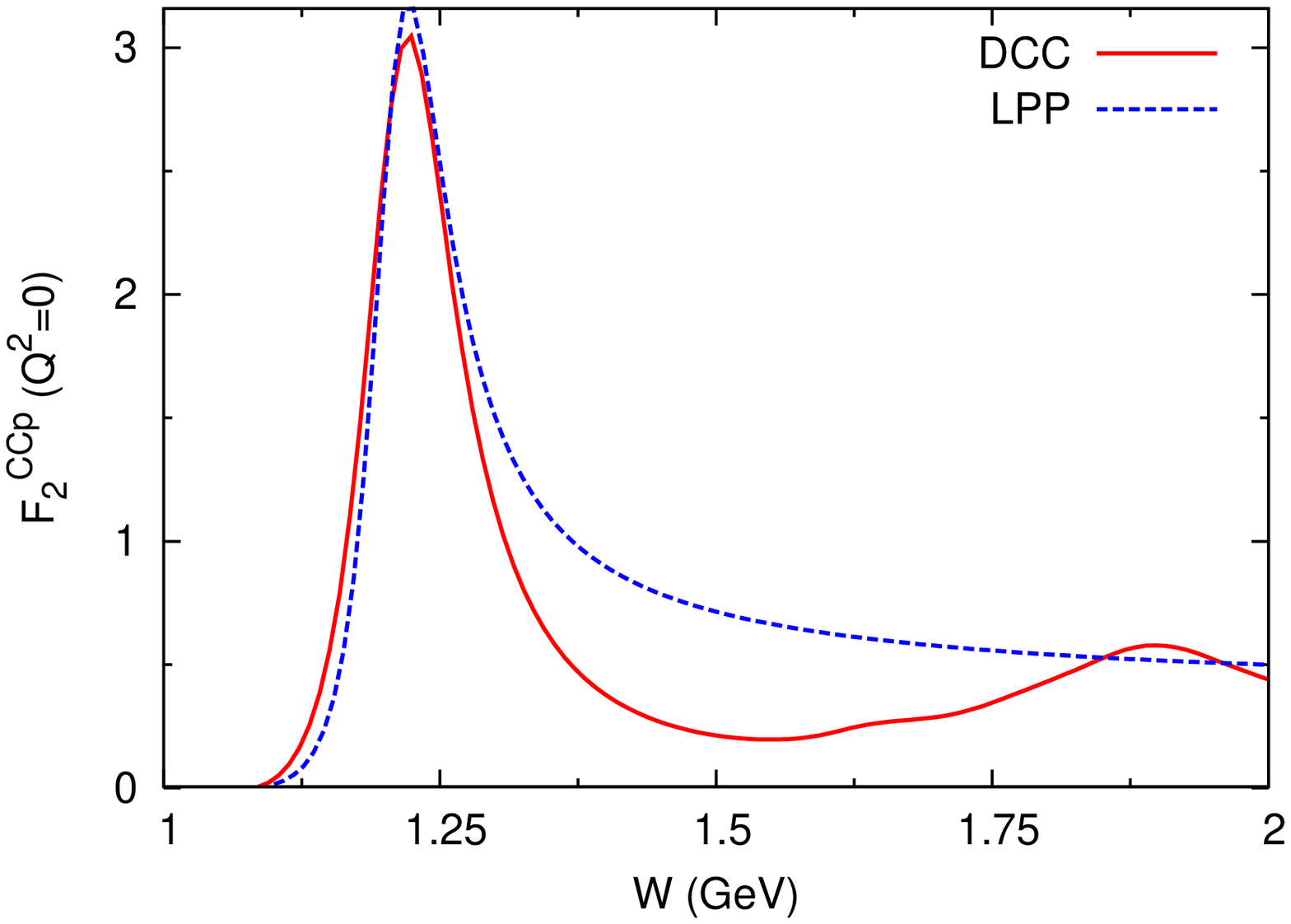}
\includegraphics[height=0.33\textwidth]{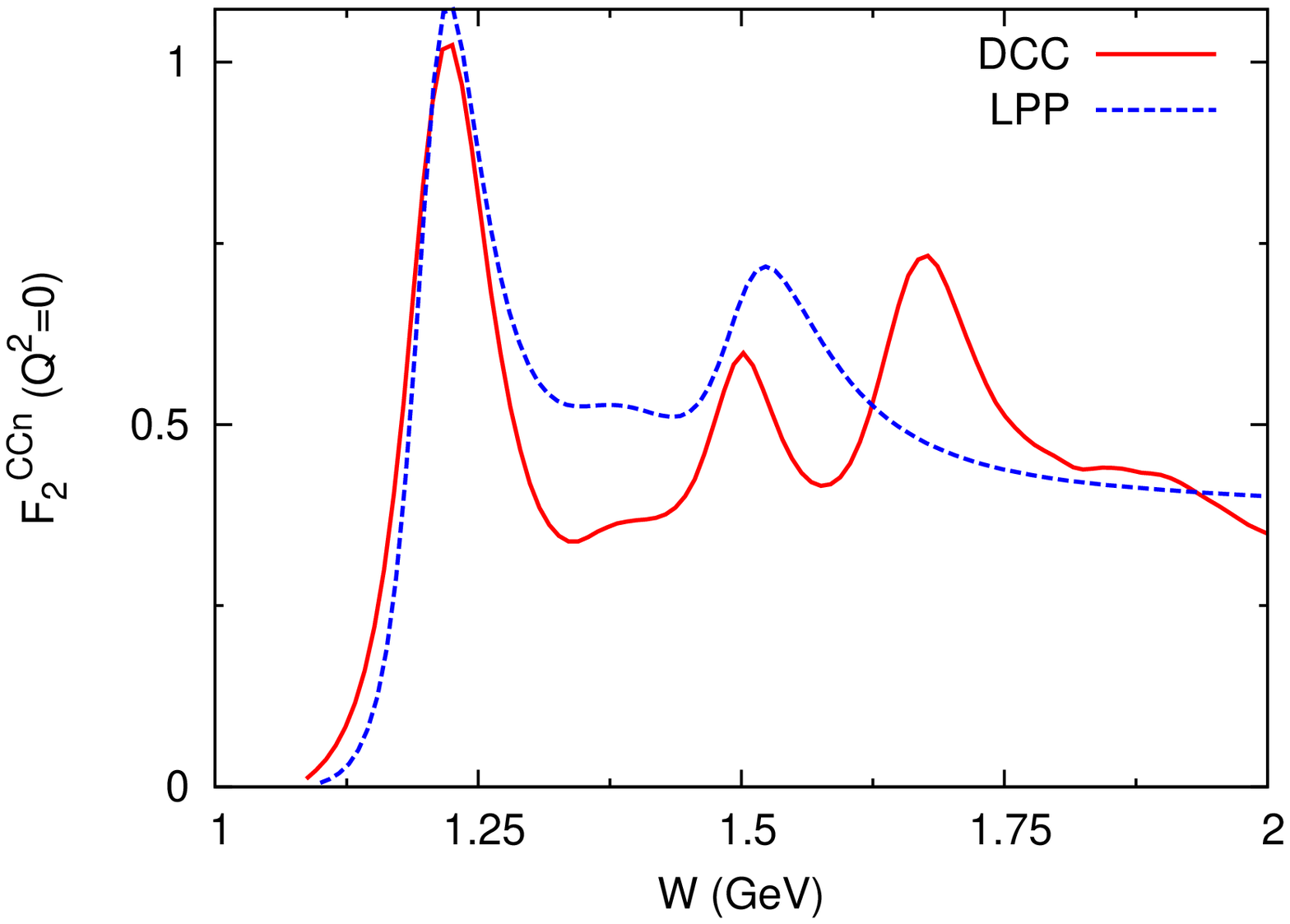}
\caption{(Color online)
The inclusive structure function $F^{\rm CC}_2$ at $Q^2=0$.
The other features are the same as those in Fig.~\ref{fig:LP}.
}
\label{fig:LP2}
\end{figure}

Since a comparison of the DCC model with other models is interesting,
we compare in Figs.~\ref{fig:LP} and \ref{fig:LP2}
the structure function 
$F^{\rm CC}_2 (Q^2=0)$ of the DCC model with
the model due to Lalakulich et al.~\cite{lalakulich} (LPP model),
and the Rein-Sehgal (RS) model~\cite{RS,RS2}.
The LPP model consists of four amplitudes of the
Breit-Wigner form
for $\Delta(1232)~3/2^+$, 
$N(1535)~1/2^-$, $N(1440)~1/2^+$ and $N(1520)~3/2^-$ resonances
with no background.
The RS model consists of 18 Breit-Wigner
terms plus a non-interfering
non-resonant background of $I=1/2$.
On the left (right) panel of Fig.~\ref{fig:LP}, we show $F^{\rm CC}_2$ for the CC
reaction on the proton (neutron) going to the $\pi N$ final state.
From the comparison, a good agreement between the DCC model and the LPP model
is found only near the $\Delta(1232)$ peak; otherwise they are rather different.
The RS model rather undershoot the $\Delta(1232)$ peak, 
as has been also pointed out in Refs.~\cite{AHN,RS-ff,leitner,nuint12}.
Near the threshold ($W\sim$ 1.1~GeV), meanwhile,
$F^{\rm CC}_2 (Q^2=0)$ of the DCC model is larger than those of the LPP
and RS models.
A similar tendency persists in the inclusive $F^{\rm CC}_2$ as shown 
in Fig.~\ref{fig:LP2}.
For the LPP model, $F_2^{{\rm CC}p}$ in the left panels of Figs.~\ref{fig:LP} and \ref{fig:LP2}
are the same because only 
the $\Delta(1232)$ contributes to the proton-target process,
and it decays almost exclusively into the $\pi N$ state.
As discussed earlier in this paper, $F^{\rm CC}_2$ at $Q^2=0$ is related to the $\pi N$ cross sections
through the PCAC relation, and thus is given almost model-independently.
We have shown in Fig.~\ref{fig:neutrino-f2} that
$F^{\rm CC}_2$ from the DCC axial current model and that from the precise $\pi N$ model agree
well in accordance with the PCAC relation.
Therefore, the difference between the DCC model and the LPP and the RS models in
$F^{\rm CC}_2(Q^2=0)$ reveals a consequence of missing the consistency between the
axial-current and the $\pi N$ interaction in the latter models.

\section{Conclusion}
\label{sec:conclusion}

In this work, we have developed a dynamical coupled-channels (DCC) model for
neutrino-nucleon reactions in the resonance region.
Our starting point is the DCC model that we have developed through a
comprehensive analysis of 
$\pi N, \gamma p\to \pi N, \eta N, K\Lambda, K\Sigma$ data
for $W\le 2.1$~GeV~\cite{knls13}.
The model has also been shown to give a reasonable description of 
$\pi N\to\pi\pi N$~\cite{kamano-pipin}.
In order to extend the DCC model of Ref.~\cite{knls13} to what works for the neutrino
reactions, we analyzed data for
the single pion photoproduction off the neutron, 
and also data for the electron scattering on both
proton and neutron targets.
Through the analysis,
we determined the $Q^2$-dependence of the
vector form factors up to $Q^2\le 3$~(GeV/$c$)$^2$. 
By combining the vector form factors for the proton and neutron, we
separated the vector form factors into the isovector and isoscalar
parts; this isospin separation is a necessary step to apply the model to
the neutrino reactions. 
We also derived the axial-current matrix elements.
An appealing point of our approach is that we can derive 
the axial-current matrix elements 
that are linked to the $\pi N$ potential of the DCC model
through the PCAC relation.
As a consequence, relative phases between the non-resonant and resonant
axial current amplitudes are uniquely determined within the DCC model.
The $Q^2$-dependences of the axial form factors are difficult to
determine with the available data.
Thus we used the same axial form factors for all 
the axial $N$-$N^*$ vertices.
Although this prescription is what we can do best for the moment,
we hope to improve this in future if more data become available.
Then,
the preparation for calculating the neutrino-induced
meson productions off the nucleon is completed.

We have presented cross sections for the neutrino-induced
meson productions for $E_\nu\le 2$~GeV.
In this energy region, the single-pion production gives the largest
contribution.
Towards $E_\nu\sim 2$~GeV, the cross section for the double-pion production
is getting larger to become 1/8 (1/4) of the single-pion production
cross section for the proton (neutron) target.
Because our DCC model has been determined by analyzing 
the $\pi N, \gamma N\to \pi N, \eta N, K\Lambda, K\Sigma$ data,
we can also make a quantitative prediction for the neutrino cross sections
for $\eta N$, $K\Lambda$, and $K\Sigma$ productions.
We found that
cross sections for $\eta N, K\Lambda$ and $K\Sigma$ productions are
$10^{-2}$-$10^{-3}$ times smaller than those for the single pion production.
We have compared our numerical results with the available experimental
data. 
For the single-pion production, our result,
for which the axial $N$-$N^*$ couplings are fixed by the PCAC relation,
is consistent with the BNL
data for $\nu_\mu p\to\mu^-\pi^+p$,
while fair agreement with both ANL
and BNL data is found for the neutron target data.
Through the comparison with the single pion production data for 
$W\ltap$1.4~GeV for which 
the $\Delta(1232)$-excitation is the dominant mechanism,
we were able to study the strength and the $Q^2$-dependence of the axial
$N$-$\Delta(1232)$ coupling.
We also calculated double-pion production cross sections by taking
account of relevant resonance contributions for the first time, and
compared them with the data.
We found a good agreement for
$\nu_\mu\, p\to \mu^- \pi^+\pi^0 p$ and 
$\nu_\mu n\to \mu^- \pi^+\pi^- p$, but not for 
$\nu_\mu p\to \mu^- \pi^+\pi^+ n$.
Because the data are statistically rather poor,
it is difficult to make a conclusive judgement on the DCC model.
We hope new high quality data will become available in near future.
We examined reaction mechanisms contributing to the single-pion
production.
For the proton target where only $I=3/2$ states contribute,
the $\Delta(1232)$ dominates. 
However, for the neutron target where 
both $I=1/2$ and 3/2 states contribute,
the $\Delta(1232)$, higher $N^*$ resonances, and non-resonant mechanisms
give comparable contributions towards $E_\nu\sim 2$~GeV.
Thus it is very important to have interference patterns among those
different mechanisms under control. 
In this regard, our DCC approach has an advantage over the other
existing models, as mentioned in the above paragraph.

Finally, we make some remarks on our future prospect.
Our DCC model should be smoothly connected to a DIS model in
the resonance-DIS overlapping region.
Because the DCC model still has degrees of freedom to vary the
$Q^2$-dependence of the axial form factors, 
we can adjust them to fit the $W$- and $Q^2$-dependences of
inclusive cross sections from the DIS model in
the overlapping region.
This needs be done in a future work.
Having developed the DCC neutrino-nucleon reaction model that covers the
whole resonance region, next task should be developing a
neutrino-nucleus reaction model
in which the DCC model describes the elementary processes.
The simplest case is the neutrino-deuteron reactions, 
and for that, we can do a relatively well-established quantum mechanical
calculation, as done in Ref.~\cite{wsl}.
It will be interesting to extend the work of Ref.~\cite{wsl} by
replacing the elementary amplitudes therein with those from the DCC
model developed here. 
Then, ANL and BNL data should be reanalyzed with this model that takes
care of not only the Fermi motion but also the final state
interactions (FSI).
The importance of FSI has been demonstrated in Ref.~\cite{wsl}.
Also, this development can serve as a preparation for a possible
T2K experiment~\cite{wilking} that utilizes heavy-water (D$_2$O) as the target.
Application to heavier nuclei will also be very important.
In the $\Delta(1232)$ region,
the well-developed $\Delta$-hole model~\cite{D-hole1,D-hole2} may give a
hint to address pion productions in the neutrino-nucleus reaction.
For even higher energy neutrino reactions, a fully quantum mechanical
calculation seems formidable.
Combining the elementary amplitudes of the DCC model
with a hadron transport model may be a possible and practical option,
as has been done in the literature~\cite{gibuu,spain-cascade}.

\begin{acknowledgments}
We thank L. Cole Smith and 
Kyungseon Joo for providing us with data for single pion
 electroproduction.
We also thank T.-S. Harry Lee for useful discussions.
We acknowledge Eric Christy for providing us with a code for the
inclusive structure functions. 
We acknowledge Wally Melnitchouk
for providing us with a code for the LPP model. 
We are also grateful for useful discussions 
at the J-PARC branch of the KEK theory center.
This work was
supported by Ministry of Education, Culture, Sports, Science and
Technology (MEXT) KAKENHI Grant Number 25105010. This work was also
supported by the Japan Society for the Promotion of Science (JSPS)
KAKENHI Grant No. 25800149 (H.K.) and No. 24540273 (T.S.). H.K.
acknowledges the support of the HPCI Strategic Program (Field 5 
``The Origin of Matter and the Universe'') of MEXT of Japan.
\end{acknowledgments}

\appendix

\section{Matrix elements of non-resonant axial currents}
\label{app:axial}

We explained in Sec.~\ref{sec:nonres}
how to derive matrix elements of non-resonant axial-currents
to be implemented in the DCC model.
Here, we present tree-level, non pion-pole (NP) part of 
the matrix elements for a single meson production,
$\langle M(k',j) B(p')|A^i_\mu (q)| N(p) \rangle$,
where $i$ and $j$ are isospin components
and the other variables in the parentheses
are four-momenta carried by
the corresponding 
nucleon ($N$),
baryon ($B$), 
meson ($M$), 
and axial-current ($A_\mu$);
labels for the spin and isospin states for the baryons are suppressed.
The following expressions in Appendices~\ref{app:axial}-\ref{app:vector-nstar}
are those in the center-of-mass frame of the hadronic system (hCM). 
The matrix elements can be expressed in the following form:
\begin{eqnarray}
\label{eq:me-current}
&&
\bra{M(k'j),B(p')} 
A^{i}_{{\rm NP,tree}}(q)\cdot \epsilon(\lambda) \ket{N(p) } 
\nonumber \\
&=& 
\frac{1}{(2\pi)^3}\sum_{n}
\sqrt{\frac{m_N m_{B}}{E_N(-\bm q) E_{B}(-\bm k') 2 E_{M}(\bm k')}}
\bar{u}_{B}(-\bm k') 
 \bar{A}(n) u_N(-\bm q) 
\,,
\end{eqnarray}
where $\epsilon^\mu(\lambda)$ is the polarization vector for the hadron
axial current with a polarization $\lambda$.
The Dirac spinor for the baryon $B$ is denoted by
$u_B(\bm p)$ that is also supposed to implicitly contain the isospin spinor.
In the following subsections, we present expressions for $\bar{A}(n)$
for each process labeled by the index $n$.
$\bar{A}(n)$ is composed of several terms, denoted as 
$\bar{A}^n_a$, $\bar{A}^n_b$,..., etc. for which we also present
diagrammatic representations in TABLE~\ref{tab:axial}.
It is noted that,
in evaluating the time component of four-momenta contained in
the propagators in the following equations, 
we follow the definite procedures defined by 
the unitary transformation method~\cite{sko,sl};
see also Appendix~C of Ref.~\cite{msl07}.

For the axial-current matrix elements shown in the following subsections,
we include at each vertex a dipole form factor in the same way as those
used for $\pi N\to MB$ potentials in Ref.~\cite{knls13}.
For a meson-baryon-baryon vertex,
we include a form factor of the form
\begin{eqnarray}
F(\bm{k},\Lambda) = \left(\Lambda^2\over \bm{k}^2 + \Lambda^2\right)^2
\label{eq:ff}
\end{eqnarray}
where $\bm{k}$ and $\Lambda$ are the meson momentum and
the cutoff, respectively.
For a $N$-$B$ transition vertex induced by the axial current,
we also include a form factor of Eq.~(\ref{eq:ff}) in which 
$\bm{k}$ is chosen to satisfy
\begin{eqnarray}
W=\sqrt{\bm{k}^2+m^2_\pi}+\sqrt{\bm{k}^2+m^2_N} \ ,
\label{eq:kk}
\end{eqnarray}
in order to satisfy the PCAC relation, Eq.~(\ref{eq:pcac}).
In a $t$-channel diagram, 
there is a meson-meson transition vertex induced by the axial-current,
and we use a dipole form factor of Eq.~(\ref{eq:ff}) where
$\bm{k}$ being the momentum of the exchanged meson.
For a contact term, we use double dipole form factor, 
$F(\bm{k}',\Lambda)F(\bm{k},\Lambda)$, 
where 
$\bm{k}'$ is the outgoing meson momentum.
In addition to this hadronic form factor, we also include the axial
form factor of the dipole form,
as discussed in Sec.~\ref{sec:nonres}, to take care of the
$Q^2$-dependence of the axial couplings.

In the following expressions, we use notations such as
$\tilde{k}=p-p'$ and $k$ that satisfies Eq.~(\ref{eq:kk}).
We also denote $\epsilon^\mu$ for $\epsilon^\mu(\lambda)$ for
simplicity.
The $i$-th component of Pauli matrix that acts on the nucleon isospin
spinors is denoted by $\tau^i$.
We denote the pseudoscalar, vector, and scalar mesons by 
$P$, $V$, and $S$, respectively, and the octet and decuplet baryons by
$B$.
With this notation, the $PBB'$, $VBB'$, and $SBB'$ coupling constants
are denoted respectively by $f_{PBB'}$, $g_{VBB'}$, and $g_{SBB'}$,
while the $PP'V$ and  $PP'S$
coupling constants are denoted by $g_{PP'V}$ and $g_{PP'S}$,
respectively.
The tensor coupling constant of a $VBB'$ coupling is denoted by $\kappa_V$.
For numerical values of couplings, masses, cutoffs appearing below in this
appendix, we use those determined in Ref.~\cite{knls13}, and listed in
TABLEs~XI-XIII of the reference.
In addition, we use $f_\pi=93$~MeV.

\begin{table}[t]
\caption{\label{tab:axial}
Diagrammatic representations for non-resonant axial-current matrix
 elements.
$\bar{A}^n_a$, $\bar{A}^n_b$,... correspond to expressions presented in 
Eqs.~(\ref{eq:v1a})-(\ref{eq:axial-last}).
The wavy lines represent the external axial-currents.
}    
\begin{ruledtabular}
\renewcommand{\arraystretch}{1.2}
\begin{center}
\begin{tabular}{rcccccc}
&
\begin{minipage}[h]{13mm} 
\scalebox{0.5}{\includegraphics{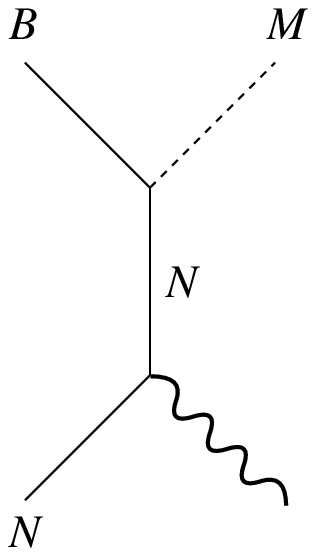}}
\end{minipage}&
\begin{minipage}{13mm} 
\scalebox{0.5}{\includegraphics{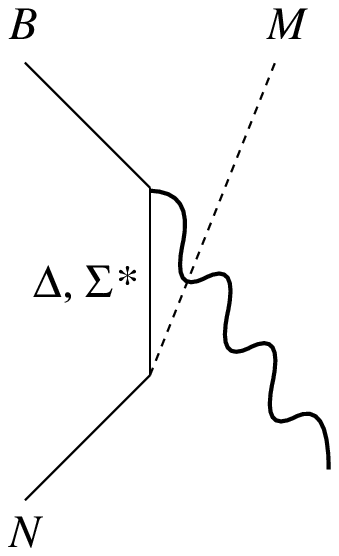}}
\end{minipage}&
\begin{minipage}{13mm} 
\scalebox{0.5}{\includegraphics{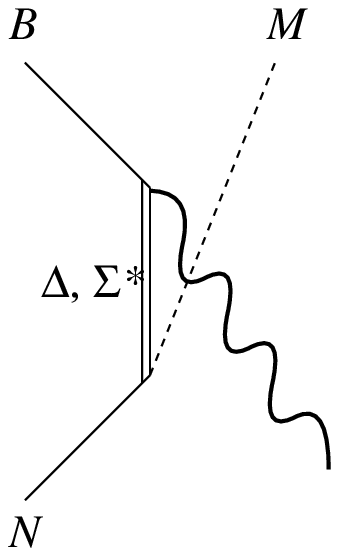}}
\end{minipage}&
\begin{minipage}{13mm} 
\scalebox{0.5}{\includegraphics{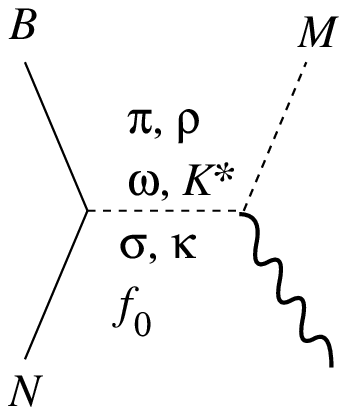}}
\end{minipage}&
\begin{minipage}{13mm} 
\scalebox{0.5}{\includegraphics{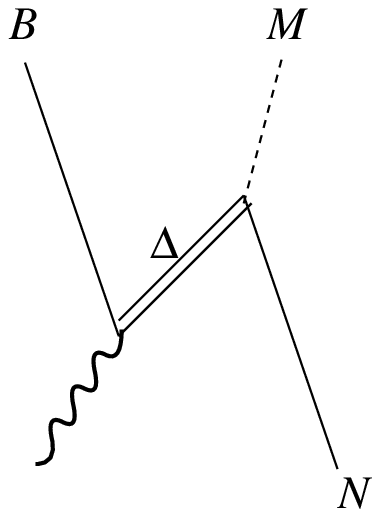}}
\end{minipage}&
\begin{minipage}{13mm} 
\scalebox{0.5}{\includegraphics{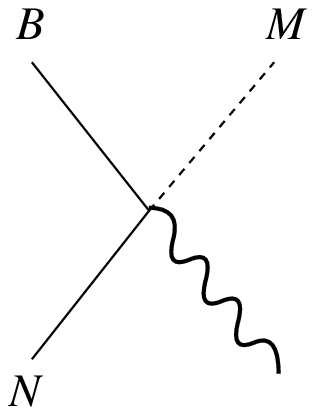}}
\end{minipage}
\\\hline
$MB=\pi N$&$\bar{A}^1_a$&$\bar{A}^1_b$&$\bar{A}^1_c$&$\bar{A}^1_d$,$\bar{A}^1_e$,$\bar{A}^1_f$
&$\bar{A}^1_g$&$\bar{A}^1_h$\\
$\eta N$&$\bar{A}^2_a$&$\bar{A}^2_b$&&&&\\
$\sigma N$&$\bar{A}^3_a$&$\bar{A}^3_b$&&$\bar{A}^3_c$&&\\
$\rho N$&$\bar{A}^4_a$&$\bar{A}^4_b$&&$\bar{A}^4_c$,$\bar{A}^4_e$&&$\bar{A}^4_d$\\
$\pi\Delta$&$\bar{A}^5_a$&$\bar{A}^5_b$&$\bar{A}^5_d$&$\bar{A}^5_c$&&\\
$K\Lambda$&$\bar{A}^6_a$&$\bar{A}^6_b$&$\bar{A}^6_c$&$\bar{A}^6_d$,$\bar{A}^6_e$&&\\
$K\Sigma$&$\bar{A}^7_a$&$\bar{A}^7_b$,$\bar{A}^7_c$&$\bar{A}^7_d$&$\bar{A}^7_e$,$\bar{A}^7_f$&&\\
\end{tabular}
\end{center}
\end{ruledtabular}
\end{table}

\subsection{$MB=\pi N$} 

\begin{eqnarray}
\bar{A}(1) = \bar{A}^1_a
 +\bar{A}^1_b+\bar{A}^1_c+\bar{A}^1_d+\bar{A}^1_e+\bar{A}^1_f+\bar{A}^1_g+\bar{A}^1_h
\ ,
\label{eq:v1}
\end{eqnarray}
with 
\begin{eqnarray}
\bar{A}^1_a &=& i 
f_\pi\left( \frac{f_{\pi NN}}{m_\pi} \right)^2
\slas{k}^\prime\gamma_5 \tau^j 
{1\over \Slash{p}'+\Slash{k}' - m_N}
\slas{\epsilon}\gamma_5 \tau^i\ , 
\label{eq:v1a}
\\
\bar{A}^1_b &=& i 
f_\pi\left( \frac{f_{\pi NN}}{m_\pi} \right)^2
  \slas{\epsilon}\gamma_5 \tau^i 
{1\over \Slash{p}-\Slash{k}' - m_N}
\slas{k}^\prime\gamma_5 \tau^j \ , 
\\
\bar{A}^1_c &=& i f_\pi\left( \frac{f_{\pi N\Delta}}{m_\pi} \right)^2 {\epsilon}_{\alpha}
(T^\dagger)^i S^{\alpha\beta}_\Delta(p-k^\prime){k}^\prime_{\beta} T^j \ , 
\label{eq:v1c}
\\
\bar{A}^1_d &=& 
-i C_{V1}\frac{f_\pi g_{\rho NN}g_{\rho\pi\pi}}{m_\rho^2}
\left[
\slas{\epsilon}+
C_{V2}\frac{\kappa_\rho}{4m_N}
(\slas{\epsilon}\slas{\tilde{k}}
-\slas{\tilde{k}}\slas{\epsilon})
\right] 
i\epsilon_{jil}
\tau^l
\nonumber \\ 
&&
+ i f_\pi g_{\rho NN}g_{\rho\pi\pi} \frac{\tilde{k}^2}{m^2_\rho(\tilde{k}^2-m^2_\rho)} 
\left[
\slas{\epsilon}+
\frac{\kappa_\rho}{4m_N}(\slas{\epsilon}\slas{\tilde{k}}
-\slas{\tilde{k}}\slas{\epsilon})
\right] 
i\epsilon_{jil}
\tau^l \ , 
\\
\label{eq:sigma-d}
\bar{A}^1_e &=& 
+i f_\pi C_S 
{\epsilon^3\over |\bm q|}
\frac{k\cdot k^\prime}{m_\pi} \delta_{ij} 
\nonumber\\
&&
-i
\left(
f_\pi g_{\sigma NN} g_{\sigma \pi\pi} \frac{\tilde{k}^2}{m_\sigma^2(\tilde{k}^2-m_\sigma^2)} 
+f_\pi g_{f_0 NN} g_{f_0 \pi\pi} \frac{\tilde{k}^2}{m_{f_0}^2(\tilde{k}^2-m_{f_0}^2)} 
\right)
{\epsilon^3\over |\bm q|}
\frac{k\cdot k^\prime}{m_\pi} \delta_{ij} \ ,
\\
\label{eq:sigma-nond}
\bar{A}^1_f &=& 
- i f_\pi {\epsilon^3\over |\bm q|}
\frac{g_{\sigma NN} \tilde g_{\sigma \pi\pi}m_\pi^2}{f_\pi} \frac{1}{\tilde{k}^2-m_\sigma^2} 
\delta_{ij} \ ,
\\
\bar{A}^1_g &=& i f_\pi\left( \frac{f_{\pi N\Delta}}{m_\pi} \right)^2 
{k}^\prime_{\alpha} (T^\dagger)^j 
\left[S^{\alpha\beta}_\Delta(p'+k')
-S^{(+)\alpha\beta}_\Delta(p'+k')\right]
{\epsilon}_{\beta}
T^i \ , 
\end{eqnarray}
where $S^{(+)\alpha\beta}_\Delta$ is the positive energy part of the
$\Delta$ propagator; in the frame where the $\Delta$ is at rest, 
\begin{eqnarray}
S^{(+) mn}_\Delta (p) = {1\over 6}
{1+\gamma^0 \over p^0-m_\Delta} 
 (3\delta_{mn}-\sigma_m\sigma_n) \ .
\end{eqnarray}
Also, the operator $T$ ($T^\dagger$) generates an
isospin transition from $I=1/2$ to 3/2 ($I=3/2$ to 1/2) states.
We note here that we include the cross diagram of $s$-channel 
$\Delta(1232)$-resonance
diagram as a part of the non-resonant mechanism, as in Eq.~(\ref{eq:v1c}).
For $S_{31}$ partial wave, we also add
\begin{eqnarray}
\label{eq:s31}
\bar{A}^1_h &=& 
i f_\pi {\epsilon^3\over |\bm q|} c_{S_{31}}
\ .
\end{eqnarray}

\subsection{$MB=\eta N$ }

\begin{eqnarray}
\bar{A}(2) = \bar{A}^2_a +\bar{A}^2_b\ ,
\end{eqnarray}
with 
\begin{eqnarray}
\bar{A}^2_a &=& i 
f_\pi \frac{f_{\pi NN}f_{\eta NN}}{m_\pi m_\eta} 
 \slas{k}^\prime\gamma_5  
{1\over \Slash{p}'+\Slash{k}' - m_N}
\slas{\epsilon}\gamma_5 \tau^i\ , \\
\bar{A}^2_b &=& i 
f_\pi \frac{f_{\pi NN}f_{\eta NN}}{m_\pi m_\eta} 
 \slas{\epsilon}\gamma_5 \tau^i 
{1\over \Slash{p}-\Slash{k}' - m_N}
\slas{k}^\prime\gamma_5 \ .
\end{eqnarray}

\subsection{$MB=\sigma N$ }

\begin{eqnarray}
\bar{A}(3) = \bar{A}^3_a +\bar{A}^3_b+\bar{A}^3_c\ ,
\end{eqnarray}
with
\begin{eqnarray}
\bar{A}^3_a&=& - 
f_\pi \frac{f_{\pi NN}}{m_\pi} 
g_{\sigma NN} {1\over \Slash{p}'+\Slash{k}' - m_N}
\slas{\epsilon}
\gamma_5\tau^i \ , \\
\bar{A}^3_b&=& - 
f_\pi \frac{f_{\pi NN}}{m_\pi} 
g_{\sigma NN}\slas{\epsilon}\gamma_5 
{1\over \Slash{p}-\Slash{k}' - m_N}\tau^i \ , \\
\bar{A}^3_c &=& - f_\pi \frac{f_{\pi NN}g_{\sigma\pi\pi}}{m_\pi^2}\slas{\tilde{k}}\gamma_5
\tau^i\frac{\tilde{k}\cdot \epsilon}{\tilde{k}^2-m^2_\pi}\ . 
\end{eqnarray}

\subsection{$MB=\rho' N$}

\begin{eqnarray}
\bar{A}(4) = \bar{A}^4_a +\bar{A}^4_b+\bar{A}^4_c+\bar{A}^4_d+\bar{A}^4_e
\ ,
\end{eqnarray}
with
\begin{eqnarray}
\bar{A}^4_a &=& - 
f_\pi \frac{f_{\pi NN}}{m_\pi} 
g_{\rho NN}\Gamma_{\rho^\prime}
{1\over \Slash{p}'+\Slash{k}' - m_N}
\slas{\epsilon} \gamma_5 \tau^i \ , \\
\bar{A}^4_b &=& - 
f_\pi \frac{f_{\pi NN}}{m_\pi} 
g_{\rho NN}\slas{\epsilon} \gamma_5 \tau^i 
{1\over \Slash{p}-\Slash{k}' - m_N}
\Gamma_{\rho^\prime}
\ , \\
\bar{A}^4_c &=& -2 i f_\pi \frac{f_{\pi NN}}{m_\pi} g_{\rho\pi\pi}
\epsilon_{ijl}\tau^l\frac{\epsilon\cdot\epsilon_{\rho^\prime}^*
\slas{\tilde{k}}\gamma_5}{\tilde{k}^2-m_\pi^2} \ , \\
\label{eq:pi-rho-contact}
\bar{A}^4_d &=& 
 - i f_\pi {\epsilon^3\over |\bm q|}
\frac{f_{\pi NN}}{m_\pi}g_{\rho NN}
\slas{\epsilon_{\rho^\prime}}^*\gamma_5 \epsilon_{jil}\tau^l
\ , \\
\bar{A}^4_e&=& i f_\pi \frac{g_{\omega NN}g_{\omega \pi\rho}}{m_\omega}\delta_{ij}
\frac{\epsilon^{\alpha\beta\gamma\delta}\epsilon_{\rho^\prime\alpha}^{*}
 k^{\prime}_\beta
\epsilon_{\gamma}}{\tilde{k}^2-m^2_\omega}
\left[
\gamma_\delta+\frac{\kappa_\omega}{4m_N}
(\gamma_\delta\slas{\tilde{k}}-\slas{\tilde{k}}\gamma_\delta)
\right]\ ,
\end{eqnarray}
where $\epsilon_{\rho^\prime}^{*\mu}$ is the polarization vector for the
$\rho$-meson, and
\begin{eqnarray}
\Gamma_{\rho^\prime}& =&\frac{\tau^j}{2}
\left[
\slas{\epsilon_{\rho'}}^*+
\frac{\kappa_\rho}{4m_N}
(\slas{\epsilon_{\rho'}}^*\slas{k}^\prime
-\slas{k}^\prime\slas{\epsilon_{\rho'}}^*)
\right] \ ,
\end{eqnarray}
and $\epsilon^{0123}=+1$ convention is taken.

\subsection{$MB= \pi\Delta$}

\begin{eqnarray}
\bar{A}(5) = \bar{A}^{5}_a +\bar{A}^{5}_b+ \bar{A}^{5}_c
+\bar{A}^{5}_d
\ ,
\end{eqnarray}
with
\begin{eqnarray}
\bar{A}^{5}_a&=& i 
f_\pi \frac{f_{\pi NN}f_{\pi N\Delta}}{m_\pi^2} 
T^j\epsilon_\Delta^*\cdot k^\prime 
{1\over \Slash{p}'+\Slash{k}' - m_N}
\slas{\epsilon}\gamma_5\tau^i \ , 
\\
\bar{A}^{5}_b&=& i f_\pi \frac{f_{\pi NN}f_{\pi N\Delta}}{m^2_\pi}
T^i\epsilon_\Delta^*\cdot \epsilon 
{1\over \Slash{p}-\Slash{k}' - m_N}
\slas{k}^\prime\gamma_5 
\tau^j \ , 
\\
\bar{A}^{5}_c&=& - 2 f_\pi \frac{f_{\rho N\Delta}f_{\rho\pi\pi}}{m_\rho}
\frac{\epsilon_{jil}T^l}{\tilde{k}^2-m^2_\rho}
[\epsilon_\Delta^*\cdot \tilde{k} \slas{\epsilon}\gamma_5
-\epsilon_\Delta^*\cdot\epsilon\slas{\tilde{k}}\gamma_5] \ , 
\\
\bar{A}^{5}_d&=& -i f_\pi \frac{f_{\pi\Delta\Delta}f_{\pi N\Delta}}{m^2_\pi}
[\epsilon_\Delta^*]_\mu \slas{\epsilon}\gamma_5 T^i_\Delta
 S^{\mu\nu}_{\Delta}(p- k^\prime) T^j k_\nu^\prime\ , 
\end{eqnarray}
where $\epsilon_{\Delta}^{*\mu}$ is the polarization vector for $\Delta$.

\subsection{$MB=K\Lambda$}

\begin{eqnarray}
\bar{A}(6) = \bar{A}^{6}_a + \bar{A}^{6}_b + \bar{A}^{6}_c + \bar{A}^{6}_d + \bar{A}^{6}_e\ ,
\end{eqnarray}
with
\begin{eqnarray}
\bar{A}^{6}_a&=& i 
f_\pi \frac{f_{\pi NN}f_{K\Lambda N}}{m_\pi m_K} 
\slas{k}^\prime
\gamma_5 
{1\over \Slash{p}'+\Slash{k}' - m_N}
\slas{\epsilon}\gamma_5\tau^i\ , 
\\
\bar{A}^{6}_b&=& i f_\pi
{f_{\pi\Lambda\Sigma}f_{K \Sigma N}\over m_K m_\pi}
\slas{\epsilon}
\gamma_5 S_\Sigma(p-k') \slas{k}^\prime\gamma_5\tau^i\ , 
\\
\bar{A}^{6}_c &=& i f_\pi \frac{f_{\pi \Lambda\Sigma^*}f_{K N\Sigma^*}}{m_K m_\pi} 
{\epsilon}_{\alpha} S^{\alpha\beta}_{\Sigma^*}(p-k^\prime){k}^\prime_{\beta} \tau^i \ , 
\\
\bar{A}^{6}_d&=& -2 i f_\pi g_{K^* N\Lambda}g_{K^*K\pi}
{-g_{\mu\rho} + \tilde{k}_\mu \tilde{k}_\rho/m^2_{K^*} \over \tilde{k}^2-m^2_{K^*}}
\left(
\gamma^\mu - i\frac{\kappa_{{K^*}N\Lambda}}{m_N+m_\Lambda} \sigma^{\mu\nu}\tilde{k}_\nu
\right)
\epsilon^\rho \tau^i\ , 
\\
\bar{A}^{6}_e &=& -i f_\pi \frac{g_{\kappa\Lambda N}g_{\kappa K\pi}}{m_\pi}
\frac{\epsilon\cdot k^\prime}{\tilde{k}^2-m_\kappa^2} \tau^i \ .
\end{eqnarray}
For the $K$ production amplitudes in this and the following subsections, 
the isospin operator $\tau$ acts on isospin spinors of
the initial nucleon and the final $K$.

\subsection{$MB=K\Sigma$}

\begin{eqnarray}
\bar{A}(7) = \bar{A}^{7}_a + \bar{A}^{7}_b + \bar{A}^{7}_c +
              \bar{A}^{7}_d + \bar{A}^{7}_e + \bar{A}^{7}_f\ ,
\end{eqnarray}
with
\begin{eqnarray}
\bar{A}^{7}_a&=& i
f_\pi \frac{f_{\pi NN}f_{K\Sigma N}}{m_\pi m_K} 
\slas{k}^\prime
\gamma_5\tau^j 
{1\over \Slash{p}'+\Slash{k}' - m_N}
\slas{\epsilon}\gamma_5\tau^i\ , 
\\
\bar{A}^{7}_b&=& i f_\pi
{f_{\pi \Lambda\Sigma}f_{K\Lambda N}\over m_K m_\pi}
\slas{\epsilon}
\gamma_5 S_\Lambda(p-k') \slas{k}^\prime\gamma_5\delta^{ij}\ , 
\\
\bar{A}^{7}_c&=& i f_\pi 
{f_{\pi \Sigma\Sigma}f_{K\Sigma N}\over m_K m_\pi}
\slas{\epsilon}
\gamma_5 S_\Sigma(p-k') \slas{k}^\prime\gamma_5 
i \epsilon^{ijk}\tau_k\ , 
\\
\bar{A}^{7}_d &=& i f_\pi
\frac{f_{\pi \Sigma\Sigma^*}f_{K N\Sigma^*}}{m_K m_\pi} 
{\epsilon}_{\alpha} S^{\alpha\beta}_{\Sigma^*}(p-k^\prime){k}^\prime_{\beta} 
i \epsilon^{ijk}\tau_k\ , 
\\
\bar{A}^{7}_e&=& -2 i f_\pi g_{K^* N\Sigma}g_{K^*K\pi}
{-g_{\mu\rho} + \tilde{k}_\mu \tilde{k}_\rho/m^2_{K^*} \over \tilde{k}^2-m^2_{K^*}} 
\left(
\gamma^\mu - i\frac{\kappa_{{K^*}N\Sigma}}{m_N+m_\Sigma} \sigma^{\mu\nu}\tilde{k}_\nu
\right)
\epsilon^\rho \tau^i\tau^j\ ,
\\
\bar{A}^{7}_f &=& -i f_\pi\frac{g_{\kappa\Sigma N}g_{\kappa K\pi}}{m_\pi}
\frac{\epsilon\cdot k^\prime}{\tilde{k}^2-m_\kappa^2} \tau^i \tau^j \ ,
\label{eq:axial-last}
\end{eqnarray}
where the suffix $j$ is the isospin state of $\Sigma$.

\section{Bare $N^*$ axial current matrix elements}
\label{app:axial-nstar}

We present 
the axial $N$-$N^*$ transition matrix element
for each of $1/2^\pm, 3/2^\pm, 5/2^\pm, 7/2^\pm$ bare $N^*$
in the helicity basis.
The matrix element is parametrized in terms of form factors
that can be determined at $Q^2=0$
by invoking the PCAC relation to
the $\pi + N\to N^*$ matrix element.
Before presenting the expressions for the axial $N$-$N^*$ matrix elements, 
we give the definition for the axial $N$-$N^*$ matrix elements
and how they are connected to the $\pi + N\to N^*$
matrix elements through the PCAC relation.
We also specify how 
the axial $N$-$N^*$ matrix elements presented in this section are
implemented in the formulae in Sec.~\ref{sec:rescatt}.

A tree-level $s$-channel bare $N^*$ 
amplitude for an axial-current induced single pion production 
is given in the plane wave basis by
\begin{eqnarray}
\label{eq:s-ch-bare}
{1\over (2\pi)^3} \sqrt{m_N^2 \over E_N(-\bm q) E_{N}(-\bm k') 2E_{\pi}(\bm k')}\ 
{\bra{\pi(k',j)N(p')} h_{N^*\to \pi N} \ket{N^*}
\bra{N^*}{A^i_{N^*}(q)\cdot\epsilon}\ket{N(p)} 
\over E-m_{N^*}}
\ ,
\end{eqnarray}
where the normalization of the amplitude is the same as that of 
Eq.~(\ref{eq:me-current}).
The axial $N$-$N^*$ and $N^*\to \pi N$ matrix elements are
$\bra{N^*}{A^i_{N^*}(q)\cdot\epsilon}\ket{N(p)}$ and
$\bra{\pi(k',j)N(p')} h_{N^*\to \pi N} \ket{N^*}$, respectively.
Let us first
present expressions for the $N^*\to \pi N$ matrix element that is
subsequently related to the axial matrix element by the PCAC relation.
The $N^*\to \pi N$ matrix element
$\bra{\pi(k',j)N(p')} h_{N^*\to \pi N} \ket{N^*}$ is parametrized by
\begin{eqnarray}
\label{eq:bare-vtx}
&&\bra{\pi(k',j)N(p')} h_{N^*\to \pi N} \ket{N^*} 
\nonumber \\
&&
=
-i\,
  g_{\pi N,N^*} (\bm k')(1 j {1\over 2} t_{N}^z | T_{N^*} T^z_{N^*})
\sum_{L^z} (L L^z {1\over 2} s_N^z | J_{N^*} M_{N^*}) Y_{LL^z}(\hat {\bm k'})
 \ ,
\end{eqnarray}
where $(1 j 1/2 t_{N}^z | T_{N^*} T^z_{N^*})$ is the isospin
Clebsch-Gordan coefficient.
The $N^*$ has the spin $J_{N^*}$
and the parity $(-1)^{L+1}$, and it decays into the $\pi N$ state with
the orbital angular momentum $L$.
The vertex function is denoted by $g_{\pi N,N^*} (k')$ that is related
to $\Gamma_{\pi N,N^*}$ in Eq.~(\ref{eq:nstargn}) by
\begin{eqnarray}
 g_{\pi N,N^*}(\bm k')  
 &=& 
i \sqrt{(2\pi)^3 E_{N}(-\bm k') 2E_{\pi}(\bm k')\over m_{N}}\,
\Gamma_{\pi N,N^*}(\bm k')
\nonumber 
\\
&=&  
 \sqrt{E_{N}(-\bm k') 2E_{\pi}(\bm k')\over m^2_{N}}\,
C_{\pi N (L,S=1/2),N^\ast}
\left(\frac{\Lambda_{N^\ast}^2}{\Lambda_{N^\ast}^2 + \bm{k}'^2}\right)^{(2+L/2)}
\left(\frac{|\bm{k}'|}{m_\pi}\right)^{L} \  ,
\label{eq:gmb}
\end{eqnarray}
where we have used the parametrization for $\Gamma_{\pi N,N^*}$ defined 
in Ref.~\cite{knls13}.
We use numerical values for the coupling $C_{\pi N(L,S=1/2),N^\ast}$ and
the cutoff $\Lambda_{N^\ast}$ presented in Ref.~\cite{knls13}.

Now we discuss the axial $N$-$N^*$ transition matrix element.
First we separate the dependence on the
isospin components from the matrix element by
\begin{eqnarray}
\bra{N^*} A^i_{N^*}(q)\cdot\epsilon \ket{N} 
= A_{J^\pm_{N^*}}(q)\cdot\epsilon\  (1 i {1\over 2} t_{N}^z | T_{N^*} T^z_{N^*})
 \ ,
 \label{eq:axial-me}
\end{eqnarray}
where $J^\pm_{N^*}$ is the spin-parity of $N^*$.
The non pion-pole part of the matrix element
$A^\mu_{J^\pm_{N^*}}(q)$
can be determined by the $\pi N N^*$ coupling constant $g_{\pi N,N^*}$
using the PCAC relation,
$\langle N^*|q\cdot A^i_{{\rm NP},N^*}|N\rangle
=if_\pi\langle N^*|h_{\pi N\to N^*}|\pi^i N\rangle$.
Since the $\pi NN^*$ coupling constants have been determined at 
$Q^2=-m_\pi^2$ from our analysis of the $\pi N$ reaction data~\cite{knls13},
we determine $A^\mu_{{\rm NP},N^*}(q)$ at $Q^2=-m_\pi^2$ 
using the PCAC relation:
\begin{eqnarray}
q\cdot A_{{\rm NP},J^\pm_{N^*}}(q)|_{Q^2=-m^2_\pi}&=&
- \sum_{M}(-1)^M f_\pi\ g_{\pi NN^*}(\bm{q})\ Y_{L,-M} (\hat{\bm q})\ (L M {1\over 2} s_N^z | J_{N^*} M+s_N^z)
\ .
\label{eq:pcac2}
\end{eqnarray}
We can simplify this equation
by taking the $z$-axis along ${\bm q}$:
\begin{eqnarray}
q\cdot A_{{\rm NP},J^\pm_{N^*}}(q)|_{Q^2=-m^2_\pi} &=&
- \sqrt{L+1\over 4\pi}
f_\pi\ g_{\pi NN^*}({\bm q}) \ ,
\label{eq:pcac_pos}
\end{eqnarray}
for $J_{N^*} = L+1/2$, and
\begin{eqnarray}
q\cdot A_{{\rm NP},J^\pm_{N^*}}(q)|_{Q^2=-m^2_\pi} &=&
\pm \sqrt{L\over 4\pi}
f_\pi\ g_{\pi NN^*}({\bm q}) \ ,
\label{eq:pcac_neg}
\end{eqnarray}
for $J_{N^*} = L-1/2$; $\pm$ is for $s_N^z=\pm 1/2$.
As will be shown in the following subsections, 
we use Eqs.~(\ref{eq:pcac_pos}) and (\ref{eq:pcac_neg}) to fix
form factors $g^{J_{N^*}^\pm}(Q^2=-m^2_\pi)$ contained in $A_{J^\pm_{N^*}}^\mu(q)$.
Then we take the PCAC hypothesis:
$g^{J_{N^*}^\pm}(Q^2=-m^2_\pi)\sim g^{J_{N^*}^\pm}(Q^2=0)$.
For the $Q^2$-dependence of $g^{J_{N^*}^\pm}(Q^2)$,
as discussed in Sec.~\ref{sec:axial-nstar},
we assume the dipole form factor that is
implicit in the expressions in the
following subsections.
We note that the axial form factor $g^{J_{N^*}^\pm}(Q^2)$ 
fixed by Eqs.~(\ref{eq:pcac_pos}) and (\ref{eq:pcac_neg}) 
at a range of $\bm q$
acquires a $W$-dependence that is the same as that of
$g_{\pi NN^*}({\bm q})$; $\bm q$ and $W$ are related by
$W=\sqrt{\bm{q}^2+m^2_\pi}+\sqrt{\bm{q}^2+m^2_N}$.
Thus we denote the axial form factors by $g^{J_{N^*}^\pm}(W,Q^2)$.
The axial-current amplitudes determined
in this way satisfy
the PCAC relation with
the $\pi N$ amplitudes
at any $W$, as shown in Fig.~\ref{fig:neutrino-f2}.
Finally, the axial-current part of
the vertex function 
$\Gamma_{N^\ast,JN}$ in Eq.~(\ref{eq:nstargn}) 
is related to the axial-current matrix element by
\begin{eqnarray}
\Gamma_{N^\ast,J N} (\lambda;W,q)
&=&  \frac{1}{(2\pi)^{3/2}}
 \sqrt{m_N\over E_N(-{\bm q})}\ 
\sqrt{4\pi\over 2J_{N^*}+1}\
{A_{J^\pm_{N^*}}(q)}\cdot \epsilon(\lambda) \ .
\end{eqnarray}

\subsection{Spin 1/2 $N^*$}
The matrix element of axial vector current between the nucleon and
spin $1/2^{\pm}$ $N^*$ 
[$A_{J^\pm_{N^*}}(q)\cdot\epsilon$ in Eq.~(\ref{eq:axial-me})]
is generally given, the induced tensor term being omitted,  by 
\begin{eqnarray}
A_{1/2^\pm}(q)\cdot \epsilon = \bar u_{1/2^\pm}(\bm 0)
\left[ g^{1/2^\pm}_A(Q^2) \Slash{\epsilon} + g^{1/2^\pm}_P(Q^2) q\cdot\epsilon
\right]
\binom{\gamma_5}{\unitfour} u_N(-{\bm q}) \ ,
\end{eqnarray}
where the upper (lower) operator is for positive (negative)
parity $N^*$;
$g^{1/2^\pm}_A(Q^2)$ and $g^{1/2^\pm}_P(Q^2)$ are form factors.
The matrix element for the divergence of the axial current is
\begin{eqnarray}
q \cdot A_{1/2^+}(q) &=& \mp {|\bm q|\over E_N(-\bm q) + m_N}
\left\{
g^{1/2^+}_A(Q^2) (q^0 + m_N + E_N(-\bm q)) + g^{1/2^+}_P(Q^2) q^2
\right\} \ ,\nonumber\\
q\cdot A_{1/2^-}(q) &=&
g^{1/2^-}_A(Q^2) \left(q^0 + {|\bm{q}|^2\over E_N(-\bm q) + m_N}\right)
+ g^{1/2^-}_P(Q^2) q^2 \ ,
\end{eqnarray}
where the sign $\mp$ is for $s_N^z=\pm 1/2$. 
Taking $Q^2=-m^2_\pi$ and dropping the small $g^{1/2^{\pm}}_P(Q^2) q^2$
term, we obtain
\begin{eqnarray}
q \cdot A_{{\rm NP},1/2^+}( q)|_{Q^2=-m^2_\pi}  &=& \mp {|{\bm q}|\over E_N(-{\bm q}) + m_N}
\left\{
g^{1/2^+}_A(Q^2=-m^2_\pi) ({q}^0 + m_N + E_N(-{\bm q})) 
\right\} \ ,\nonumber\\
q\cdot A_{{\rm NP},1/2^-}(q)|_{Q^2=-m^2_\pi}  &=&
g^{1/2^-}_A(Q^2=-m^2_\pi) \left( q^0 + {|{\bm{q}}|^2\over E_N(-{\bm q}) + m_N}\right)
\ .
\end{eqnarray}
Comparing this with Eqs.~(\ref{eq:pcac_pos}) and (\ref{eq:pcac_neg}),
we find
\begin{eqnarray}
g^{1/2^+}_A(W,Q^2=-m^2_\pi) &=&
- {1\over \sqrt{4\pi}} {E_N(-{\bm q}) + m_N\over  q^0 + m_N + E_N(-{\bm q})} 
{f_\pi\over |{\bm q}|}
g^{1/2^+}_{\pi NN^*}({\bm q})\ ,
\nonumber \\
g^{1/2^-}_A(W,Q^2=-m^2_\pi) &=&
- {1\over \sqrt{4\pi}}
{f_\pi\over  q^0 + {|{\bm{q}}|^2\over E_N(-{\bm q}) + m_N} }
g^{1/2^-}_{\pi NN^*}({\bm q}) \ ,
\end{eqnarray}
where we have introduced the $W$-dependence in the form factors.
The PCAC hypothesis dictates
$g^{1/2^\pm}_A(W,Q^2=-m^2_\pi)\sim g^{1/2^\pm}_A(W,Q^2=0)$.
The $g_P$ term can be understood as the pion pole term, and thus is
given by
\begin{eqnarray}
g^{1/2^+}_P(W,Q^2) &=& {
   q^0 + m_N + E_N(-{\bm q})  \over Q^2 + m_\pi^2} g^{1/2^+}_A(W,Q^2) 
\ , 
\nonumber \\
g^{1/2^-}_P(W,Q^2) &=& {
 q^0 + {|{\bm{q}}|^2\over E_N(-{\bm q}) + m_N} \over Q^2 + m_\pi^2} 
g^{1/2^-}_A(W,Q^2) 
\ .
\end{eqnarray}
The axial-current amplitudes for $1/2^+ N^*$
in the helicity basis are 
\begin{eqnarray}
A_{1/2^+}(q)\cdot \epsilon^{(t)} &=& \mp {|{\bm{q}}|\over E_N(-{\bm q}) + m_N} 
\left[g^{1/2^+}_A(W,Q^2) + q^0 g^{1/2^+}_P(W,Q^2)\right]\ ,
\nonumber \\
A_{1/2^+}(q)\cdot \epsilon^{(0)} &=& \mp \left[g^{1/2^+}_A(W,Q^2) - 
{|\bm{q}|^2\over E_N(-{\bm q}) + m_N} g^{1/2^+}_P(W,Q^2)\right]\ ,
\nonumber \\
A_{1/2^+}(q)\cdot \epsilon^{(+1)} &=& 
\left\{
\begin{array}{l}
0
\qquad\qquad\qquad\qquad (s_N^z=+1/2) \ , \\
\sqrt{2}\, g^{1/2^+}_A(W,Q^2) 
\qquad (s_N^z=-1/2) \ ,
\end{array}
\right.
\end{eqnarray}
where the helicity of the axial current is indicated by
$\epsilon^{(\lambda)}$ with $\lambda=+1,0,t$ in the spherical basis.
The sign $\mp$ is for $s_N^z=\pm 1/2$.
For the $1/2^- N^*$ state, the helicity amplitudes are
\begin{eqnarray}
A_{1/2^-}(q)\cdot \epsilon^{(t)} &=& g^{1/2^-}_A(W,Q^2) 
+ q^0 g^{1/2^-}_P(W,Q^2) \ .
\nonumber \\
A_{1/2^-}(q)\cdot \epsilon^{(0)} &=&  
{|{\bm{q}}|\over E_N(-{\bm q}) + m_N} g^{1/2^-}_A(W,Q^2) 
- |\bm{q}|\, g^{1/2^-}_P(W,Q^2) \ ,
\nonumber\\
A_{1/2^-}(q)\cdot \epsilon^{(+1)} &=& 
\left\{
\begin{array}{l}
0
\qquad\qquad\qquad\qquad\qquad\qquad\quad (s_N^z=+1/2) \ , \\
 {|{\bm{q}}|\over E_N(-{\bm q}) + m_N} \sqrt{2}\, g^{1/2^-}_A(W,Q^2) 
\qquad (s_N^z=-1/2) \ .
\end{array}
\right.
\end{eqnarray}

\subsection{Spin 3/2 $N^*$}

The matrix element of axial vector current between the nucleon and
$3/2^{\pm}$ $N^*$ 
[$A_{J^\pm_{N^*}\, \mu}(q)$ in Eq.~(\ref{eq:axial-me})]
is generally given by 
\begin{eqnarray}
A_{3/2^\pm\, \mu}(q)& =& \bar u^\alpha_{3/2^\pm}(\bm 0)
\left[
g^{3/2^\pm}_1(Q^2)( g_{\alpha\mu}\Slash{q}-q_\alpha\gamma_\mu)
+g^{3/2^\pm}_2(Q^2)( g_{\alpha\mu}q\cdot p - q_\alpha p_\mu)
\right.
\nonumber\\
&&\left.
+g^{3/2^\pm}_3(Q^2) g_{\alpha\mu}
+g^{3/2^\pm}_4(Q^2) q_\alpha q_\mu
\right]
\binom{\unitfour}{\gamma_5} u_N(-{\bm q}) \ ,
\end{eqnarray}
where $g^{3/2^\pm}_n(Q^2)\ (n=1...4)$ are form factors.
The matrix element of the divergence of the axial current is
\begin{eqnarray}
q\cdot A_{3/2^\pm}(q)  = \bar u^\alpha_{3/2^\pm}(\bm 0)
\left[
g^{3/2^\pm}_3(Q^2) q_\alpha
+g^{3/2^\pm}_4(Q^2) q_\alpha q^2
\right]
\binom{\unitfour}{\gamma_5} u_N(-{\bm q}) \ .
\end{eqnarray}
Taking $Q^2=-m^2_\pi$ and dropping the small 
$g^{3/2^{\pm}}_4(Q^2) q_\alpha q^2$ term, we obtain
\begin{eqnarray}
q\cdot A_{{\rm NP},3/2^+}( q)|_{Q^2=-m^2_\pi} &=&
-\sqrt{2\over 3} |\bm{ q}| g^{3/2^+}_3(Q^2=-m^2_\pi) \ , 
\nonumber \\
q\cdot A_{{\rm NP},3/2^-}( q)|_{Q^2=-m^2_\pi}   &=& 
\pm \sqrt{2\over 3} {|\bm{ q}|^2\over 2 m_N} g^{3/2^-}_3(Q^2=-m^2_\pi) \ .
\end{eqnarray}
Comparing this with Eqs.~(\ref{eq:pcac_pos}) and (\ref{eq:pcac_neg}),
we find
\begin{eqnarray}
g^{3/2^+}_3(W,Q^2=-m^2_\pi) &=&
 \sqrt{3\over 4\pi} {f_\pi\over |\bm{ q}|} 
g^{3/2^+}_{\pi NN^*}({\bm q}) \ ,
\nonumber\\
g^{3/2^-}_3(W,Q^2=-m^2_\pi) &=&
 \sqrt{3\over 4\pi} {2 m_N f_\pi\over |\bm{ q}|^2} 
g^{3/2^-}_{\pi NN^*}({\bm q}) \ ,
\label{eq:ANN*(P33)}
\end{eqnarray}
where we have introduced the $W$-dependence in the form factors.
The PCAC hypothesis dictates
$g^{3/2^\pm}_3(W,Q^2=-m^2_\pi)\sim g^{3/2^\pm}_3(W,Q^2=0)$.
From the PCAC and the pion dominance, we have a pion-pole term:
\begin{eqnarray}
g^{3/2^\pm}_4(W,Q^2) = {1\over Q^2+m_\pi^2} g^{3/2^\pm}_3(W,Q^2) \ .
\end{eqnarray}
Considering $g_3$ and $g_4$ terms,
the helicity amplitudes for $3/2^+ N^*$ are 
\begin{eqnarray}
A_{3/2^+}(q)\cdot \epsilon^{(t)} &=&   -\sqrt{2\over 3} 
|\bm{q}| q^0\, g^{3/2^+}_4(W,Q^2) 
\ , 
\nonumber \\
A_{3/2^+}(q)\cdot \epsilon^{(0)} &=& -\sqrt{2\over 3} 
\left[g^{3/2^+}_3(W,Q^2) - |\bm{q}|^2\, g^{3/2^+}_4(W,Q^2) \right] \ , \nonumber \\
A_{3/2^+}(q)\cdot \epsilon^{(+1)} &=&
\left\{
\begin{array}{l}
- g^{3/2^+}_3(W,Q^2)\qquad\quad (s_N^z=+1/2) \ , \\
 -{1\over \sqrt{3}}  g^{3/2^+}_3(W,Q^2) \qquad (s_N^z=-1/2) \ ,
\end{array}
\right.
\end{eqnarray}
and for $3/2^- N^*$,
\begin{eqnarray}
A_{3/2^-}(q)\cdot \epsilon^{(t)} &=& 
\pm \sqrt{2\over 3} {|{\bm{q}}|^2 q^0 \over 2
m_N}  g^{3/2^-}_4(W,Q^2) \ , 
\nonumber \\
A_{3/2^-}(q)\cdot \epsilon^{(0)} &=& \pm \sqrt{2\over 3} {|{\bm{q}}| \over 2
m_N} 
\left[ g^{3/2^-}_3(W,Q^2) - |\bm{q}|^2\, g^{3/2^-}_4(W,Q^2) \right] \ , \nonumber \\
A_{3/2^-}(q)\cdot \epsilon^{(+1)} &=&
\left\{
\begin{array}{l}
 {|{\bm{q}}| \over 2 m_N} g^{3/2^-}_3(W,Q^2)
\qquad\qquad (s_N^z=+1/2) \ , \\
- {1\over \sqrt{3}}   {|{\bm{q}}| \over 2 m_N} g^{3/2^-}_3(W,Q^2)
\qquad (s_N^z=-1/2) \ ,
\end{array}
\right.
\end{eqnarray}
where the sign $\pm$ is for $s_N^z=\pm 1/2$.

\subsection{Spin 5/2 $N^*$}

The matrix element of axial vector current between the nucleon and
$5/2^{\pm}$ $N^*$ 
[$A_{J^\pm_{N^*}\, \mu}(q)$ in Eq.~(\ref{eq:axial-me})]
is generally given by 
\begin{eqnarray}
A_{5/2^\pm\, \mu}(q) &=& \bar u^{\alpha\beta}_{5/2^\pm}(\bm 0) q_\beta
\left[
g^{5/2^\pm}_1(Q^2)( g_{\alpha\mu}\Slash{q}-q_\alpha\gamma_\mu)
+g^{5/2^\pm}_2(Q^2)( g_{\alpha\mu}q\cdot p - q_\alpha p_\mu)
\right.
\nonumber\\
&&\left.
+g^{5/2^\pm}_3(Q^2) g_{\alpha\mu}
+g^{5/2^\pm}_4(Q^2) q_\alpha q_\mu
\right]
\binom{\gamma_5}{\unitfour} u_N(-{\bm q}) \ ,
\end{eqnarray}
where $g^{5/2^\pm}_n(Q^2)\ (n=1...4)$ are form factors.
The matrix element of the divergence of the axial current is
\begin{eqnarray}
q\cdot A_{5/2^\pm}(q) = \bar u^{\alpha\beta}_{5/2^\pm}(\bm 0) q_\beta
\left[
g^{5/2^\pm}_3(Q^2) q_\alpha
+g^{5/2^\pm}_4(Q^2) q_\alpha q^2
\right]
\binom{\gamma_5}{\unitfour}
u_N(-{\bm q}) \ .
\end{eqnarray}
Taking $Q^2=-m^2_\pi$ and dropping the small 
$g^{5/2^{\pm}}_4(Q^2) q_\alpha q^2$ term, we obtain
\begin{eqnarray}
q\cdot A_{{\rm NP},5/2^+}( q)|_{Q^2=-m^2_\pi} &=& \mp \sqrt{2\over 5} {|{\bm{q}}|^3\over 2 m_N}
 g^{5/2^+}_3(Q^2=-m^2_\pi)\ ,
 \nonumber \\
q\cdot A_{{\rm NP},5/2^-}( q)|_{Q^2=-m^2_\pi}   &=&  \sqrt{2\over 5}  |{\bm{q}}|^2 g^{5/2^-}_3(Q^2=-m^2_\pi)\ .
\end{eqnarray}
Comparing this with Eqs.~(\ref{eq:pcac_pos}) and (\ref{eq:pcac_neg}),
we find
\begin{eqnarray}
g^{5/2^+}_3(W,Q^2=-m^2_\pi) &=&
- \sqrt{15\over 8\pi} {2 m_N f_\pi\over |{\bm{q}}|^3}
  g^{5/2^+}_{\pi NN^*}({\bm{q}}) \ ,
\nonumber\\
g^{5/2^-}_3(W,Q^2=-m^2_\pi) &=&
- \sqrt{15\over 8\pi} {f_\pi\over |{\bm{q}}|^2} 
g^{5/2^-}_{\pi NN^*}({\bm{q}}) \ ,
\end{eqnarray}
where we have introduced the $W$-dependence in the form factors.
The PCAC hypothesis dictates
$g^{5/2^\pm}_3(W,Q^2=-m^2_\pi)\sim g^{5/2^\pm}_3(W,Q^2=0)$.
From the PCAC and the pion dominance, we have a pion-pole term:
\begin{eqnarray}
g^{5/2^\pm}_4(W,Q^2) = {1\over Q^2+m_\pi^2} g^{5/2^\pm}_3(W,Q^2) \ .
\end{eqnarray}
Considering $g_3$ and $g_4$ terms,
the helicity amplitudes for $5/2^+ N^*$ are 
\begin{eqnarray}
A_{5/2^+}(q)\cdot \epsilon^{(t)} &=& \mp \sqrt{2\over 5} q^0 
{|{\bm{q}}|^3 \over 2 m_N} g^{5/2^+}_4(W,Q^2)
 \ , 
\nonumber \\
A_{5/2^+}(q)\cdot \epsilon^{(0)} &=& 
 \mp \sqrt{2\over 5} 
{|{\bm{q}}|^2 \over 2 m_N} \left[g^{5/2^+}_3(W,Q^2) - |\bm{q}|^2\, g^{5/2^+}_4(W,Q^2) \right]
\ , \nonumber \\
A_{5/2^+}(q)\cdot \epsilon^{(+1)} &=&
\left\{
\begin{array}{l}
 -\sqrt{2\over 5} {|{\bm{q}}|^2 \over 2 m_N} g^{5/2^+}_3(W,Q^2)
\qquad (s_N^z=+1/2) \ , \\
 {1\over \sqrt{5} }
{|{\bm{q}}|^2 \over 2 m_N} g^{5/2^+}_3(W,Q^2)
\quad\qquad (s_N^z=-1/2) \ ,
\end{array}
\right.
\end{eqnarray}
where the sign $\mp$ is for $s_N^z=\pm 1/2$, 
and for $5/2^- N^*$,
\begin{eqnarray}
A_{5/2^-}(q)\cdot \epsilon^{(t)} &=& \sqrt{2\over 5} q^0 
|\bm{q}|^2 g^{5/2^-}_4(W,Q^2)
 \ , 
\nonumber \\
A_{5/2^-}(q)\cdot \epsilon^{(0)} &=& 
 \sqrt{2\over 5} |{\bm{q}}| \left[g^{5/2^-}_3(W,Q^2) - |\bm{q}|^2\, g^{5/2^-}_4(W,Q^2)\right]
 \ , \nonumber \\
A_{5/2^-}(q)\cdot \epsilon^{(+1)} &=&
\left\{
\begin{array}{l}
 \sqrt{2\over 5} 
 |{\bm{q}}| g^{5/2^-}_3(W,Q^2)
\qquad (s_N^z=+1/2) \ , \\
{1\over  \sqrt{5}}
|{\bm{q}}| g^{5/2^-}_3(W,Q^2)
\qquad (s_N^z=-1/2) \ .
\end{array}
\right.
\end{eqnarray}

\subsection{Spin 7/2 $N^*$}

The matrix element of axial vector current between the nucleon and
$7/2^{\pm}$ $N^*$ 
[$A_{J^\pm_{N^*}\, \mu}(q)$ in Eq.~(\ref{eq:axial-me})]
is generally given by 
\begin{eqnarray}
A_{7/2^\pm\, \mu}(q) &=& \bar u^{\alpha\beta\gamma}_{7/2^\pm}(\bm 0) q_\beta q_\gamma
\left[
g^{7/2^\pm}_1(Q^2)( g_{\alpha\mu}\Slash{q}-q_\alpha\gamma_\mu)
+g^{7/2^\pm}_2(Q^2)( g_{\alpha\mu}q\cdot p - q_\alpha p_\mu)
\right.\nonumber \\
&&\left.+g^{7/2^\pm}_3(Q^2) g_{\alpha\mu}
+g^{7/2^\pm}_4(Q^2) q_\alpha q_\mu
\right]
\binom{\unitfour}{\gamma_5} u_N(-{\bm q}) \ ,
\end{eqnarray}
where $g^{7/2^\pm}_n(Q^2)\ (n=1...4)$ are form factors.
The matrix element of the divergence of the axial current is
\begin{eqnarray}
q\cdot A_{7/2^\pm}(q) = \bar u^{\alpha\beta\gamma}_{7/2^\pm}(\bm 0) q_\beta q_\gamma
\left[
g^{7/2^\pm}_3(Q^2) q_\alpha
+g^{7/2^\pm}_4(Q^2) q_\alpha q^2
\right]
\binom{\unitfour}{\gamma_5}
u_N(-{\bm q})
 \ .
\nonumber\\
\end{eqnarray}
Taking $Q^2=-m^2_\pi$ and dropping the small 
$g^{7/2^{\pm}}_4(Q^2) q_\alpha q^2$ term, we obtain
\begin{eqnarray}
q\cdot A_{{\rm NP},7/2^+}( q)|_{Q^2=-m^2_\pi}  &=&  - \sqrt{8\over 35} |{\bm{q}}|^3
 g^{7/2^+}_3(Q^2=-m^2_\pi)\ , \\
q\cdot A_{{\rm NP},7/2^-}( q)|_{Q^2=-m^2_\pi} &=&  \pm \sqrt{8\over 35}  {|{\bm{q}}|^4\over 2 m_N}  g^{7/2^-}_3(Q^2=-m^2_\pi)\ . 
\end{eqnarray}
Comparing this with Eqs.~(\ref{eq:pcac_pos}) and (\ref{eq:pcac_neg}),
we find
\begin{eqnarray}
g^{7/2^+}_3 (W,Q^2=-m^2_\pi)&=&
 \sqrt{35\over 8\pi} {f_\pi\over |{\bm{q}}|^3}
g^{7/2^+}_{\pi NN^*}({\bm q})
\ ,
\nonumber\\
g^{7/2^-}_3(W,Q^2=-m^2_\pi) &=&
 \sqrt{35\over 8\pi} {2 m_N f_\pi\over |{\bm{q}}|^4}
g^{7/2^-}_{\pi NN^*}({\bm q}) \ ,
\end{eqnarray}
where we have introduced the $W$-dependence in the form factors.
The PCAC hypothesis dictates
$g^{7/2^\pm}_3(W,Q^2=-m^2_\pi)\sim g^{7/2^\pm}_3(W,Q^2=0)$.
From the PCAC and the pion dominance, we have a pion-pole term:
\begin{eqnarray}
g^{7/2^\pm}_4(W,Q^2) = {1\over Q^2+m_\pi^2} g^{7/2^\pm}_3(W,Q^2) \ .
\end{eqnarray}
Considering $g_3$ and $g_4$ terms,
the helicity amplitudes for $7/2^+$ are 
\begin{eqnarray}
A_{7/2^+}(q)\cdot \epsilon^{(t)} &=& -\sqrt{8\over 35} q^0 
|{\bm{q}}|^3 g^{7/2^+}_4(W,Q^2)
 \ , 
\nonumber \\
A_{7/2^+}(q)\cdot \epsilon^{(0)} &=& 
- \sqrt{8\over 35} |{\bm{q}}|^2
\left[g^{7/2^+}_3(W,Q^2) - |\bm{q}|^2\,
g^{7/2^+}_4(W,Q^2) \right]
 \ , \nonumber \\
A_{7/2^+}(q)\cdot \epsilon^{(+1)} &=&
\left\{
\begin{array}{l}
-{2\over \sqrt{21}}
|{\bm{q}}|^2 g^{7/2^+}_3 (W,Q^2)
\qquad (s_N^z=+1/2) \ , \\
-{2\over \sqrt{35} }
 |{\bm{q}}|^2 g^{7/2^+}_3 (W,Q^2)
\qquad (s_N^z=-1/2) \ ,
\end{array}
\right.
\end{eqnarray}
and for $7/2^-$ (the sign $\pm$ is for $s_N^z=\pm 1/2$),
\begin{eqnarray}
A_{7/2^-}(q)\cdot \epsilon^{(t)} &=& \pm \sqrt{8\over 35} q^0 
{|{\bm{q}}|^4 \over 2 m_N} g^{7/2^-}_4(W,Q^2)
 \ , 
\nonumber \\
A_{7/2^-}(q)\cdot \epsilon^{(0)} &=& 
 \pm \sqrt{8\over 35} 
{|{\bm{q}}|^3 \over 2 m_N} 
\left[g^{7/2^-}_3(W,Q^2) - |\bm{q}|^2\, g^{7/2^-}_4(W,Q^2) \right]
\ , \nonumber \\
A_{7/2^-}(q)\cdot \epsilon^{(+1)} &=&
\left\{
\begin{array}{l}
{2\over \sqrt{21} }
{|{\bm{q}}|^3 \over 2 m_N} g^{7/2^-}_3(W,Q^2)
\quad\qquad (s_N^z=+1/2) \ , \\
 - {2\over \sqrt{35} }
{|{\bm{q}}|^3 \over 2 m_N} g^{7/2^-}_3(W,Q^2)
\qquad (s_N^z=-1/2) \ .
\end{array}
\right.
\end{eqnarray}

\section{Bare $N^*$ vector current matrix elements}
\label{app:vector-nstar}

Following the formulation of Ref.~\cite{knls13}, we parametrize a bare
$\gamma^{(*)} N \to N^*$ vertex matrix element
in the helicity representation as
\begin{eqnarray}
\Gamma_{N^\ast,\gamma^{(*)} N}(q) = \frac{1}{(2\pi)^{3/2}}
\sqrt{4\pi\over 2 J_{N^*}+1}
\sqrt{\frac{m_N}{E_N(-\bm q)}}\sqrt{\frac{q_R}{|q_0|}} G^{NN^\ast}_\lambda(q)
\delta_{\lambda,(\lambda_\gamma-\lambda_N)},
\label{eq:nstar-gn}
\end{eqnarray}
where $\lambda_\gamma$ and $\lambda_N$ are the helicity quantum numbers of
the (virtual) photon and the nucleon, respectively, 
and $q_R$ and $q_0$ are defined by $M_{N^\ast}=q_R+E_N(\bm q_R)$ and
$W=q_0+E_N(-\bm q)$, respectively. 
For $N^*$ of $I=1/2$,
this vertex can be separated into the isovector ($\Gamma^{IV}$)
 and isoscalar ($\Gamma^{IS}$) parts as follows:
\begin{eqnarray}
\Gamma^{IV}_{N^\ast,V N}  &=& 
{1\over 2}
{\Gamma_{N^\ast,\gamma^{(*)} p} - \Gamma_{N^\ast,\gamma^{(*)} n} \over
 (10{1\over 2}{1\over 2}|{1\over 2}{1\over 2}) } \ , \\
\Gamma^{IS}_{N^\ast,V N}  &=& 
{\Gamma_{N^\ast,\gamma^{(*)} p} + \Gamma_{N^\ast,\gamma^{(*)} n} \over 2}
\ ,
\end{eqnarray}
where $(10{1\over 2}{1\over 2}|{1\over 2}{1\over 2})$ is the isospin
Clebsch-Gordan coefficient.
Regarding $N^*$ of $I=3/2$, they have only the isovector part as given by
\begin{eqnarray}
\Gamma^{IV}_{N^\ast,V N}  &=& 
{\Gamma_{N^\ast,\gamma^{(*)} p} \over
 (10{1\over 2}{1\over 2}|{3\over 2}{1\over 2}) }
={\Gamma_{N^\ast,\gamma^{(*)} n} \over
 (10{1\over 2}{1\over 2}|{3\over 2}{1\over 2}) } \ , 
\end{eqnarray}
The quantities introduced above 
($\Gamma^{IV}_{N^\ast,V N}$, $\Gamma^{IS}_{N^\ast,V N}$)
correspond to the vector part of 
$\Gamma_{N^\ast,J N}$ in Eq.~(\ref{eq:nstargn}).
The helicity amplitudes $G^{NN^*}_\lambda(q)$ in
Eq.~(\ref{eq:nstar-gn}) are 
\begin{eqnarray}
G^{NN^*}_\lambda(q)&=&A^{NN^*}_\lambda(q)\qquad {\rm for\ transverse\ current,}\\
&=&S^{NN^*}_\lambda(q)\qquad {\rm for\ longitudinal\ current.}
\end{eqnarray}
The helicity amplitudes can be
written with the multipole amplitudes of
the vector $N$-$N^*$ transition such as 
$E^{NN^\ast}_{l\pm}$,
$M^{NN^\ast}_{l\pm}$ and $S^{NN^\ast}_{l\pm}$
[${l\pm}$ is related to the spin ($J_{N^*}$) and parity ($P$)
of $N^*$ by $J_{N^*}=l\pm 1/2$ and $P=(-1)^{l+1}$]
as
\begin{eqnarray}
A^{NN^*}_{3/2}(q)&=& \frac{\sqrt{l(l+2)}}{2}[-M^{NN^\ast}_{l+}(q) + E^{NN^\ast}_{l+}(q)],
\label{eq:a32-me+}
\\
A^{NN^*}_{1/2}(q)&=&-\frac{1}{2}[lM^{NN^\ast}_{l+}(q) + (l+2) E^{NN^\ast}_{l+}(q)],
\label{eq:a12-me+}
\\
S^{NN^*}_{1/2}(q)&=&S^{NN^\ast}_{l+}(q),
\end{eqnarray}
for $J_{N^*} = l + 1/2$, and
\begin{eqnarray}
A^{NN^*}_{3/2}(q)&=& -\frac{\sqrt{(l-1)(l+1)}}{2}[M^{NN^\ast}_{l-}(q) + E^{NN^\ast}_{l-}(q)],
\label{eq:a32-me-}
\\
A^{NN^*}_{1/2}(q)&=&+\frac{1}{2}[(l+1)M^{NN^\ast}_{l-}(q) - (l-1) E^{NN^\ast}_{l-}(q)],
\label{eq:a12-me-}
\\
S^{NN^*}_{1/2}(q)&=&S^{NN^\ast}_{l-}(q),
\end{eqnarray}
for $J_{N^*} = l - 1/2$. 
The multipole amplitudes are parametrized as
\begin{eqnarray}
M^{NN^\ast}_{l\pm}(q)&=& \left(\frac{q}{m_\pi}\right)^l
\left(\frac{(\Lambda^{\text{e.m.}}_{N^\ast})^2 + m_\pi^2}{(\Lambda_{N^\ast}^{\text{e.m.}})^2 + q^2}\right)^{(2+l/2)} \tilde M^{NN^\ast}_{l\pm}(Q^2),
\label{eq:M-para}
\\
E^{NN^\ast}_{l\pm}(q)&=& \left(\frac{q}{m_\pi}\right)^{(l\pm 1)}
\left(\frac{(\Lambda^{\text{e.m.}}_{N^\ast})^2 + m_\pi^2}{(\Lambda_{N^\ast}^{\text{e.m.}})^2 + q^2}\right)^{[2+(l\pm 1)/2]} \tilde E^{NN^\ast}_{l\pm}(Q^2),
\label{eq:E-para}
\\
S^{NN^\ast}_{l\pm}(q)
&=& \left(\frac{q}{m_\pi}\right)^{(l\pm 1)}
\left(\frac{(\Lambda^{\text{e.m.}}_{N^\ast})^2 +
 m_\pi^2}{(\Lambda_{N^\ast}^{\text{e.m.}})^2 + q^2}\right)^{[2+(l\pm
1)/2]} \tilde S^{NN^\ast}_{l\pm}(Q^2) \ .
\label{eq:long-amp}
\end{eqnarray}
One exception for the above parametrization is applied to 
the first bare state in $P_{33}$ for which we use the following forms:
\begin{eqnarray}
A^{\text{1st}P_{33}}_{3/2}(q) &=& -x_{A_{3/2}}(Q^2)\frac{\sqrt{3}}{2} A\left[G^{\rm SL}_{M}(Q^2) - (1-N)G_E^{\rm SL}(Q^2)\right] ,
\label{eq:1p33-a32}
\\
A^{\text{1st}P_{33}}_{1/2}(q) &=&-x_{A_{1/2}}(Q^2)\frac{1       }{2} A\left[G^{\rm SL}_{M}(Q^2) - (1+N)G_E^{\rm SL}(Q^2)\right] ,
\label{eq:1p33-a12}
\\
S^{\text{1st}P_{33}}_{1/2}(q) &=& x_{S_{1/2}}(Q^2) B\, G^{\rm SL}_{C}(Q^2)
,
\label{eq:1p33-s12}
\end{eqnarray}
with
\begin{eqnarray}
A   &=& 
e W |\bm{q}| \left(m_\Delta+m_N\over m_N\right)
\sqrt{ {1\over 2 q_R} {E_N(q_R)\over m_N}
{E_N(-\bm{q})+m_N\over E_N(-\bm{q})}}
{1\over (m_\Delta+m_N)^2+Q^2}\ ,
\\
N   &=&
{4 W |\bm{q}|^2\over E_N(-\bm{q})+m_N} {1\over (m_\Delta-m_N)^2+Q^2}\ ,
\\
B   &=& 
-e
\sqrt{ {1\over 2 q_R} {E_N(q_R)\over m_N}}
{4W |\bm{q}|^4\over \sqrt{2E_N(-\bm{q}) [E_N(-\bm{q})+m_N]}}
\left( m_\Delta+m_N \over 2 m_N \right)
\nonumber \\
&&\times
{1\over [(m_\Delta+m_N)^2+Q^2] [(m_\Delta-m_N)^2+Q^2]}
\ ,
\\
G_\xi(Q^2) &=&
 G_\xi(0)
\left( \frac{1}{1 + Q^2/0.71({\rm GeV}/c)^2} \right)^2 
(1+aQ^2) \exp(-bQ^2), 
\end{eqnarray}
with $\xi=M,E,C$ and
$G^{\rm SL}_M(0) = 1.85$, $G^{\rm SL}_E(0) = 0.025$, $G^{\rm SL}_C(0) = -0.238$~\cite{sl,sl2};
$a=0.154$ and $b=0.166$ (GeV/c)$^{-2}$~\cite{sl2}.
The cutoff $\Lambda^{\text{e.m.}}_{N^\ast}$ and the coupling constants
$\tilde M_{l\pm}^{NN^*}(Q^2)$, $\tilde E_{l\pm}^{NN^*}(Q^2)$ and $\tilde S_{l\pm}^{NN^*}(Q^2)$
in Eqs.~(\ref{eq:M-para})-(\ref{eq:long-amp}), as well as 
the factors $x_{A_{3/2}}(Q^2)$, $x_{A_{1/2}}(Q^2)$ and $x_{S_{1/2}}(Q^2)$
in Eqs.~(\ref{eq:1p33-a32})-(\ref{eq:1p33-s12})
are determined by fitting data.
In Ref.~\cite{knls13}, we have done a combined analysis of 
$\pi N,\gamma p\to \pi N, \eta N, K\Lambda, K\Sigma$ reaction data, 
and fixed these parameters at $Q^2=0$ for the proton target. 
The numerical values for these parameters are also presented in the reference.

\section{Parameters $c^N_n$ in Eq.~(\ref{eq:q2-fit})}
\label{app:cn}

\begin{table}[t]
\caption{\label{tab:param-cn} 
Parameters $c^p_n$ defined in Eq.~(\ref{eq:q2-fit})
for the proton target.
The first column shows bare $N^*$ specified by their partial waves;
(1) [(2)] indicates the first [second] bare state in the specified partial wave.
For 
${ F}^{V}_{pN^*}(Q^2)$=$\tilde M^{N^*}_{l\pm}(Q^2)$,
$\tilde E^{N^*}_{l\pm}(Q^2)$, and
$\tilde S^{N^*}_{l\pm}(Q^2)$,
the unit for a parameter 
$c^p_n$ is [$10^{-3}$ GeV$^{-1/2}$ (GeV/$c$)$^{-2n}$].
Only for $P_{33}(1)$ for which we present 
${ F}^{V}_{pN^*}(Q^2)$=
$x_{A_{3/2}}(Q^2)$, $x_{A_{1/2}}(Q^2)$, and $x_{S_{1/2}}(Q^2)$ from the
 left to right,
the unit for a parameter 
$c^p_n$ is [(GeV/$c$)$^{-2n}$].
}    
\begin{ruledtabular}
\tiny
\renewcommand{\arraystretch}{1.2}
\begin{tabular}{r|rrrrrr|rrrrrr|rrrrrr}
${ F}^{V}_{pN^*}$           & \multicolumn{6}{c|}{$\tilde M^{N^*}_{l\pm}(Q^2)$}& \multicolumn{6}{c|}{$\tilde E^{N^*}_{l\pm}(Q^2)$}& \multicolumn{6}{c}{$\tilde S^{N^*}_{l\pm}(Q^2)$}\\
           &  $c^p_0$ &  $c^p_1$ &  $c^p_2$ &  $c^p_3$ &  $c^p_4$ &  $c^p_5$ &  $c^p_0$ &  $c^p_1$ &  $c^p_2$ &  $c^p_3$ &  $c^p_4$ &  $c^p_5$ &  $c^p_0$ &  $c^p_1$ &  $c^p_2$ &  $c^p_3$ &  $c^p_4$ &  $c^p_5$ \\\hline
$S_{11}(1)$&   ---  &   ---  &   ---  &   ---  &   ---  &   ---  &    -47.&   -429.&    754.&   -648.&    237.&    -31.&    -0.5&     9.0&   -22.3&    21.5&    -8.3&     1.1\\
$S_{11}(2)$&   ---  &   ---  &   ---  &   ---  &   ---  &   ---  &    -13.&    113.&   -182.&    141.&    -50.&      7.&    0.29&   -2.55&   -0.74&    3.64&   -1.91&    0.29\\
$P_{11}(1)$&      4.&   -101.&     60.&     -7.&     -1.&     ---&   ---  &   ---  &   ---  &   ---  &   ---  &   ---  &    -0.7&     6.6&   -16.0&    13.0&    -4.6&     0.6\\
$P_{11}(2)$&     66.&   -303.&    530.&   -417.&    143.&    -18.&   ---  &   ---  &   ---  &   ---  &   ---  &   ---  &     -4.&   -455.&    307.&    -86.&      8.&     ---\\
$P_{11}(3)$&    -28.&    161.&   -873.&    971.&   -420.&     62.&   ---  &   ---  &   ---  &   ---  &   ---  &   ---  &    13.4&   -32.0&    35.4&   -18.8&     4.7&    -0.4\\
$P_{13}(1)$&     29.&   -133.&   -498.&    629.&   -240.&     30.&    -11.&   -158.&    437.&   -385.&    136.&    -17.&     1.7&     7.5&    11.2&   -23.0&    11.0&    -1.6\\
$P_{13}(2)$&    -57.&    339.&   -689.&    544.&   -190.&     24.&   -101.&    802.&  -2074.&   1799.&   -638.&     80.&    -24.&    203.&   -182.&     59.&     -6.&     ---\\
$D_{13}(1)$&     20.&    -38.&    205.&   -176.&     59.&     -7.&     44.&   -503.&   1526.&  -1253.&    414.&    -49.&     36.&     28.&   -246.&    429.&   -210.&     31.\\
$D_{13}(2)$&    -2.8&    15.3&   -45.9&    42.0&   -15.1&     1.9&     -7.&   -195.&    420.&   -348.&    121.&    -15.&     40.&   -103.&    131.&    -59.&      9.&     ---\\
$D_{15}(1)$&    -2.7&    15.6&    -3.7&     9.0&    -5.5&     0.9&     2.8&   -18.0&    33.1&   -24.9&     7.8&    -0.9&    19.4&   -65.2&    98.6&   -67.6&    20.8&    -2.4\\
$D_{15}(2)$&    -19.&   -365.&    920.&   -803.&    294.&    -38.&     -4.&    -34.&    103.&    -94.&     35.&     -5.&     3.2&   -24.3&    53.9&   -48.0&    17.1&    -2.1\\
$F_{15}(1)$&    1.56&   -5.27&    7.86&   -5.42&    1.71&   -0.20&     1.6&     0.7&    17.0&   -24.9&    11.1&    -1.6&     4.0&   -13.5&    21.8&   -10.6&     1.6&     ---\\
$F_{17}(1)$&    0.75&   -4.30&    5.53&   -2.76&    0.44&     ---&   -0.13&    0.67&   -0.73&    0.26&   -0.03&     ---&    1.22&   -0.24&   -1.74&    1.73&   -0.61&    0.07\\
$G_{17}(1)$&     3.0&   -27.6&    32.2&   -13.5&     1.9&     ---&     3.6&   -16.7&    20.7&    -9.5&     1.4&     ---&    12.2&   -32.9&    34.3&   -14.1&     2.0&     ---\\
$G_{19}(1)$&    -0.3&   -21.4&    17.5&    -7.1&     1.1&     ---&   -0.11&    2.78&   -3.60&    1.70&   -0.25&     ---&    0.72&   -1.73&    3.58&   -1.95&    0.32&     ---\\
$H_{19}(1)$&    0.06&     ---&     ---&     ---&     ---&     ---&   -0.06&     ---&     ---&     ---&     ---&     ---&     ---&     ---&     ---&     ---&     ---&     ---\\
$S_{31}(1)$&   ---  &   ---  &   ---  &   ---  &   ---  &   ---  &    260.&  -1903.&   4324.&  -3470.&   1189.&   -146.&   -0.61&    2.67&   -3.59&    8.07&   -4.36&    0.69\\
$S_{31}(2)$&   ---  &   ---  &   ---  &   ---  &   ---  &   ---  &   -256.&  -6013.&  17817.& -17166.&   6567.&   -872.&    271.&   -181.&    -88.&    175.&    -78.&     11.\\
$P_{31}(1)$&     22.&    336.&   -673.&    499.&   -161.&     19.&   ---  &   ---  &   ---  &   ---  &   ---  &   ---  &   -2.12&   -2.01&    0.79&   -0.13&     ---&     ---\\
$P_{31}(2)$&    -73.& -11741.&  30223.& -27142.&   9778.&  -1236.&   ---  &   ---  &   ---  &   ---  &   ---  &   ---  &     97.&   -393.&    754.&   -560.&    180.&    -21.\\
$P_{33}(1)$&    0.99&    0.54&   -0.40&    0.16&   -0.03&     ---&    1.22&    1.43&   -3.62&    3.51&   -1.43&    0.20&    -0.2&     3.5&   -12.9&    13.1&    -5.0&     0.7\\
$P_{33}(2)$&     94.&   -352.&    576.&   -435.&    149.&    -19.&   -1.22&   -1.58&    6.42&   -5.28&    1.79&   -0.22&   -0.25&    2.98&   -1.63&    0.27&     ---&     ---\\
$D_{33}(1)$&    -3.3&   -28.3&    69.3&   -55.1&    18.3&    -2.2&    19.9&    29.7&   -74.1&    67.2&   -28.3&     4.3&    0.21&   -1.60&    3.51&   -3.06&    1.10&   -0.14\\
$D_{33}(2)$&      0.&   -192.&    486.&   -448.&    167.&    -22.&    109.&    -93.&     12.&    178.&   -100.&     15.&    184.&   -108.&   -246.&    433.&   -191.&     26.\\
$D_{35}(1)$&   -3.13&   -1.06&    4.01&   -0.62&   -0.09&     ---&   -0.59&    0.39&    0.31&   -0.26&    0.05&     ---&    4.83&   -6.69&    0.16&    3.56&   -1.77&    0.25\\
$D_{35}(2)$&   -10.3&    -1.6&    -2.3&     4.0&    -1.0&     ---&   -9.01&    8.03&   -8.87&    4.70&   -0.80&     ---&     39.&    143.&    -44.&      1.&      1.&     ---\\
$F_{35}(1)$&   -0.44&    1.41&   -1.29&    0.49&   -0.06&     ---&   -0.73&    0.52&   -0.07&   -0.25&    0.07&     ---&     5.6&     6.0&   -10.9&     4.7&    -0.6&     ---\\
$F_{35}(2)$&    -0.5&   -26.1&    42.1&   -21.2&     3.4&     ---&    40.1&   -76.4&    69.4&   -27.3&     3.8&     ---&    0.02&    0.11&   -0.43&    0.16&   -0.02&     ---\\
$F_{37}(1)$&    0.48&   -2.84&    4.59&   -3.15&    0.98&   -0.11&    0.02&    0.06&   -0.10&    0.05&   -0.01&     ---&   -0.04&    0.20&   -0.27&    0.13&   -0.02&     ---\\
$F_{37}(2)$&    11.2&    34.7&   -16.2&    12.5&    -3.2&     ---&    -1.8&   -57.3&    83.5&   -43.6&     7.2&     ---&     6.2&   -17.8&    26.6&   -12.2&     1.8&     ---\\
$G_{37}(1)$&   -0.30&    0.63&   -0.58&    0.42&   -0.09&     ---&    0.89&   -2.53&    1.95&   -0.65&    0.08&     ---&     3.1&    -7.2&    11.9&    -5.2&     0.7&     ---\\
$G_{39}(1)$&     0.2&    -9.2&    10.2&    -4.2&     0.6&     ---&   -0.03&   -0.37&   -0.45&    0.52&   -0.12&     ---&    0.89&   -1.48&    1.78&   -1.24&    0.40&   -0.05\\
$H_{39}(1)$&   -0.09&    1.44&   -2.16&    1.07&   -0.17&     ---&   -0.22&   -1.44&    2.16&   -1.07&    0.17&     ---&     ---&     ---&     ---&     ---&     ---&     ---\\
\end{tabular}
\end{ruledtabular}
\end{table}

\begin{table}[t]
\caption{\label{tab:param-cn-n} 
Parameters $c^n_n$ defined in Eq.~(\ref{eq:q2-fit}) for the neutron target.
The parameters for $I$=1/2 $N^*$ are presented while those
for $I$=3/2 $N^*$ are the same as those for the proton target.
The other features are the same as those in TABLE~\ref{tab:param-cn}. 
}    
\begin{ruledtabular}
\tiny
\renewcommand{\arraystretch}{1.2}
\begin{tabular}{r|rrrrrr|rrrrrr|rrrrrr}
${ F}^{V}_{nN^*}$           & \multicolumn{6}{c|}{$\tilde M^{N^*}_{l\pm}(Q^2)$}& \multicolumn{6}{c|}{$\tilde E^{N^*}_{l\pm}(Q^2)$}& \multicolumn{6}{c}{$\tilde S^{N^*}_{l\pm}(Q^2)$}\\
           &  $c^n_0$ &  $c^n_1$ &  $c^n_2$ &  $c^n_3$ &  $c^n_4$ &  $c^n_5$ &  $c^n_0$ &  $c^n_1$ &  $c^n_2$ &  $c^n_3$ &  $c^n_4$ &  $c^n_5$ &  $c^n_0$ &  $c^n_1$ &  $c^n_2$ &  $c^n_3$ &  $c^n_4$ &  $c^n_5$ \\\hline
$S_{11}(1)$&   ---  &   ---  &   ---  &   ---  &   ---  &   ---  &     28.&    730.&  -1439.&   1092.&   -344.&     39.&    180.&   -944.&   1672.&  -1067.&    279.&    -25.\\
$S_{11}(2)$&   ---  &   ---  &   ---  &   ---  &   ---  &   ---  &     38.&   -250.&    456.&   -343.&    114.&    -14.&     16.&    -67.&    105.&    -69.&     20.&     -2.\\
$P_{11}(1)$&   -30.2&    25.3&    42.9&   -59.2&    23.0&    -2.9&   ---  &   ---  &   ---  &   ---  &   ---  &   ---  &    -13.&    156.&   -282.&    188.&    -47.&      4.\\
$P_{11}(2)$&   -118.&    222.&   -345.&    256.&    -83.&     10.&   ---  &   ---  &   ---  &   ---  &   ---  &   ---  &   -11.9&    34.5&   -42.4&    23.1&    -5.7&     0.5\\
$P_{11}(3)$&    107.&   -527.&    939.&   -701.&    230.&    -27.&   ---  &   ---  &   ---  &   ---  &   ---  &   ---  &     2.4&   -32.7&    -1.1&    27.8&   -13.8&     1.9\\
$P_{13}(1)$&     -2.&    -76.&    126.&    -98.&     36.&     -5.&      5.&    -85.&    115.&    -78.&     27.&     -4.&   -13.2&    33.8&   -32.4&    13.9&    -2.9&     0.2\\
$P_{13}(2)$&    124.&    152.&    -30.&    -26.&      7.&     -0.&     34.&     70.&   -186.&    116.&    -21.&     -0.&    25.4&   -20.7&   -21.3&    39.6&   -16.9&     2.3\\
$D_{13}(1)$&    -35.&    216.&   -521.&    392.&   -120.&     13.&    -21.&    343.&   -731.&    564.&   -180.&     21.&    139.&   -379.&    426.&   -113.&     -3.&      3.\\
$D_{13}(2)$&     20.&   -109.&    193.&   -139.&     44.&     -5.&    -14.&     90.&   -164.&    119.&    -38.&      4.&     6.6&   -48.5&    70.6&   -42.6&    11.6&    -1.2\\
$D_{15}(1)$&     22.&   -117.&    208.&   -154.&     50.&     -6.&     0.9&    -8.0&    17.4&   -13.3&     4.3&    -0.5&     6.1&   -24.6&    35.3&   -22.9&     6.9&    -0.8\\
$D_{15}(2)$&    -30.&    154.&   -271.&    193.&    -60.&      7.&    -2.1&    11.3&   -20.0&    15.1&    -5.0&     0.6&    13.8&   -49.4&    67.6&   -42.8&    12.6&    -1.4\\
$F_{15}(1)$&    -1.8&    14.6&   -18.5&    10.0&    -2.7&     0.3&     0.6&   -10.6&    44.9&   -44.0&    16.2&    -2.0&    40.5&   -88.5&    48.1&    15.3&   -14.9&     2.5\\
$F_{17}(1)$&   -1.37&    1.45&    0.80&   -4.85&    2.84&   -0.46&     3.9&   -16.3&    26.9&   -17.3&     4.9&    -0.5&    -2.0&    -5.3&    16.0&   -14.0&     4.9&    -0.6\\
$G_{17}(1)$&    2.77&   -8.29&    7.62&   -6.36&    3.11&   -0.52&    -19.&    112.&   -202.&    147.&    -47.&      6.&   -4.32&    2.07&   -5.28&    5.77&   -2.16&    0.26\\
$G_{19}(1)$&    -2.7&    14.4&   -25.1&    18.0&    -5.7&     0.7&    0.13&    0.73&   -2.75&    2.06&   -0.54&    0.04&    16.8&   -46.8&    49.0&   -23.7&     6.0&    -0.6\\
$H_{19}(1)$&    0.84&   -3.22&    4.25&   -3.14&    1.17&   -0.16&    -3.3&    20.0&   -36.5&    26.5&    -8.3&     1.0&    10.7&   -28.9&    29.6&   -13.7&     3.4&    -0.4\\
\end{tabular}
\end{ruledtabular}
\end{table}

\begin{table}[t]
\caption{\label{tab:param-cutoff} 
Parameters $\Lambda_{N^*}^{\rm e.m.}$ defined in 
Eqs.~(\ref{eq:M-para})-(\ref{eq:long-amp})
for the neutron target.
The parameters for $I$=1/2 $N^*$ are presented while those
for $I$=3/2 $N^*$ are the same as those for the proton target and are
 given in Ref.~\cite{knls13}.
}    
\begin{ruledtabular}
\tiny
\renewcommand{\arraystretch}{1.2}
\begin{tabular}{rc}
&  $\Lambda_{N^*}^{\rm e.m.}$ (MeV) \\\hline
$S_{11}(1)$& 571.35 \\
$S_{11}(2)$& 1455.5 \\
$P_{11}(1)$& 1925.4 \\
$P_{11}(2)$& 1685.8 \\
$P_{11}(3)$& 1975.7 \\
$P_{13}(1)$& 902.49 \\
$P_{13}(2)$& 500.13 \\
$D_{13}(1)$& 647.46 \\
$D_{13}(2)$& 1053.7 \\
$D_{15}(1)$& 994.51 \\
$D_{15}(2)$& 815.99 \\
$F_{15}(1)$& 647.65 \\
$F_{17}(1)$& 503.27 \\
$G_{17}(1)$& 500.27 \\
$G_{19}(1)$& 500.31 \\
$H_{19}(1)$& 553.28 \\
\end{tabular}
\end{ruledtabular}
\end{table}

\end{document}